\documentclass[article]{./macros/elsarticle_qh}

\usepackage[hmarginratio=1:1,top=32mm,left=20mm,columnsep=1cm]{geometry} %
\usepackage{bm}
\usepackage{amsmath}
\usepackage{relsize} % e.g. used for \mathsmaller
\usepackage[svgnames]{xcolor} 
\usepackage{grffile} % allows to have dots in the name of graphic files
\usepackage{graphicx}
\usepackage{hyperref}
\hypersetup{
	colorlinks=true,                          
	linkcolor=DarkRed,
	citecolor=DarkRed,
	urlcolor=DarkRed  }

\usepackage{bm}
\usepackage{amsmath}
\usepackage{booktabs}
\usepackage{amssymb}
\usepackage{etoolbox}
\usepackage{mathtools}
\usepackage{gensymb}
\usepackage[section]{placeins} %Float Barriers
\usepackage{float}
\usepackage{subfig}
\usepackage[thinc]{esdiff}

\DeclareMathAlphabet{\mathsfbi}{OT1}{\sfdefault}{bx}{sl}
\DeclareMathVersion{sfletters}
\SetSymbolFont{letters}{sfletters}{OML}{ntxsfmi}{b}{it}

\makeatletter
\newcommand{\mathbfsbilow}[1]{%
	\text{\mathversion{sfletters}$\m@th#1$}%
}
\DeclareRobustCommand{\tensor}[1]{%
	\begingroup
	\ifcat\noexpand #1\relax
	\edef\greek@test{\detokenize{#1}}%
	\edef\greek@test{\expandafter\@cdr\greek@test\@nil}%
	\edef\greek@test{\expandafter\@car\greek@test\@nil}%
	\edef\x{\the\lccode\expandafter`\greek@test}%
	\edef\y{\number\expandafter`\greek@test}%
	\ifnum\x=\y\relax
	\mathbfsbilow{#1}%
	\else
	\mathsfbi{#1}%
	\fi
	\else
	\mathsfbi{#1}%
	\fi
	\endgroup
}
\makeatother
\makeatletter
\newcommand{\sbullet}{%
	\hbox{\fontfamily{lmr}\fontsize{.4\dimexpr(\f@size pt)}{0}\selectfont\textbullet}}

\makeatother

\usepackage[numbers]{natbib}
\begin{document}
	
\begin{frontmatter}
		
\title{{\Large \textbf{Nearly constant \textit{Q} models of the generalized standard linear solid type and the corresponding wave equations}}
}

\cortext[mycorrespondingauthor]{Corresponding author}

\address[KFUPM]{CPG, KFUPM, Dhahran, 31261, Saudi Arabia}
\address[ETHZ]{Institute of Geophysics, ETH Zurich, Zurich, 8092, Switzerland}

\author[KFUPM]{Qi Hao\corref{mycorrespondingauthor}} 
\ead{{xqi.hao@gmail.com, qi.hao@kfupm.edu.sa}}

\author[ETHZ]{Stewart Greenhalgh}
\ead{gstewart@retired.ethz.ch}

% ABSTRACT
\begin{abstract}
Time-domain seismic forward and inverse modeling for a dissipative medium is a vital research topic to investigate the attenuation structure of the Earth. 
Constant $Q$, also called frequency independence of the quality factor, is a common assumption for seismic $Q$ inversion. We propose the first- and second-order nearly constant $Q$ dissipative models of the generalized standard linear solid type, using a novel $Q$-independent weighting function approach. The two new models, which originate from the Kolsky model (a nearly constant $Q$ model) and the Kjartansson model (an exactly constant $Q$ model), result in the corresponding wave equations in differential form. Even for extremely strong attenuation (e.g., $Q=5$), the quality factor  and phase velocity for the two new models are close to those for the Kolsky and Kjartansson models, in a frequency range of interest. The wave equations for the two new models involve explicitly a specified $Q$ parameter and have compact and simple forms. We provide a novel perspective on how to build a nearly constant $Q$ dissipative model which is beneficial for time-domain large scale wavefield forward and inverse modeling. This perspective could also help obtain other dissipative models with similar advantages. We also discuss the extension beyond viscoacousticity and other related issues, for example, extending the two new models to viscoelastic anisotropy.
\end{abstract}


\end{frontmatter}

%\linenumbers

\section{Introduction}
Mechanical wave propagation through dissipative media such as the Earth is characterized by energy absorption and velocity dispersion. As a consequence of the causality principle, the energy absorption and the velocity dispersion are linked to each other by the Kramers-Kronig relations \cite[e.g.,][]{kronig:1926,futterman:1962,carcione:2014}. This means that understanding energy absorption is helpful for deducing the velocity dispersion, and vice versa. 

The quality factor (i.e., $Q$) is an important dimensionless physical quantity, whose inverse $1/Q$ is a measure of the degree of energy absorption for a dissipative medium. Various definitions of the quality factor can be found in the literature \citep[e.g.][]{green:1955,knopoff:1958,buchen:1971,hamilton:1972,connell:1978,toksoz:1981,carcione:2014}. In this paper, the quality factor is defined as $4\pi$ times the ratio of the \textit{averaged} energy of a non-dissipative harmonic plane wave over a cycle to the energy loss of a dissipative harmonic plane wave in the same cycle. This gives rise to a quite simple expression for the quality factor: the ratio between the real and imaginary parts of the complex modulus. This definition of the quality factor is suggested by \cite{connell:1978} who followed \cite{dain:1962} to modify the classic definition of the quality factor in \cite{knopoff:1958}. 
 
The term ``constant $Q$'', which appears frequently in the literature to describe the frequency-independent quality factor, implicitly corresponds to a specific definition of the quality factor. Strictly speaking, there is no physical significance in pursuing an exactly constant $Q$ dissipative model in seismology, because (1) a number of experimental seismic studies show the frequency dependence of the quality factor for the Chandler wobble, tidal and free oscillation data at frequency range $[10^{-8}, 10^{-2}]$~Hz \cite[]{anderson:1979}, for teleseismic waves at frequency range $[0.05, 0.5]$~Hz \cite[]{flanagan:1998}, for earthquake waves in the upper crust at frequency range $[25, 102]$~Hz \cite[]{yoshimoto:1998}, and for normal modes and surface waves at frequency range $[3.3\times 10^{-4}, 1.25\times 10^{-2}]$~Hz \cite[]{lekic:2009}; (2) in the weak attenuation case, it is hard to distinguish between pulse  propagation in an exactly constant $Q$ model and a nearly constant $Q$ model, because they both have similar velocity dispersion behavior; (3) most theoretical mechanisms for energy loss in a wave, e.g., internal friction, relative fluid displacement, and scattering, show it is strongly frequency dependent. Despite these facts, the constant $Q$ assumption is useful for developing simple and feasible methods, such as the spectral ratio method \citep[e.g.,][]{tonn:1991} and the central frequency shift method \citep[e.g.,][]{quan:1997}, to measure the quality factor in practice. These methods can be further developed for frequency dependence of $Q$. For example, the spectral ratio method is incorporated with a frequency power law for $Q$ to estimate the frequency variation of the quality factor \citep[e.g.,][]{lekic:2009,beckwith:2017}.

Multiple dissipative models have been developed for constant $Q$. As the classic dissipative models, the \cite{kolsky:1956} and the \cite{kjartansson:1979} models are nearly constant $Q$ and exactly constant $Q$, respectively, under the definition of the quality factor suggested by \cite{connell:1978}. Although the attenuation power law model proposed by \cite{strick:1967} is constant $Q$ under the definition that quality factor is half the ratio of the wavenumber to the attenuation coefficient, it can be transformed to the Kjartansson model under the low-loss condition. 
Despite having non-physical behavior at zero and infinite frequencies, all these models are widely used to theoretically interpret practical observations about the quality factor being independent of frequency. The Kolsky model is the weak-dissipation approximation of the Kjartansson model (see the section ``The Kjartansson and Kolsky models''). The phase velocity and quality factor for Kolsky model can also be reached from one of the absorption-dispersion pairs in \cite{futterman:1962} and a continuous distribution of relaxation mechanisms given in \cite{liu:1976}, \cite{kanamori:1977} and \cite{aki.richards:1980}. In addition, constant $Q$ can be approximately modeled by applying the generalized standard-linear-solid (SLS) model (sometimes referred to as the generalized Zener model) to fit a given quality factor over a specified frequency range of interest. 

Wavefield numerical modeling based on wave equations is a vital research method to understand wave propagation phenomena and is an essential part of developing an inverse method based on the wave equation. However, it meets multiple challenges in particular for the nearly constant $Q$ models, as elaborated below. 

The Kolsky and Kjartansson models have logarithmic and power-law forms for the complex modulus, respectively, as shown later. Such forms of modulus mean that the time-domain constitutive relation between the stress and the strain, which is expressed by a special convolution, cannot be expressed in differential equation form by introducing auxiliary variables. Although it is argued that for the Kjartansson model the convolution in the constitutive relation can be rewritten in fractional differential form, it is essentially an integral operation  \cite[]{kjartansson:1979,carcione.cavallini:2002,carcione:2010}. Computing the wave equation with a convolution requires the complete time history of the wavefield, which is much more computationally costly than that in differential form. The inconvenience of temporal convolution can be overcome by using the dispersion relation, which expresses the frequency in terms of the wavenumber, to formulate the pseudo wave equation with fractional order spatial derivatives \citep[]{carcione:2010,carcione:2014}, where these derivatives can be calculated by the Fourier transform technique \citep[e.g.,][]{carcione:2010,zhu:2014} and the truncated finite-difference method \cite[]{song:2020}. 
The frequency-domain methods \citep[e.g.,][]{stekl:1998,operto:2009} to model the dissipative wavefields require solving a complex-coefficient linear equation system for each frequency, which is computationally prohibitive in the large-scale 3-D case. 

Unlike the Kolsky model and the Kjartansson model, the generalized SLS model can lead to the wave equation in differential form, which can be solved by multiple time-domain numerical methods such as the finite difference method \citep[e.g.,][]{carcione:1988b}, the staggered-grid finite-difference method \citep[e.g.,][]{bohlen:2002,bai.tsvankin:2016}, the rotated-staggered-grid finite-difference method \citep[e.g.,][]{saenger:2004}, the pseudospectral method \citep[e.g.,][]{carcione:1993}, the finite-element method \citep[e.g.,][]{ham:2012} and the spectral-element method \citep[e.g.,][]{komatitsch.trump:1999}. A variety of techniques have been developed to make the generalized SLS model accurately represent a quality factor, such as \cite{liu:1976}, \cite{emmerich:1987}, \cite{blanch:1995} and \cite{blanc:2016}. Since all these techniques rely on fitting the quality factor, they are called collectively a class of $Q$-fitting methods for convenience. These $Q$-fitting methods require numerically solving a highly nonlinear optimization problem about the unknown parameters in the generalized SLS model. These parameters are implicit functions of the quality factor, which means that once a new quality factor is given one will have to invert for these parameters again. Indeed, the $\tau$-method \cite[]{blanch:1995} as a representative of the $Q$-fitting methods can overcome this drawback but it imposes an extra assumption of $1 + \tau \approx 1$ on the quality factor expression of the GSLS model, in addition to forcing all the SLS elements to share the same unknown parameter $\tau$ ($\tau$ is dimensionless, and it is distinct from the similar symbols $\tau_{\epsilon l}$ and $\tau_{\sigma l}$ in the remainder of this paper, which represents the strain and stress relaxation times for the $l$th mechanism in a weighting function of the generalized SLS type). 
A further improvement of the $\tau$-method, proposed by \cite{fichtner:2014}, gives rise to the generalized SLS model for nearly constant $Q$ and a power law $Q$ function. The generalized SLS model from their method involves an explicit $Q$ parameter, which is also true for the corresponding dissipative wave equations. This method facilitates seismic inverse modeling \cite[]{fichtner:2014} and imaging \cite[]{guo:2018}.   

In this paper, we propose a weighting function method to build the nearly constant $Q$ dissipative models suitable for time-domain wavefield forward and inverse modeling. The weighting function, which is dimensionless and independent of $Q$, has a similar form as the complex modulus for the generalized SLS model. Determination of this weighting function requires only the frequency range of interest. We use the weighting function to represent the moduli for the Kolsky and Kjartansson models, whereby we build the first- and second-order nearly constant $Q$ models of the generalized SLS type. The two new models exhibit an accurate constant $Q$ behavior comparable with the Kolsky and Kjartansson models. Of importance is that the two new models can always yield the corresponding wave equations in differential form, which involve explicitly a specified $Q$ parameter. The wave equations for the two new models have simple and compact form. Especially for the first-order nearly constant $Q$ model, its wave equation is as simple in form as that of the generalized SLS model. Because of the above advantages, the wave equations for the two new models are quite suitable for large-scale 3D constant $Q$ seismic wavefield forward and inverse modeling. It is straightforward and easy to extend the two new models and their wave equations to the viscoacoustic anisotropic situation and the viscoelastic isotropic or anisotropic situation.

The structure of the rest of this paper is as follows. First, we introduce some essential preliminaries. Next, we show the time- and frequency-domain constitutive relations for a general dissipative model. Then, we give the properties and relationship between the Kolsky and Kjartansson models. This is followed by the derivation of a $Q$-independent weighting function and the determination of its optimized coefficients. Next, we show the complex moduli, the relaxation functions and the creep functions for the first- and second-order nearly constant $Q$ models, and compare the two new models with the Kolsky and Kjartansson models. We then show the wave equations in differential form for the two new models. This is followed by the use of numerical examples to analyze and compare the nearly constant $Q$ dissipative wave propagation. Finally, we discuss the possible extension and other related issues before drawing conclusions and providing technical appendices on the mathematical details. 

\section{Essential preliminaries}
To facilitate the description of dissipative wave propagation and as an essential lead-in to what follows, in this section we stipulate our convention for the Fourier transform and its inverse, specify the complex modulus, define the quality factor and give the formula for the phase velocity. 

The Fourier transform of a temporal signal $f(t)$ is written as:
\begin{equation} \label{eq:Fourier}
\hat{f}(\omega) = 
                  \int_{-\infty}^{\infty} 
                  f(t) e^{i \omega t}
                  \text{d}t ,
\end{equation}
where $t$ is time and $\omega$ is angular frequency.

The inverse Fourier transform of the frequency-domain signal $\hat{f}(\omega)$ is written as:
\begin{equation} \label{eq:invFourier}
f(t) = \frac{1}{2\pi} 
       \int_{-\infty}^{\infty} 
       \hat{f}(\omega) 
       e^{- i \omega t}
       \text{d}\omega .
\end{equation}

As a consequence of the Fourier transform definition, the first temporal derivative ``$d/dt$'' corresponds to ``$-i\omega$'' in 
the frequency domain. 
For the dissipative models shown in the following sections, the Fourier transform (equation \ref{eq:Fourier}) suggests that the complex modulus can be generally expressed as $M(\omega) = M_{R}(\omega) - i \text{sgn}(\omega) M_{I}(\omega)$, where $M_{R}$ and $M_{I}$ denote the real part and the magnitude of the imaginary part, respectively. The symbol $\text{sgn}(.)$ denotes the sign function. The minus sign ``$-$'' in front of the imaginary unit ``$i$'' corresponds to the sign convention in the exponential term of the Fourier transform.       

We adopt the definition of the quality factor suggested by \cite{connell:1978} throughout the paper, namely
\begin{equation} \label{eq:Qdef}
Q \equiv \frac{4\pi E}{\Delta E} = \frac{M_{R}}{M_{I}} ,
\end{equation}
where $E$ denotes the time-averaged energy of a harmonic nondissipative plane wave over a cycle. $\Delta E$ denotes the averaged energy loss of a dissipative plane wave over the same cycle. The inverse of the quality factor is interpreted as $1/4\pi$ times the ratio of the fractional average energy dissipated per cycle. The $Q$ definition is valid for both homogeneous and inhomogeneous plane waves. As a consequence of the $Q$ definition, the ratio between the real and imaginary parts of the complex modulus on the far right side of equation \ref{eq:Qdef} is valid only for homogeneous plane waves. 

Referring to \cite{knopoff:1964,knopoff:1965}, the phase velocity for a harmonic dissipative wave is given by:
\begin{equation} \label{eq:Vdef}
V = \frac{v_{R}^2 + v_{I}^2}{v_{R}} ,
\end{equation}
where $v_{R}$ and $v_{I}$ denote the real part and the magnitude of the imaginary part of the complex velocity $v = v_{R} - i\text{sgn}(\omega) v_{I}$. Here, the negative sign in front of the imaginary unit corresponds to the sign convention in the exponent of the Fourier transform (equation \ref{eq:Fourier}), which is consistent with the similar treatment for the complex modulus $M$. 

\section{Time- and frequency-domain constitutive relations}
The relationship between stress and strain is referred to as the constitutive equation. In a dissipative medium, the constitutive relationship is described physically by the Boltzmann superposition principle \cite[]{zener:1956,lakes:2009}. The time-domain constitutive relationship is characterized by the relaxation and creep functions, whereas the frequency-domain one is characterized by the complex modulus and compliance. In this section, we show the time- and frequency-domain constitutive relations for a general dissipative model. For convenience, we omit the spatial coordinate dependence in the constitutive equations. 

\subsection{The time-domain equations}
In a 1-D dissipative medium, the time-domain constitutive relationship for the stress as a function of strain is expressed by the Riemann-Stieltjes convolution integral \cite[]{gurtin:1962,apostol:1974}, namely
\begin{equation} \label{eq:const_t1}
\sigma(t) = \psi(t) \odot \epsilon(t) ,
\end{equation}
where $\sigma$ and $\epsilon$ denote stress and strain, respectively. Quantity $t$ denotes time. Quantity $\psi$ denotes relaxation function. The operation $\odot$ is defined as:
\begin{equation} \label{eq:phitheta}
\psi(t) \odot \epsilon(t)  \equiv 
\int_{-\infty}^{\infty} \psi(t-\tau) \text{d}\epsilon(\tau) .
\end{equation}
Until now, we have not taken account of causality on the relaxation function. Such a definition of this operation is used below to explain the anti-causality problem of a failed nearly constant $Q$ model. It is noteworthy that the upper bound of the integral in equation \ref{eq:phitheta} is different from that in equation 3 of \cite{hao.alkhalifah:2019}, because their definition already implies that the relaxation function is causal, viz., zero for negative time.

As a consequence of equation \ref{eq:const_t1}, the relaxation function physically means the stress response corresponding to the unit step function (the Heaviside step function) in strain, starting at zero time. If the dissipative medium is designated to start moving at $t=0$, the stress and strain in equation \ref{eq:const_t1} are nonzero for a positive time ($t>0$) and zero for a negative time ($t<0$). Hence, the constitutive equation \ref{eq:const_t1} can be rewritten as \cite[]{gurtin:1962,hudson:1980,hao.alkhalifah:2019}:
\begin{equation} \label{eq:calconst}
\psi(t) \odot \epsilon(t) = 
\breve{\psi}(0+) \epsilon(t) + \int_{0}^{t} \dot{\psi}(t-\tau) \epsilon(\tau) \text{d}\tau , 
\end{equation}
\noindent where $0+$ means that time approaches zero from the positive axis. The dot above $\psi$ denotes temporal derivative. $\breve{\psi}(0+)$ denotes the result after excluding the singularity term in $\psi(0+)$. 
If $\psi(0+)$ has no singularity, for example, for the standard-linear-solid model, then $\breve{\psi}(0+) = \psi(0+)$. In the case that $\psi(0+)$ is singular, for example, for the Kjartansson model, then $\breve{\psi}(0+) = 0$.

As an inverse of equation \ref{eq:const_t1}, the constitutive relation for the strain as a function of the stress is written as:
\begin{equation} \label{eq:const_t2}
\epsilon(t) = \chi(t) \odot \sigma(t) ,
\end{equation}
where $\chi$ denotes the creep function. It physically means the strain response corresponding to a unit step function in stress, starting at $t=0$. 

A combination of the physical meaning of creep function and the constitutive relation \ref{eq:const_t1} leads to the relation between the relaxation and creep functions:
\begin{equation}
\psi(t) \odot \chi(t) = H(t) ,
\end{equation}
where $H(.)$ denotes the Heaviside step function.

\subsection{The frequency-domain equations}
The Fourier transform of equation \ref{eq:const_t1} gives rise to the frequency-domain constitutive equation for the stress as a function of the strain:
\begin{equation} \label{eq:const_f1}
\hat{\sigma}(\omega) = M(\omega) \hat{\epsilon}(\omega) ,
\end{equation}
where $M(\omega)$ denotes the complex modulus given by:
\begin{equation} \label{eq:Momega_gen}
M(\omega) =  -i \omega \int_{-\infty}^{\infty} \psi(t) e^{i\omega t} \text{d}t .
\end{equation}
The complex modulus physically means the frequency-domain stress response corresponding to a sinusoidal strain of frequency $\omega$ and amplitude unity. 
Corresponding to the relaxation function in equation \ref{eq:calconst}, the modulus is written as:
\begin{equation}  \label{eq:Momega}
M(\omega) =  \breve{\psi}(0+) 
+ \int_{0}^{\infty} \dot{\psi}(t) e^{i\omega t} \text{d}t .
\end{equation}

Transforming equation \ref{eq:const_t2} into the frequency domain, we obtain the constitutive equation for the strain as a function of the stress:
\begin{equation} \label{eq:const_f2}
\hat{\epsilon}(\omega) = J(\omega) \hat{\sigma}(\omega) ,
\end{equation} 
where $J$ denotes the complex compliance. It physically means the strain response due to a sinusoidal stress of frequency $\omega$ and unit amplitude. 

By analogy with the complex modulus \ref{eq:Momega}, the complex compliance is expressed in terms of the creep function as:
\begin{equation}  \label{eq:Comega}
J(\omega) =  \breve{\chi}(0+) 
+ \int_{0}^{\infty} \dot{\chi}(t) e^{i\omega t} \text{d}t .
\end{equation}

The relationship between the complex modulus and compliance is expressed by:
\begin{equation} \label{eq:MC}
M(\omega) J(\omega) = 1 .
\end{equation}

\section{The Kjartansson and Kolsky models}
Referring to \cite{kolsky:1956} and \cite{kjartansson:1979}, we summarize the Kolsky and Kjartansson models and show their relations below. 

The relaxation function for the Kjartansson model is given by:
\begin{equation} \label{eq:psi_kjar}
\psi(t) = \frac{M_{0}}{\Gamma (1- 2\gamma)} 
               \left( \frac{t}{t_{0}} \right)^{-2\gamma} H(t) ,
\end{equation}
with
\begin{equation}
\gamma = \frac{1}{\pi} \text{tan}^{-1} \left(\frac{1}{Q_{0}} \right) ,
\end{equation}
where $H(.)$ denotes the Heaviside function and $\Gamma(.)$ denotes the Gamma function \cite[]{arfken:2013}. Quanity $Q_{0}$ denotes the reference quality factor. Quantity $M_{0} = \rho v_{0}^2$ denotes the reference modulus corresponding to $Q_{0} = \infty$, where $\rho$ and $v_{0}$ denote the medium density and the reference velocity, respectively. Quantity $t_{0}$ denotes the reference time.  

The creep function for the Kjartansson model is given by:
\begin{equation} \label{eq:chi_kjar}
\chi(t) = \frac{J_{0}}{\Gamma (1 + 2\gamma)} 
\left( \frac{t}{t_{0}} \right)^{2\gamma} H(t) ,
\end{equation}
where $J_{0} = 1/M_{0}$ denotes the reference compliance corresponding to $Q_{0} = \infty$.

The complex modulus in the Kjartansson model is given by:
\begin{equation} \label{eq:Mkjar}
M(\omega) = M_{0} \left(-i\frac{\omega}{\omega_{0}} \right)^{2\gamma} ,
\end{equation}
where $\omega$ and $\omega_{0}=1/t_{0}$ denote angular frequency and reference angular frequency, respectively. The minus sign in front of the imaginary unit ``$i$'' corresponds to the definition of the Fourier transform in equation \ref{eq:Fourier}. The phase velocities for the Kjartansson model at $\omega=0$ and $\omega=\infty$ are zero and infinity, respectively, which implies that this model is non-physical. However, this model can be used to interpret the constant $Q$ phenomenon of dissipative waves in a frequency range of interest.

The Maclaurin series expansion of equation \ref{eq:Mkjar} with respect to $1/Q_{0}$ is given by:
\begin{equation} \label{eq:Mkjar_appr}
\frac{M}{M_{0}} = 1 + 
\frac{1}{Q_{0}}
\left[\frac{2}{\pi} \text{ln} \left|\frac{\omega}{\omega_{0}}\right| - i \text{sgn}(\omega) \right] 
+ \frac{1}{2Q_{0}^2} 
\left[\frac{2}{\pi} \text{ln}\left|\frac{\omega}{\omega_{0}}\right| - i \text{sgn}(\omega) \right]^2  +  
O\left(\frac{1}{Q_{0}^3}\right) .
\end{equation}
Truncating the above series up to the first order accuracy, we obtain the complex modulus for the Kolsky model:
\begin{equation} \label{eq:Mkolsky}
M(\omega) = M_{0} \left \{
1 + \frac{1}{Q_{0}}
\left[\frac{2}{\pi} \text{ln}\left|\frac{\omega}{\omega_{0}}\right| - i \text{sgn}(\omega) \right]  
\right \} .
\end{equation}
Although the above complex modulus expression is not mentioned in \cite{kolsky:1956}, he obtained the corresponding phase velocity and attenuation coefficient, from which we derive equation \ref{eq:Mkolsky} (see Appendix A). Although the Kolsky model is non-physical as the frequency approaches $\omega=0$ or $\omega=\infty$, it may be used to interpret the nearly constant-$Q$ behavior of a dissipative wave in a frequency range of interest. Kolsky actually assumed a linear relationship between the attenuation coefficient and frequency in his model (see Appendix A), which implies an almost constant $Q$.  

To derive the relaxation function for the Kolsky model, we take into account the fact that the relaxation function $\psi(t)$ in equation \ref{eq:Momega_gen} is identical to the inverse Fourier transform of $M(\omega)/(-i \omega)$. Using the complex modulus (equation \ref{eq:Mkolsky}) and the inverse Fourier transform (equation \ref{eq:invFourier}), we derive the following expression for the relaxation function corresponding to the Kolsky model:
\begin{equation} \label{eq:psi_kols}
\psi(t) = M_{0} \left[
1 - \frac{2}{\pi Q_{0}}\left(\gamma_{E} + 
\text{ln}\left| \frac{t}{t_{0}} \right| \right)  
\right] H(t),
\end{equation}
where $\gamma_{E} \approx 0.577216$ is the Euler-Mascheroni constant \cite[]{arfken:2013}, $t_{0}$ is the reciprocal of $\omega_{0}$, and $H(t)$ denotes the Heaviside step function. 

As an alternative, equation \ref{eq:psi_kols} can also be derived from the relaxation function (equation \ref{eq:psi_kjar}) for the Kjartansson model. Observing equation \ref{eq:Momega_gen}, we note that the complex modulus and the relaxation function satisfy a correspondence relation, that is, a linear combination of two complex moduli yields the same combination of the corresponding relaxation functions. The complex modulus (equation \ref{eq:Mkolsky}) can be viewed as a linear combination with respect to $1/Q_{0}$. Hence, the relaxation function must be a linear function of $1/Q_{0}$. On the other hand, we already know that the complex modulus for the Kolsky model is the first-order Maclaurin series expansion of the complex modulus for the Kjartansson model with respect to $1/Q_{0}$. Hence, the relaxation function for the Kolsky model must be identical to the first-order Maclaurin series expansion of the relaxation function for the Kjartansson model with respect to $1/Q_{0}$, whereby we may obtain equation \ref{eq:psi_kols}.

The creep function should be obtained from the complex compliance using equation \ref{eq:Comega}, where the complex compliance is given as the reciprocal of the complex modulus via equation \ref{eq:Mkolsky}. However, it is hard to imitate the aforementioned scheme for the relaxation function to derive the creep function, because the complex compliance for the Kolsky model involves the logarithmic function $\text{ln}|\omega/\omega_{0}|$ appearing in the denominator of a fraction. In addition, the logarithmic function multiplied by $1/Q_{0}$ as an unbounded function cannot enable us to expand the complex compliance into the Maclaurin series with respect to $1/Q_{0}$, from which we may apply the inverse Fourier transform to derive the creep function. We recall again the fact that the complex modulus for the Kolsky model is the first-order approximation of the one for the Kjartansson model. The first-order Maclaurin series expansion of equation \ref{eq:chi_kjar} with respect to $1/Q_{0}$ results in the approximate creep function for the Kolsky model, namely
\begin{equation} \label{eq:chi_kols}
\chi(t) \approx J_{0} \left[
1 + \frac{2}{\pi Q_{0}}\left(\gamma_{E} + 
\text{ln}\left| \frac{t}{t_{0}} \right| \right)
\right] H(t) .
\end{equation} 

\section{\textit{Q}-independent weighting function}
By analogy with the complex modulus for the generalized SLS model \cite[]{carcione:2014,hao.greenhalgh:2019}, we define a weighting function, which is dimensionless and independent of the quality factor, as follows:
\vspace{-1ex}
\begin{equation} \label{eq:W}
W(\omega)
\equiv \sum_{l=1}^{L} \frac{1-i\omega\tau_{\epsilon l}}{1-i\omega\tau_{\sigma l}} ,
\end{equation}
where $\tau_{\epsilon l}$ and $\tau_{\sigma l}$ are $Q$-independent strain and stress relaxation times in the $l$-th term in the summation for the weighting function, respectively. 

We further express $W(\omega)$ as $W(\omega) = W_{R}(\omega) - i W_{I}(\omega)$, where $W_{R}$ and $W_{I}$ correspond to the real and imaginary parts, and the minus sign in front of ``$i$'' follows the sign convention of the exponential term in the Fourier transform.  
We use $W(\omega)-W_{R}(\omega_{0})$ to fit the term inside the square brackets in equation \ref{eq:Mkjar_appr}, and then split the result into the real and imaginary parts, namely
% %
\begin{align}
\label{eq:WWRr}
& W_{R}(\omega) - W_{R}(\omega_{0}) = 
\sum_{l=1}^{L}
\frac{1 + \omega^2 \tau_{\epsilon l}\tau_{\sigma l}}
{1+\omega^2 \tau_{\sigma l}^2} 
-
\sum_{l=1}^{L}
\frac{1 + \omega_{0}^2 \tau_{\epsilon l}\tau_{\sigma l}}
{1+\omega_{0}^2 \tau_{\sigma l}^2} 
\approx
\frac{2}{\pi} \text{ln}\left|\frac{\omega}{\omega_{0}}\right| , \\
\label{eq:WWRi}
& W_{I}(\omega) = 
\sum_{l=1}^{L}
\frac{\omega (\tau_{\epsilon l}-\tau_{\sigma l})}{1+\omega^2 \tau_{\sigma l}^2} \approx \text{sgn}(\omega) .
\end{align}
% %
Equation \ref{eq:WWRr} is always valid at $\omega = \omega_{0}$, whatever the values of $\tau_{\epsilon_l}$ and $\tau_{\sigma_l}$ are. The involvement of $\omega_{0}$ in equation \ref{eq:WWRr} can be eliminated by taking the first derivative with respect to $\omega$. Finally, the  cost function is formulated as:
% %
\begin{equation} \label{eq:costF}
\displaystyle
G = \frac{1}{2(\omega_{U}-\omega_{L})} \int_{\omega_{L}}^{\omega_{U}} 
\left[
\left(\pi \sum_{l=1}^{L} \frac{\omega^2 \tau_{\sigma l} \Delta \tau_{l}}
{(1+\omega^2 \tau_{\sigma l}^2)^2} - 1 \right)^2 +
\left(
\sum_{l=1}^{L}
\frac{\omega \Delta \tau_{l}}{1+\omega^2 \tau_{\sigma l}^2} - 1
\right)^2 
\right]
d\omega ,
\end{equation}
% %
where $\Delta \tau_{l}=\tau_{\epsilon l} - \tau_{\sigma l}$, and $\omega_{L}$ and $\omega_{U}$ are the lower and upper bounds of the positive frequency range of interest. Equation \ref{eq:costF} measures the mean squared error, which may eliminate the effect of the interval length of the frequency range on the cost function. 

Minimizing equation \ref{eq:costF} is a nonlinear optimization problem. 
We combine the dual-annealing method \citep[e.g.,][]{xiang:1997} (a global optimization method) and the Broyden-Fletcher-Goldfarb-Shanno (BFGS) method \citep[e.g.,][]{nocedal:2006} (a localized optimization method) to find the best solution. We implement individually the dual-annealing method 20,000 times and obtain a set of globally optimized solutions. From these solutions, we select the optimal solution (i.e., the one that minimizes the cost function) and then use it as the initial value for the BFGS method to find the final optimized solution. The BFGS method requires the first partial derivatives of the cost function with respect to the unknown parameters. These are provided in Appendix B.

Tables \ref{tab:tabl1}-\ref{tab:tabl8} show the optimal values of the parameters $\tau_{\sigma l}$ and $\Delta \tau_l$ obtained for the five- and six-element weighting functions with different frequency ranges of interest. A comparison between these tables shows that (1) the optimal values of the parameters in a weighting function of a fixed number of elements will decrease with a widening of the frequency range of interest (e.g., Tables \ref{tab:tabl1}-\ref{tab:tabl4}); (2) for a fixed frequency range of interest the optimal values of the parameters are varied widely by increasing the number of elements in the weighting function (e.g., Tables \ref{tab:tabl1} and \ref{tab:tabl5}).
Table \ref{tab:tabl9} shows that (1) for a weighting function with a fixed number of elements increasing the frequency range of interest decreases the accuracy of the optimal parameters; (2) for a fixed frequency range of interest increasing the number of elements in a weighting function improves the accuracy of the optimal parameters.

\begin{table}[H]
\centering
\caption{The optimal parameters for the five-element weighting function in the frequency range $[1,50]$~Hz.}
\label{tab:tabl1}
\begin{tabular}{c c c}
\toprule
$l$ & $\tau_{\sigma l}$ (s) & $\Delta \tau_{l} = \tau_{\epsilon l}-\tau_{\sigma l}$ (s) \\
\midrule
1 &	3.5513403 $\times 10^{-1}$ &	5.5479304 $\times 10^{-1}$ \\
2 &	6.4907438 $\times 10^{-2}$ &	5.5691466 $\times 10^{-2}$ \\
3 &	1.8510729 $\times 10^{-2}$ &	1.4094923 $\times 10^{-2}$ \\
4 &	5.6320673 $\times 10^{-3}$ &	4.4188133 $\times 10^{-3}$ \\
5 &	1.1429090 $\times 10^{-3}$ &	1.7382742 $\times 10^{-3}$ \\
\bottomrule   
\end{tabular}
\end{table}

\begin{table}[H]
\centering
\caption{The optimal parameters for the five-element weighting function in the frequency range $[1,100]$~Hz.}
\label{tab:tabl2}
\begin{tabular}{c c c}
\toprule
$l$ & $\tau_{\sigma l}$ (s) & $\Delta \tau_{l} = \tau_{\epsilon l}-\tau_{\sigma l}$ (s) \\
\midrule
1 &	2.8834448 $\times 10^{-1}$ &	4.4811122 $\times 10^{-1}$ \\
2 &	4.7554203 $\times 10^{-2}$ &	4.5510704 $\times 10^{-2}$ \\
3 &	1.1745042 $\times 10^{-2}$ &	9.8954582 $\times 10^{-3}$ \\
4 &	3.2170335 $\times 10^{-3}$ &	2.6901902 $\times 10^{-3}$ \\
5 &	6.2054849 $\times 10^{-4}$ &	9.5122738 $\times 10^{-4}$ \\
\bottomrule     
\end{tabular}
\end{table}

\begin{table}[H]
\centering
\caption{The optimal parameters for the five-element weighting function in the frequency range $[1,150]$~Hz.}
\label{tab:tabl3}
\begin{tabular}{c c c}
\toprule
$l$ & $\tau_{\sigma l}$ (s) & $\Delta \tau_{l} = \tau_{\epsilon l}-\tau_{\sigma l}$ (s) \\
\midrule
1 &	2.2340486 $\times 10^{-1}$ &	3.2107169 $\times 10^{-1}$ \\
2 &	3.7233817 $\times 10^{-2}$ &	3.7322062 $\times 10^{-2}$ \\
3 &	8.6301965 $\times 10^{-3}$ &	7.5762611 $\times 10^{-3}$ \\
4 &	2.2599473 $\times 10^{-3}$ &	1.9393628 $\times 10^{-3}$ \\
5 &	4.2652419 $\times 10^{-4}$ &	6.5631993 $\times 10^{-4}$ \\
\bottomrule     
\end{tabular}
\end{table}

\begin{table}[H]
\centering
\caption{The optimal parameters for the five-element weighting function in the frequency range $[1,200]$~Hz.}
\label{tab:tabl4}
\begin{tabular}{c c c}
\toprule
$l$ & $\tau_{\sigma l}$ (s) & $\Delta \tau_{l} = \tau_{\epsilon l}-\tau_{\sigma l}$ (s) \\
\midrule
1 &	1.4388052 $\times 10^{-1}$ &	1.8931948 $\times 10^{-1}$ \\
2 &	2.6506214 $\times 10^{-2}$ &	2.6022735 $\times 10^{-2}$ \\
3 &	6.2887118 $\times 10^{-3}$ &	5.4548056 $\times 10^{-3}$ \\
4 &	1.6688598 $\times 10^{-3}$ &	1.4214801 $\times 10^{-3}$ \\
5 &	3.1668719 $\times 10^{-4}$ &	4.8742543 $\times 10^{-4}$ \\
\bottomrule     
\end{tabular}
\end{table}

\begin{table}[H]
\centering
\caption{The optimal parameters for the six-element weighting function in the frequency range $[1,50]$~Hz.}
\label{tab:tabl5}
\begin{tabular}{c c c}
\toprule
$l$ & $\tau_{\sigma l}$ (s) & $\Delta \tau_{l} = \tau_{\epsilon l}-\tau_{\sigma l}$ (s)\\
\midrule
1 &	4.4915262 $\times 10^{-1}$ &	6.8664148 $\times 10^{-1}$ \\
2 &	9.2934004 $\times 10^{-2}$ &	6.9600103 $\times 10^{-2}$ \\
3 &	3.1659618 $\times 10^{-2}$ &	2.0500434 $\times 10^{-2}$ \\
4 &	1.1748298 $\times 10^{-2}$ &	7.3165182 $\times 10^{-3}$ \\
5 &	4.2770492 $\times 10^{-3}$ &	2.9788159 $\times 10^{-3}$ \\
6 &	9.4659276 $\times 10^{-4}$ &	1.4201223 $\times 10^{-3}$ \\
\bottomrule   
\end{tabular}
\end{table}

\begin{table}[H]
\centering
\caption{The optimal parameters for the six-element weighting function in the frequency range $[1,100]$~Hz.}
\label{tab:tabl6}
\begin{tabular}{c c c}
\toprule
$l$ & $\tau_{\sigma l}$ (s) & $\Delta \tau_{l} = \tau_{\epsilon l}-\tau_{\sigma l}$ (s)\\
\midrule
1 &	3.8705303 $\times 10^{-1}$ & 6.0103005 $\times 10^{-1}$ \\
2 &	7.3380142 $\times 10^{-2}$ & 6.0613810 $\times 10^{-2}$ \\
3 &	2.2067095 $\times 10^{-2}$ & 1.5991205 $\times 10^{-2}$ \\
4 &	7.3318632 $\times 10^{-3}$ & 5.0255261 $\times 10^{-3}$ \\
5 &	2.4579583 $\times 10^{-3}$ & 1.8124172 $\times 10^{-3}$ \\
6 &	5.2254525 $\times 10^{-4}$ & 7.8877463 $\times 10^{-4}$ \\
\bottomrule   
\end{tabular}
\end{table}

\begin{table}[H]
\centering
\caption{The optimal parameters for the six-element weighting function in the frequency range $[1,150]$~Hz.}
\label{tab:tabl7}
\begin{tabular}{c c c}
\toprule
$l$ & $\tau_{\sigma l}$ (s) & $\Delta \tau_{l} = \tau_{\epsilon l}-\tau_{\sigma l}$ (s) \\
\midrule
1 &	3.5583900 $\times 10^{-1}$ & 5.5705567 $\times 10^{-1}$ \\
2 &	6.3570120 $\times 10^{-2}$ & 5.5796953 $\times 10^{-2}$ \\
3 &	1.7663115 $\times 10^{-2}$ & 1.3626618 $\times 10^{-2}$ \\
4 &	5.4969651 $\times 10^{-3}$ & 3.9678196 $\times 10^{-3}$ \\
5 &	1.7573930 $\times 10^{-3}$ & 1.3369547 $\times 10^{-3}$ \\
6 &	3.6512446 $\times 10^{-4}$ & 5.5311137 $\times 10^{-4}$ \\
\bottomrule 
\end{tabular}
\end{table}

\begin{table}[H]
\centering
\caption{The optimal parameters for the six-element weighting function in the frequency range $[1,200]$~Hz.}
\label{tab:tabl8}
\begin{tabular}{c c c}
\toprule
$l$ & $\tau_{\sigma l}$ (s) & $\Delta \tau_{l} = \tau_{\epsilon l}-\tau_{\sigma l}$ (s)\\
\midrule
1 &	3.3462365 $\times 10^{-1}$ & 5.2512642 $\times 10^{-1}$ \\
2 &	5.7203494 $\times 10^{-2}$ & 5.2461629 $\times 10^{-2}$ \\
3 &	1.4998295 $\times 10^{-2}$ & 1.2071172 $\times 10^{-2}$ \\
4 &	4.4582319 $\times 10^{-3}$ & 3.3304998 $\times 10^{-3}$ \\
5 &	1.3793789 $\times 10^{-3}$ & 1.0715824 $\times 10^{-3}$ \\
6 &	2.8209314 $\times 10^{-4}$ & 4.2837752 $\times 10^{-4}$ \\
\bottomrule   
\end{tabular}
\end{table}

\begin{table}[H]
\centering
\caption{The variation of the minimum of the cost function (equation \ref{eq:costF}) with the number of the weighting function elements $L$ and the frequency range of interest.}
\label{tab:tabl9}
\begin{tabular}{c c c c c}
\toprule
$L$ & $[1,50]$~Hz & $[1,100]$~Hz & $[1,150]$~Hz & $[1,200]$~Hz \\
\midrule
$5$ & 9.181 $\times 10^{-7}$  &	3.693 $\times 10^{-6}$  &	6.811  $\times 10^{-6}$ & 8.659 $\times 10^{-6}$ \\
$6$ & 4.561 $\times 10^{-8}$ & 2.376 $\times 10^{-7}$ & 5.472 $\times 10^{-7}$ & 9.264 $\times 10^{-7}$ \\
\bottomrule     
\end{tabular}
\end{table}

\section{The nearly constant \textit{Q} models of the generalized SLS type}
Adopting the method shown in the previous section, we find the optimal relaxation times $\tau_{\epsilon l}$ and $\tau_{\sigma l}$ and determine the weighting function $W(\omega)$ (equation \ref{eq:W}) in a frequency range of interest. The weighting function is similar in form to the complex modulus for the generalized SLS model, but this function is dimensionless and independent of medium parameters (i.e., the reference modulus and quality factor). We next use this weighting function to represent approximately the complex moduli for the Kolsky and Kjartansson models, which yield the nearly constant $Q$ models of the generalized SLS type. The complex moduli for these two new models are first- and second-order polynomials with respect to the reference quality factor, and hence we call them the first- and second-order nearly constant $Q$ models for convenience throughout the remainder of this paper. 

\subsection{The first-order nearly constant \textit{Q} model}
The first-order nearly constant $Q$ model is an approximation for the Kolsky model. It results from retaining just the first two terms (zeroth and first orders in $1/Q$) in the Maclaurin series expansion for the Kjartansson model (equation \ref{eq:Mkolsky}). The substitution of equations \ref{eq:WWRr} and \ref{eq:WWRi} into equation \ref{eq:Mkolsky} leads to the complex modulus for the first-order nearly constant $Q$ model:
% %
\begin{equation} \label{eq:M1st}
M(\omega) = M_{0} \left \{
1 + \frac{1}{Q_{0}} \left[W(\omega) - W_{R}(\omega_{0}) \right] 
\right \} ,
\end{equation}
where $W(\omega) - W_{R}(\omega_{0})$ is given by
\begin{equation} \label{eq:W_Wr}
W(\omega) - W_{R}(\omega_{0})
= \sum_{l=1}^{L} \frac{1-i\omega\tau_{\epsilon l}}{1-i\omega\tau_{\sigma l}}
-
\sum_{l=1}^{L}
\frac{1 + \omega_{0}^2 \tau_{\epsilon l}\tau_{\sigma l}}
{1+\omega_{0}^2 \tau_{\sigma l}^2} .
\end{equation}
This complex modulus can be rewritten as the complex modulus of the generalized SLS model:
\begin{equation} \label{eq:Mgsls}
M(\omega) = \frac{M_{R}'}{L} \sum_{l=1}^{L} \frac{1-i\omega\tau_{\epsilon l}'}{1-i\omega\tau_{\sigma l}} ,
\end{equation}
where $M_{R}'$ and $\tau_{\epsilon l}'$ are given by
\begin{align}
& M_{R}' = M_{0} \left(\kappa + \frac{L}{Q_{0}} \right) , \\
& \tau_{\epsilon l}' = \frac{\kappa Q_{0} \tau_{\sigma l} + L\tau_{\epsilon l}}{\kappa Q_{0} + L} ,
\end{align}
with
\begin{equation}
\kappa = 1 - \frac{1}{Q_{0}} 
\sum_{l=1}^{L}
\frac{1 + \omega_{0}^2 \tau_{\epsilon l}\tau_{\sigma l}}
{1+\omega_{0}^2 \tau_{\sigma l}^2} .
\end{equation}
Hence, the first-order nearly constant Q model is identical to the generalized SLS model.

Referring to the quality factor definition (equation \ref{eq:Qdef}), the quality factor for the first-order nearly constant $Q$ model is written as:
\begin{equation} \label{eq:Q1st}
\frac{1}{Q} = \frac{W_{I}(\omega)}{Q_{0}} + O(\frac{1}{Q_{0}^2})
\approx \frac{1}{Q_{0}} ,  
\end{equation}
where we already apply equation \ref{eq:WWRi} for a positive frequency and a weak attenuation case ($1/Q_{0} \ll 1$) to the term on the far right side of equation \ref{eq:Q1st}. Equation \ref{eq:Q1st} indicates that the first-order nearly constant $Q$ model will become closer to being constant $Q$ as $Q_{0}$ increases. 

Referring to equation \ref{eq:Vdef}, the phase velocity for the first-order nearly constant $Q$ model is written as:
\begin{equation} \label{eq:V1st}
V \approx v_{0} \left\{
1 + \frac{1}{2Q_{0}} \left[ W_{R}(\omega) - W_{R} (\omega_{0}) \right] 
\right\} ,
\end{equation}
where $v_{0} = \sqrt{M_{0}/\rho}$ denotes the reference velocity corresponding to $Q_{0} = \infty$, and $\rho$ denotes density. 
Here, we already take into account the first two terms in the Maclaurin series expansion of the phase velocity with respect to $1/Q_{0}$.   
Equation \ref{eq:V1st} indicates that in a weak attenuation case ($1/Q_{0} \ll 1$) the frequency variation of the phase velocity for the first-order nearly constant $Q$ model is characterized by the real part of the weighting function. 

The relaxation function is linked to the complex modulus through equation \ref{eq:Momega}. As illustrated in equation \ref{eq:W}, the weighting function $W(\omega)$ is similar in form to the modulus of the generalized SLS model \cite[]{hao.greenhalgh:2019}. Hence, we may ascertain the relaxation function corresponding to the complex modulus defined as the weighting function. Since equation \ref{eq:Momega}, as an integral-differential equation for $\psi(t)$, is linear, the relaxation function corresponding to the complex modulus $M(\omega) = W(\omega) - W_{R}(\omega_{0})$, where we ignore the physical dimension between the complex modulus and the weighting function, can be obtained by the same combination of the relaxation functions corresponding to the complex moduli $M(\omega) = W(\omega)$ and $M(\omega) = W_{R}(\omega_{0})$, respectively. Furthermore, we may determine the relaxation function corresponding to the complex modulus in equation \ref{eq:M1st}.

The creep function is linked to the complex compliance through equation \ref{eq:Comega}. The complex compliance is the reciprocal of the complex modulus, as shown in equation \ref{eq:MC}. From equation \ref{eq:M1st}, we may obtain the complex compliance for the first-order nearly constant $Q$ model. The Maclaurin series of the complex compliance with respect to $1/Q_{0}$ involves the term $W(\omega) - W_{R}(\omega_{0})$ as a common factor in the series coefficients. We note that equation \ref{eq:Comega} is mathematically identical to equation \ref{eq:Momega}. Following a similar idea as to how we deal with the relaxation function, we may derive the creep function for the first-order nearly constant $Q$ model. The only difference is that here the weighting function $W(\omega)$ is interpreted as the complex compliance, whereas it is viewed as the complex modulus when we deal with the relaxation function. The complete derivations of the relaxation and creep functions are given in Appendix C. The results are summarized below. 

The relaxation function for the first-order nearly constant $Q$ model is given by:
\begin{equation} \label{eq:psi1st}
\psi(t) = M_{0} \left[
H(t) + \frac{1}{Q_{0}} \zeta(t)
\right] ,
\end{equation}
where 
\begin{equation}
\zeta(t) = - W_{R}(\omega_{0}) H(t) 
+ 
\sum_{l=1}^{L} 
\left[
1 - \left(1 - 
\frac{\tau_{\epsilon_l}}{\tau_{\sigma_l}} 
\right)
e^{-\frac{t}{\tau_{\sigma_l}}}
\right] H(t). 
\end{equation}

The creep function for the first-order model is given by:
\begin{equation} \label{eq:chi1st}
\chi(t) = J_{0} \left[
H(t) + 
\sum_{n=1}^{\infty} \frac{(-1)^{n}}{Q_{0}^n} \zeta^{\langle n \rangle}(t)  
\right] ,
\end{equation}
where $J_{0} = 1/M_{0}$ denotes the reference compliance, and
the function $\zeta^{\langle n \rangle}(t)$ is defined as:
\begin{equation}
\zeta^{\langle n \rangle}(t) =
\begin{cases}
\underbrace{\zeta(t) \odot \zeta(t) \cdots \odot \zeta(t)}_{n}, & \text{if } n > 1, \\
\zeta(t), & \text{if } n = 1 .
\end{cases}
\end{equation}

\subsection{The second-order nearly constant \textit{Q} model}
The complex modulus of the second-order nearly constant $Q$ model is an approximation for the second-order Maclaurin series expansion of the complex modulus for the Kjartansson model. Replacing the term inside the brackets in equation \ref{eq:Mkjar_appr} by $W_{R}(\omega)-W_{R}(\omega_{0})$, the complex modulus for the second-order nearly constant $Q$ model is written as:
\begin{equation} \label{eq:M2nd}
M(\omega) = M_{0} \left \{
1 + \frac{1}{Q_{0}} 
\left[W(\omega) - W_{R}(\omega_{0}) \right] 
+ \frac{1}{2Q_{0}^2} 
\left[W(\omega) - W_{R}(\omega_{0}) \right]^2
\right \}.
\end{equation}
This complex modulus involves the second-order term with respect to $W(\omega) - W_{R}(\omega_{0})$, which makes it distinct in form from the complex modulus for the generalized SLS model. Because the first-order nearly constant $Q$ model is identical to the generalized SLS model, we may call the second-order nearly constant $Q$ model the quasi generalized SLS model of nearly constant $Q$ . 

We may use the same way to derive the relaxation and creep functions for the second-order nearly constant $Q$ model as we did for the first-order nearly constant $Q$ model. Their derivation can be found in Appendix C. The results are summarized below. 

The relaxation function for the second-order nearly constant $Q$ model is given by:
\begin{equation} \label{eq:psi2nd}
\psi(t) = M_{0} \left \{
H(t) + \frac{1}{Q_{0}} \zeta(t) 
+ \frac{1}{2Q_{0}^2} \zeta^{\langle 2 \rangle}(t) 
\right \}.
\end{equation}

The creep function for the second-order nearly constant $Q$ model is given by:
\begin{equation} 
\begin{aligned} \label{eq:chi2nd}
\chi(t) & = J_{0} H(t) + 
J_{0} \sum_{n=1}^{\infty} 
\frac{(-1)^n}{2^{2n} Q_{0}^{4n}} \zeta^{\langle 4n \rangle}(t) 
+ J_{0} \sum_{n=0}^{\infty} 
\frac{(-1)^{n+1}}{2^{2n} Q_{0}^{4n+1}} \zeta^{\langle 4n+1 \rangle}(t)  \\
& + J_{0} \sum_{n=0}^{\infty} 
\frac{(-1)^n}{2^{2n+1} Q_{0}^{4n+2}} \zeta^{\langle 4n+2 \rangle}(t) .
\end{aligned}
\end{equation}

\subsection{Relaxed and unrelaxed moduli}
Unlike the Kolsky and Kjartansson models, the complex moduli for the first- and second-order nearly constant $Q$ models are physically plausible (i.e., bounded) at zero and infinite frequencies. We now consider these two special cases: (1) both models are fully relaxed, which corresponds to $\omega=0$; (2) both models are completely unrelaxed, which corresponds to $\omega=\infty$. 

In the first case, the relaxed modulus for the second-order nearly constant $Q$ model is given by:
\begin{equation} \label{eq:Mrelax}
M_{relax} = M_{0} \left[ 1 - 
\frac{1}{Q_{0}}
\left( \sum_{l=1}^{L} 
\frac{\omega_{0}^2\tau_{\sigma l}\Delta \tau_{l}}{1 + \omega_{0}^2 \tau_{\sigma l}^2}
\right)  
+ \frac{1}{2Q_{0}^2} 
\left( \sum_{l=1}^{L} \frac{\omega_{0}^2\tau_{\sigma l}\Delta \tau_{l}}{1 + \omega_{0}^2 \tau_{\sigma l}^2}
\right)^2 
\right] .
\end{equation}

In the second case, the unrelaxed modulus for the second-order nearly constant $Q$ model is given by:
\begin{equation} \label{eq:Munrelax}
M_{unrelax} = M_{0} \left\{ 1 + 
\frac{1}{Q_{0}}
\left[ \sum_{l=1}^{L} \frac{\Delta \tau_{l}}
{\tau_{\sigma l}(1 + \omega_{0}^2 \tau_{\sigma l}^2)}
\right]
+ \frac{1}{2Q_{0}^2}
\left[ \sum_{l=1}^{L} \frac{\Delta \tau_{l}}
{\tau_{\sigma l}(1 + \omega_{0}^2 \tau_{\sigma l}^2)}
\right]^2 
\right \} ,
\end{equation}
where $M_{relax}$ and $M_{unrelax}$ denote the relaxed and unrelaxed moduli, respectively. 

The relaxed and unrelaxed moduli for the first-order nearly constant $Q$ model are the result of truncating equations \ref{eq:Mrelax} and \ref{eq:Munrelax}, respectively, up to the first-order accuracy with respect to $1/Q_{0}$.

\subsection{Scaling the valid frequency range}
An important property of the first- and second-order nearly constant $Q$ models is that of scaling. We assume that $\tau_{\epsilon l}^{(b)}$ and $\tau_{\sigma l}^{(b)}$ are the relaxation times defined in the frequency range $[\omega_{L}, \omega_{U}]$ before scaling. By introducing a scaling factor $\xi$ ($\xi>0$), the valid frequency range is scaled to $[\xi \omega_{L}, \xi \omega_{U}]$ by using the following relaxation times:
\begin{equation} \label{eq:tau_scale}
\begin{split}
& \tau_{\epsilon l}^{(a)} = \frac{\tau_{\epsilon l}^{(b)}}{\xi} , \\ 
& \tau_{\sigma l}^{(a)} = \frac{\tau_{\sigma l}^{(b)}}{\xi}   ,
\end{split}
\end{equation}
where $\tau_{\epsilon l}^{(a)}$ and $\tau_{\sigma l}^{(a)}$ denote the relaxation times in the frequency range $[\xi \omega_{L}, \xi \omega_{U}]$ after scaling. This scaling enables us to use the coefficients (Tables \ref{tab:tabl1} - \ref{tab:tabl8}, and \ref{tab:tabl11}) over other frequency ranges that might be encountered in practice.

\subsection{A comparison with the Kolsky and Kjartansson models}
Here, we show a numerical example to compare the first- and second-order nearly constant $Q$ models with the Kolsky and Kjartansson models and the generalized SLS model for nearly constant $Q$, determined by the $\tau$-method \cite[]{blanch:1995,bohlen:2002}. We investigate four attenuation cases: the weak attenuation case ($Q_{0}=100$), the moderate attenuation case ($Q_{0} = 60$), the strong attenuation case ($Q_{0} = 30$) and the extremely strong attenuation case ($Q_{0} = 5$), where $Q_{0}$ denotes the reference quality factor in a considered model. The frequency range of interest is taken as [$1, 200$]~Hz. The reference frequency is $f_{0}=40$~Hz, which is used to determine the reference angular frequency $\omega_{0}$ in the complex moduli for all these models. We assume the density to be $\rho = 10^3~\text{kg/m}^{3}$, set the reference velocity as $v_{0}=3$~km/s and use the parameters shown in Table \ref{tab:tabl4} to determine the complex moduli, the relaxation functions, and the creep functions for the first- and second-order nearly constant $Q$ models. 

\begin{figure}[H]
\centering
\subfloat[$Q_{0} = 100$]
{\includegraphics[width=0.23\textwidth]{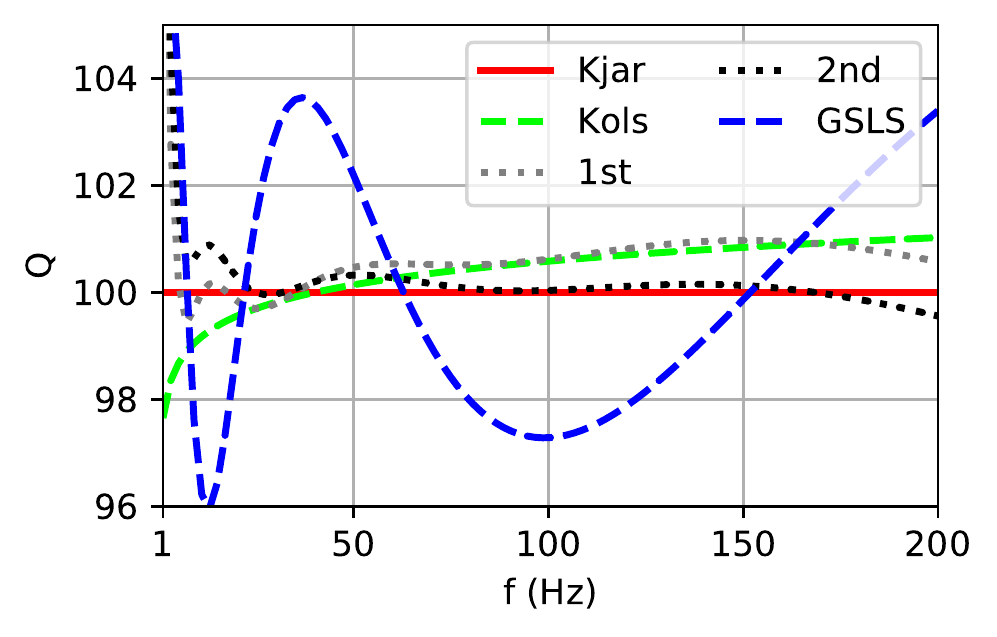}}
\subfloat[$Q_{0} = 60$]
{\includegraphics[width=0.23\textwidth]{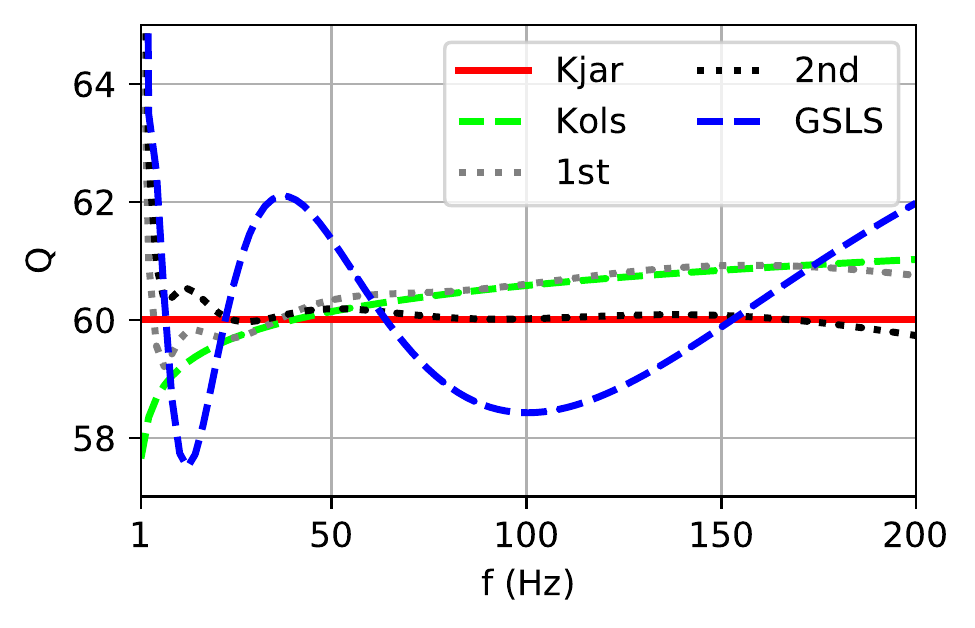}}
\subfloat[$Q_{0} = 30$]
{\includegraphics[width=0.23\textwidth]{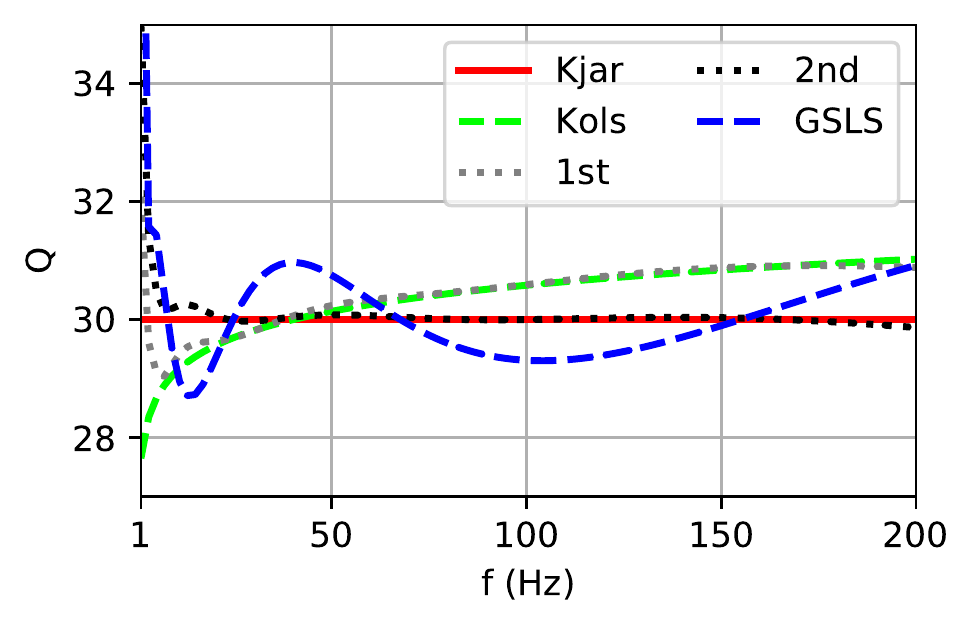}}
\subfloat[$Q_{0} = 5$]
{\includegraphics[width=0.23\textwidth]{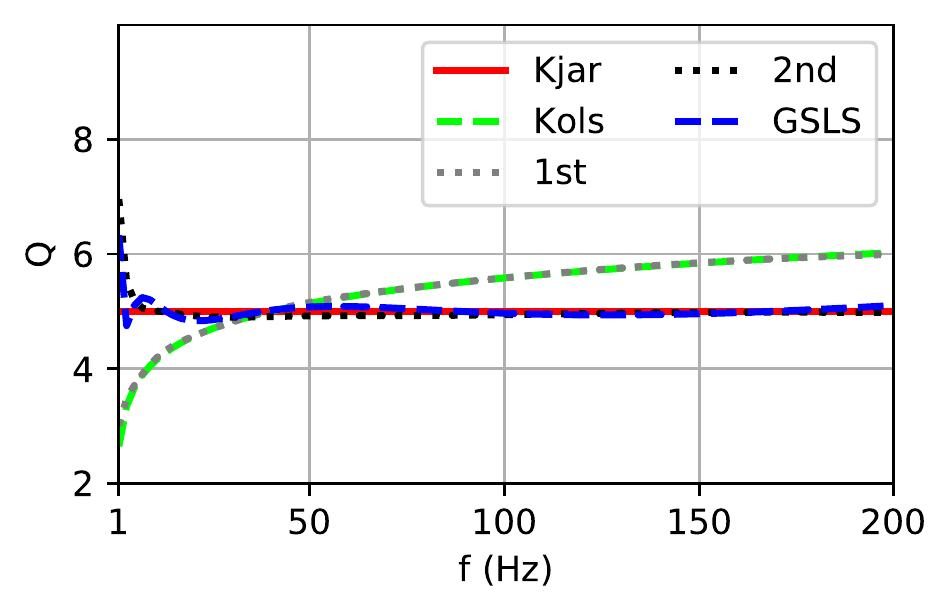}}
\caption{
The variation of the quality factor with frequency in (a) the weak attenuation case ($Q_{0} = 100$), (b) the moderate attenuation case ($Q_{0} = 60$), (c) the strong attenuation case ($Q_{0} = 30$) and (d) the extremely strong attenuation case ($Q_{0} = 5$).  The legend abbreviations ``Kols'', ``Kjar'', ``1st'' and ``2nd'' denote the Kolsky model, the Kjartansson model and the first- and second-order nearly constant-$Q$ models, respectively. The legend abbreviation ``GSLS'' denotes the generalized SLS model for nearly constant Q, determined by the $\tau$ method, and the optimized coefficients in the complex modulus formula (\ref{eq:Mgsls}) for this models can be known from Table \ref{tab:tabl10}. 
}
\label{fig:fig1}
\end{figure}

\begin{figure}[H]
\centering
\subfloat[$Q_{0} = 100$]
{\includegraphics[width=0.23\textwidth]{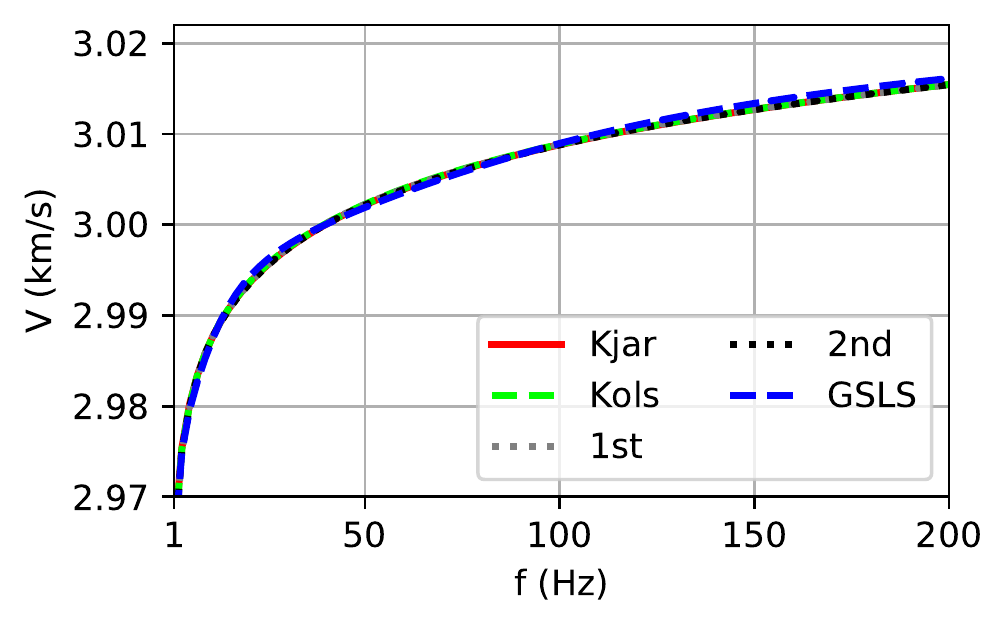}}
\subfloat[$Q_{0} = 60$]
{\includegraphics[width=0.23\textwidth]{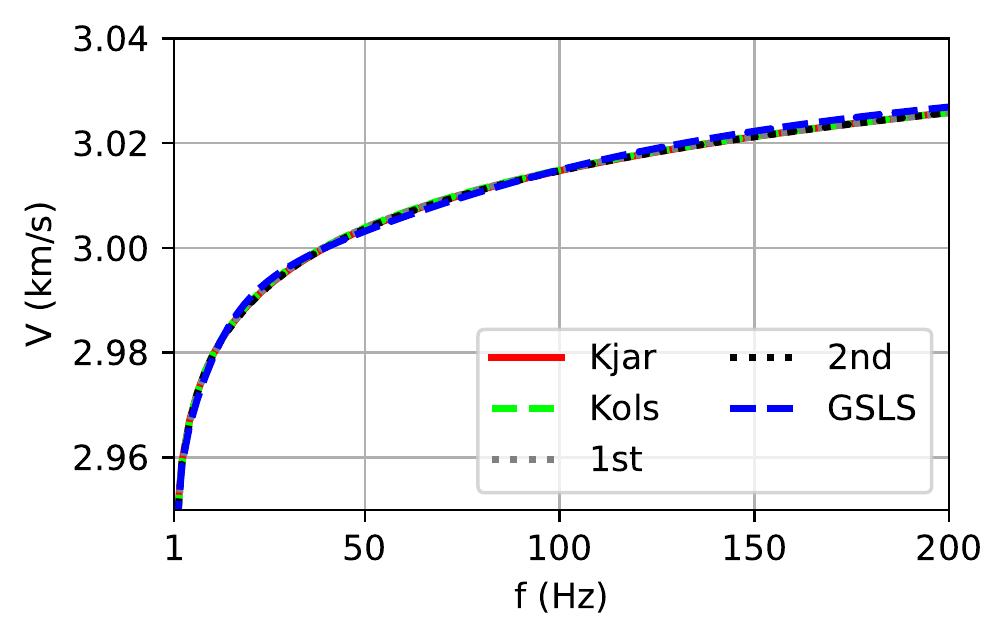}}
\subfloat[$Q_{0} = 30$]
{\includegraphics[width=0.23\textwidth]{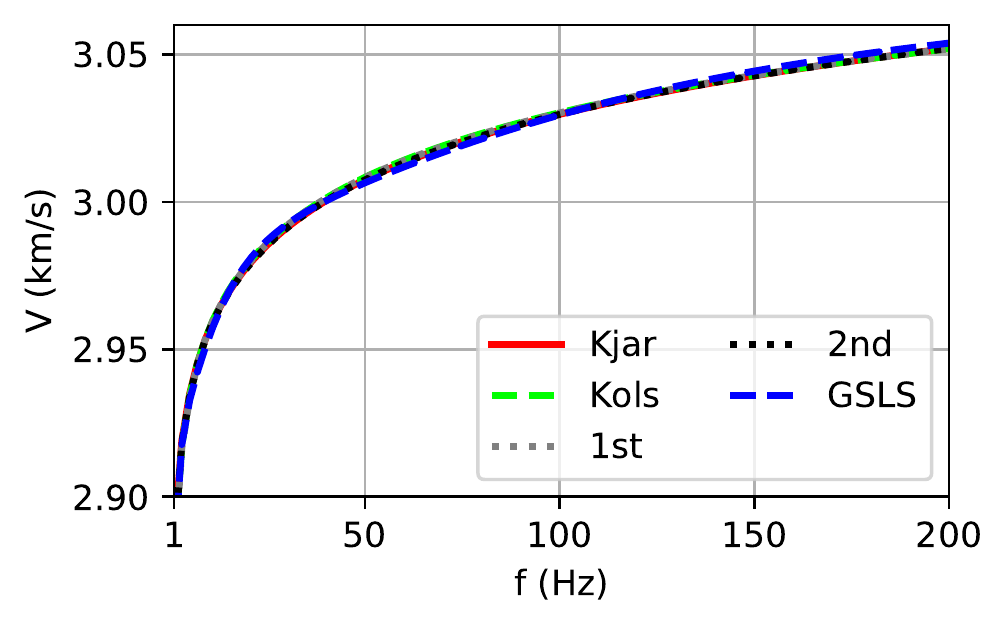}}
\subfloat[$Q_{0} = 5$]
{\includegraphics[width=0.23\textwidth]{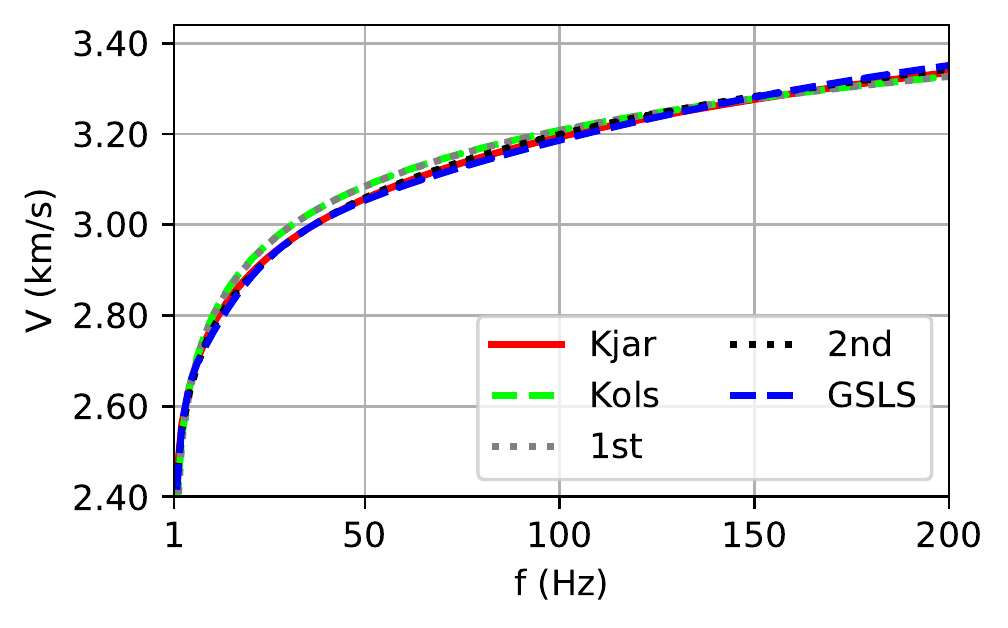}}
\caption{
The variation of the phase velocity with frequency. The plot order, the legend abbreviations and the parameters are the same as those in Figure \ref{fig:fig1}.  
}
\label{fig:fig2}
\end{figure}

We first analyze the quality factor and velocity for these dissipative models. 
The complex moduli for the Kolsky and Kjartansson models and the first- and second-order models, from which the quality factor and the phase velocity are calculated, are given in equations \ref{eq:Mkolsky}, \ref{eq:Mkjar}, \ref{eq:M1st} and \ref{eq:M2nd}, respectively. For the first- and second-order nearly constant-$Q$ models, the relaxation times in the weighting function are shown in Table \ref{tab:tabl4}. 
The complex modulus for the generalized SLS model for nearly constant $Q$, the relaxation times of which are determined by the $\tau$-method (see Table \ref{tab:tabl10}), has the same form as equation \ref{eq:Mgsls} but parameter $M_{R}'$ is determined by fitting at $\omega=\omega_{0}$ the real part of equation \ref{eq:Mgsls} with the real part of equation \ref{eq:Mkjar}.   
As illustrated in Figure \ref{fig:fig1}, except for frequencies quite close to $1$~Hz (the lower bound of the frequency range of interest), the quality factors for the first- and second-order nearly constant $Q$ models match well with those for the Kolsky model and the Kjartansson model, respectively, and their respective maximum deviations are less than one in all the attenuation cases. From the perspective of approximation, the term $1/Q_{0}$ governs the deviation of the complex moduli for the first- and second-order nearly constant $Q$ models from those for the Kolsky and Kjartansson models, as illustrated in equations \ref{eq:M1st} and \ref{eq:M2nd}. With the increase in $Q_{0}$, the complex moduli for the first- and second-order nearly constant $Q$ models become closer to those for the Kolsky and Kjartansson models. The quality factor curves (blue dashed lines) from the generalized SLS model are of oscillatory shape except for the extremely strong attenuation case ($Q_{0} = 5$). Figure \ref{fig:fig2} shows that except for the extremely strong attenuation case the first- and second-order nearly constant Q models, the Kolsky model and the Kjartansson model have almost the same velocity variation in the frequency range of interest. It implies that the effect of the second term $1/Q_{0}^2$ on the phase velocity is negligible even for the strong attenuation case ($Q_{0}=30$). In the extremely strong attenuation case ($Q_{0}=5$), the velocities from the first- and second-order nearly constant Q models are quite close to those from the Kolsky and Kjartansson models, respectively. However, a difference is observable between the velocities from the Kolsky and Kjartansson models. In all four attenuation cases, the velocity from the generalized SLS model is close to that for the Kjartansson model, although a slight difference between them can be found in particular for high frequencies ($[150,200]$~Hz).

\begin{table}[H]
\centering
\caption{The optimal parameters for the generalized SLS model (equation \ref{eq:Mgsls}) with five-elements, determined by the $\tau$ method \cite[]{blanch:1995,bohlen:2002} for various constant Q values in the frequency range $[1,200]$~Hz. In the following table, $\tau$ is given by $\tau=\tau_{\epsilon l}^\prime / \tau_{\sigma l} - 1$, from which $\tau_{\epsilon l}^\prime$ can be known.}
\label{tab:tabl10}
\small{
\begin{tabular}{c c c c c c c}
\toprule
$Q_{0}$ & $\tau$ & $\tau_{\sigma 1}$~(s) & $\tau_{\sigma 2}$~(s) & $\tau_{\sigma 3}$~(s)& $\tau_{\sigma 4}$~(s) & $\tau_{\sigma 5}$~(s)  \\
\midrule
5   & 2.0521     & 6.652208 $\times 10^{-2}$ & 6.964370 $\times 10^{-3}$ & 1.084894 $\times 10^{-3}$ & 4.007076 $\times 10^{-4}$ & 2.43423489 $\times 10^{-5}$  \\
30  & 2.433 $\times 10^{-1}$	 & 8.985713 $\times 10^{-2}$ & 1.147818 $\times 10^{-2}$ & 1.445100 $\times 10^{-3}$ & 1.991529 $\times 10^{-4}$ & 1.59155135 $\times 10^{-5}$ \\
60  & 1.164 $\times 10^{-1}$	 & 9.043722 $\times 10^{-2}$ & 1.229250 $\times 10^{-2}$ & 1.524493 $\times 10^{-3}$ & 1.990469 $\times 10^{-4}$ & 1.59155038 $\times 10^{-5}$ \\
100 & 6.860 $\times 10^{-2}$	 & 9.046149 $\times 10^{-2}$ & 1.261133  $\times 10^{-2}$ & 1.555575 $\times 10^{-3}$ & 1.990116 $\times 10^{-4}$ & 1.59155013 $\times 10^{-5}$ \\
\bottomrule   
\end{tabular}
}
\end{table}

\begin{figure}[H]
\centering
\subfloat[$Q_{0} = 100$]
{\includegraphics[width=0.23\textwidth]{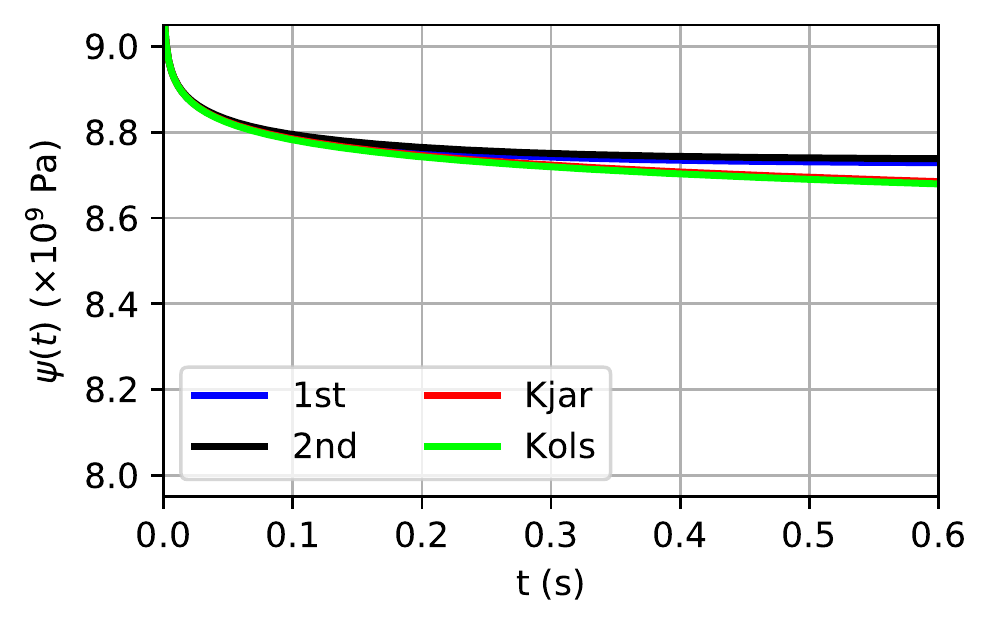}}
\subfloat[$Q_{0} = 60$]
{\includegraphics[width=0.23\textwidth]{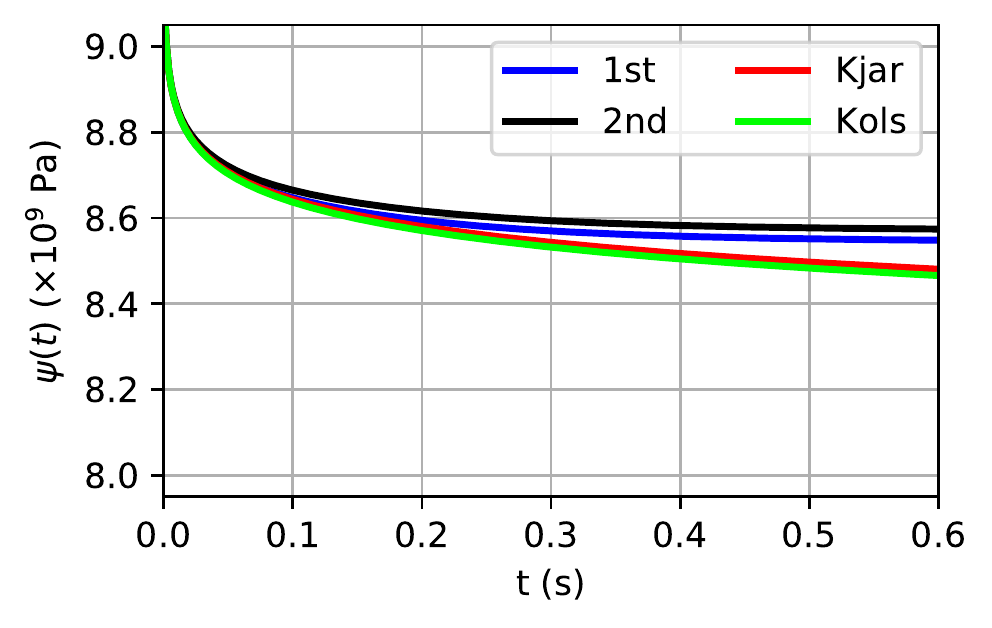}}
\subfloat[$Q_{0} = 30$]
{\includegraphics[width=0.23\textwidth]{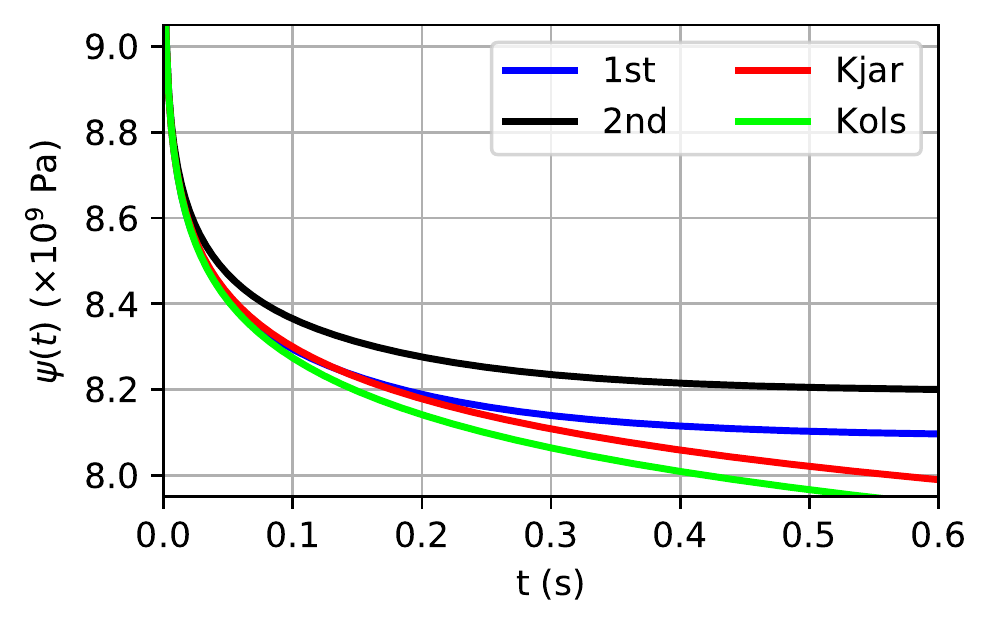}}
\subfloat[$Q_{0} = 5$]
{\includegraphics[width=0.23\textwidth]{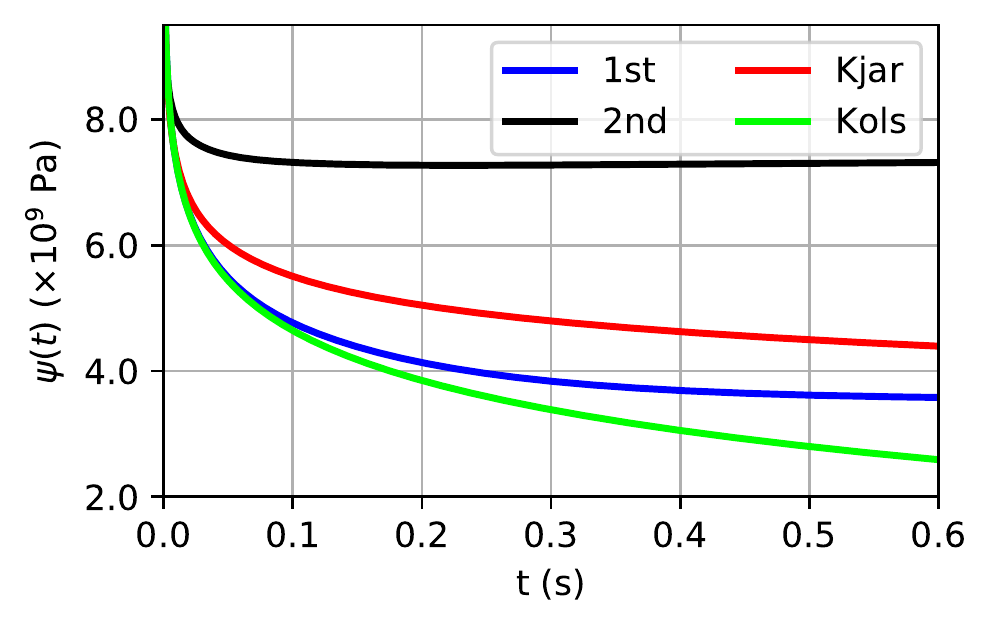}}
\caption{
The variation of the relaxation function with frequency. The plot order, the legend abbreviations and the parameters are the same as those in Figure \ref{fig:fig1}. 
}
\label{fig:fig3}
\end{figure}

\begin{figure}[H]
\centering
\subfloat[$Q_{0} = 100$]
{\includegraphics[width=0.23\textwidth]{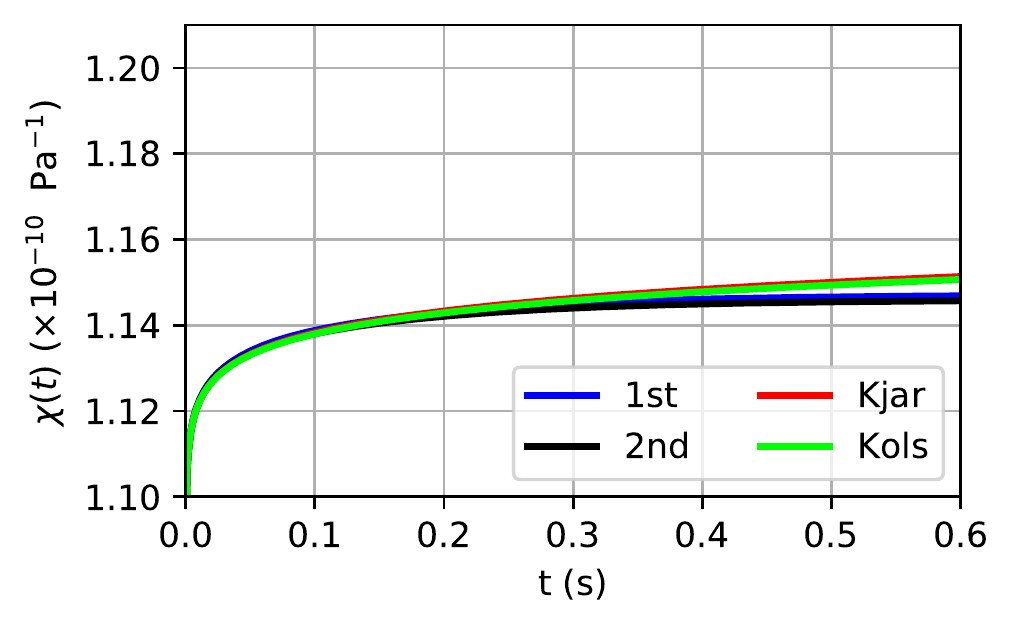}}
\subfloat[$Q_{0} = 60$]
{\includegraphics[width=0.23\textwidth]{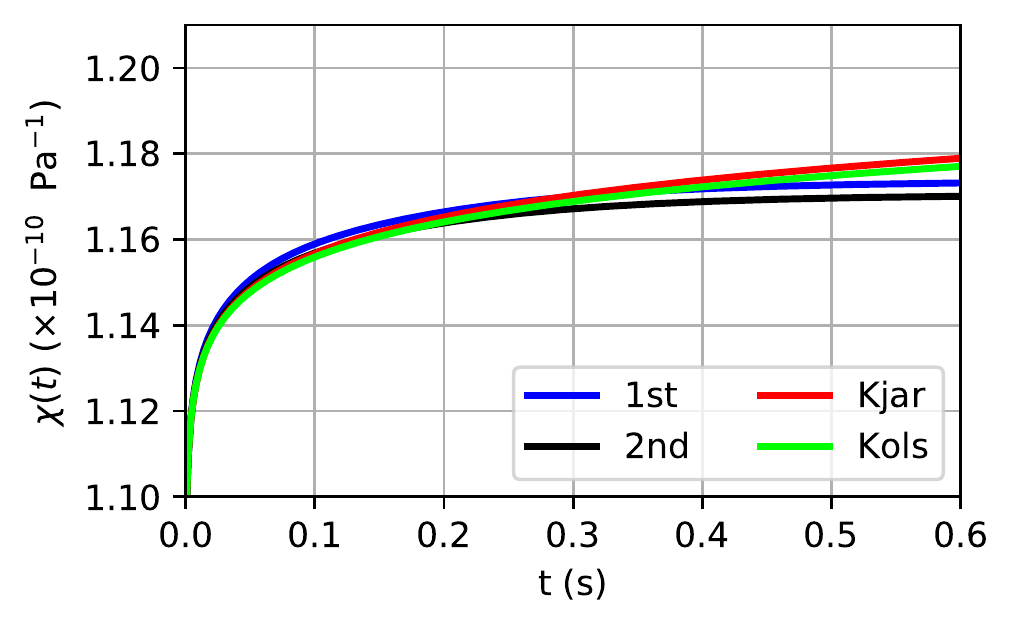}}
\subfloat[$Q_{0} = 30$]
{\includegraphics[width=0.23\textwidth]{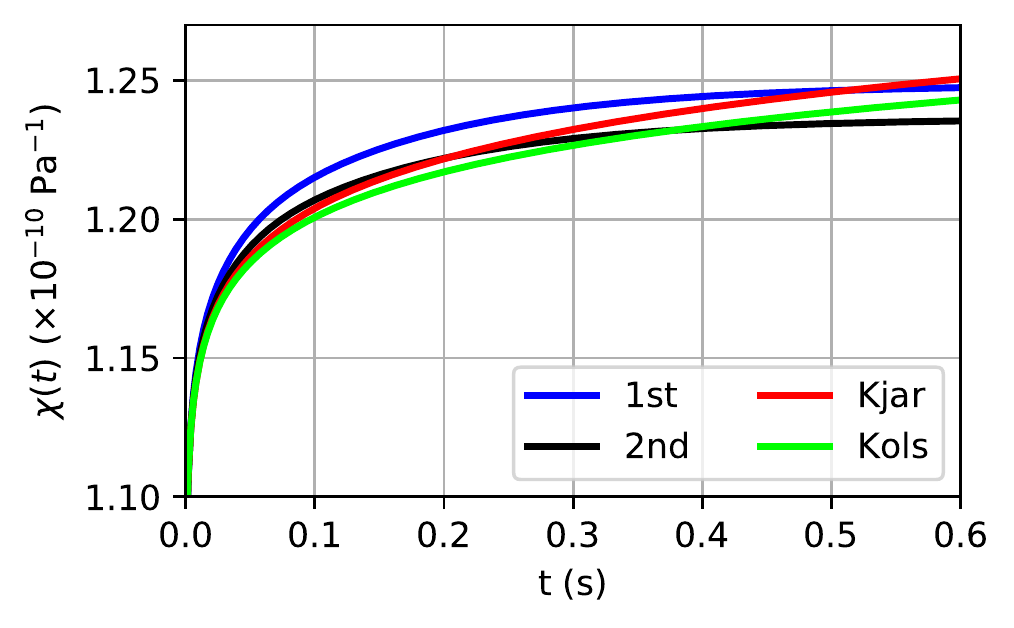}}
\subfloat[$Q_{0} = 5$]
{\includegraphics[width=0.23\textwidth]{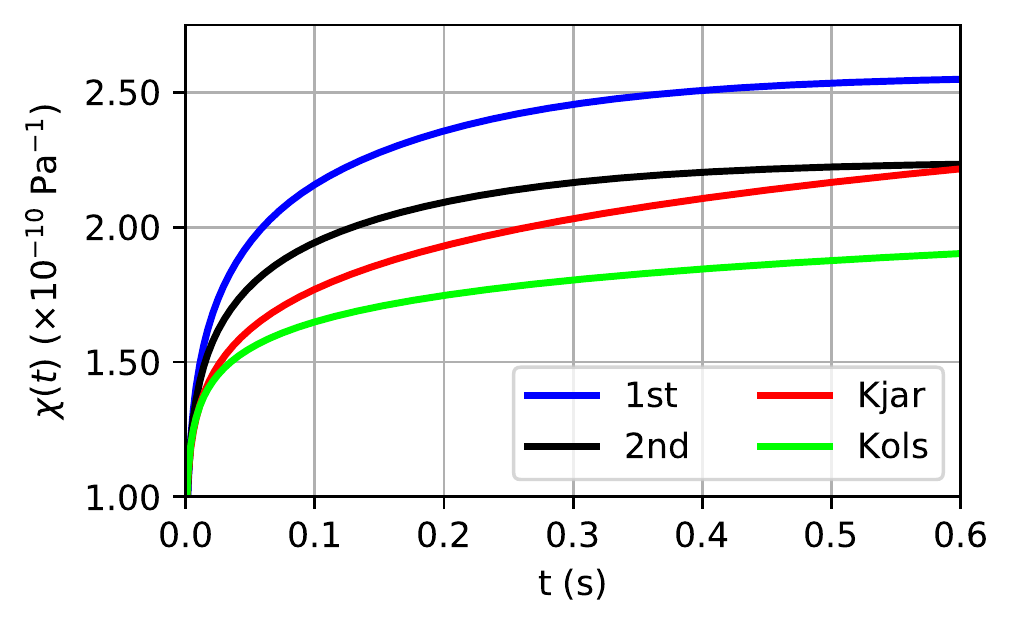}}
\caption{
The variation of the creep function with frequency. The plot order, the legend abbreviations and the parameters are the same as those in Figure \ref{fig:fig1}. 
}
\label{fig:fig4}
\end{figure}

We next analyze the relaxation and creep functions. The relaxation functions for the Kolsky and Kjartansson models and the first- and second-order nearly constant $Q$ models are given in equations \ref{eq:psi_kols}, \ref{eq:psi_kjar}, \ref{eq:psi1st} and \ref{eq:psi2nd}, respectively. Figure \ref{fig:fig3} shows a comparison between the relaxation functions for all these dissipative models except the generalized SLS model. Unlike the results for quality factor and velocity, only in the weak attenuation case do the relaxation functions for the first- and second-order nearly constant $Q$ models match with those for the Kolsky and the Kjartansson model, respectively. Their difference increases with $1/Q_{0}$. However, the relaxation functions for the first- and second-order constant $Q$ models have a similar frequency variation trend. The relaxation function for the Kjartansson model is close to that for the Kolsky model in the strong attenuation case ($Q_{0} = 30$). The difference between the relaxation functions for all the models decreases as the quality factor parameter $Q_{0}$ increases. The creep functions for the Kolsky and Kjartansson models and the first- and second-order nearly constant $Q$ models are given in equations \ref{eq:chi_kols}, \ref{eq:chi_kjar}, \ref{eq:chi1st} and \ref{eq:chi2nd}, respectively. 
The series in equations \ref{eq:chi1st} and \ref{eq:chi2nd} are truncated up to $n=36$, which ensures that the truncated series approach the exact results as much as possible. As shown in Figure \ref{fig:fig4}, we reach a similar conclusion for the creep functions. 

We last compare the first- and second-order nearly constant $Q$ models with the Kolsky and Kjartansson models, from a theoretical perspective.  
Comparing equations \ref{eq:M1st} and \ref{eq:M2nd} with equations \ref{eq:Mkolsky} and \ref{eq:Mkjar} at zero and infinite frequencies, we may find that (1) the relaxed and unrealxed moduli (corresponding to $\omega=0$ and $\omega=\infty$, respectively) for the first- and second-order nearly constant $Q$ models are finite and physically plausible; (2) the relaxed moduli for the Kolsky and Kjartansson models are negative infinity and zero, respectively; (3) the unrelaxed moduli for the Kolsky and Kjartansson models are complex infinity. Zero or infinite modulus does not exist for real rocks. Besides, the relaxation and creep functions for the first- and second-order nearly constant $Q$ models are quite different from those for the Kolsky and Kjartansson models. Equations \ref{eq:psi1st} and \ref{eq:psi2nd} show that the relaxation functions for the first- and second-order nearly constant $Q$ models are finite and positive at infinite time. However, equations \ref{eq:psi_kols} and \ref{eq:psi_kjar} show that the relaxation functions for the Kolsky and Kjartansson model are negative infinity and zero at infinite time, respectively. These two cases cannot happen in real rocks. From equation \ref{eq:chi1st} and \ref{eq:chi2nd}, it is hard to know the values of the creep functions for the first- and second-order nearly constant $Q$ models at $t=\infty$. Our numerical testing shows that the creep functions for the first- and second-order nearly constant $Q$ models tend to finite values as time increases. 
Equations \ref{eq:chi_kols} and \ref{eq:chi_kjar} show that the creep functions for the Kolsky and Kjartansson models increase with time and finally approach infinity. 
The physical meaning of the creep function is the strain response of a unit step function in stress, starting at zero time. 
For real rocks, it cannot happen that this strain response becomes infinitely large.

Overall, the first- and second-order nearly constant $Q$ models are distinct from the Kolsky and Kjartansson models, although the complex moduli for these two new models are quite close to the complex moduli for these two existing models in a frequency range of interest. 
The Kolsky and Kjartansson models exhibit non-physical behavior at very low and very high frequency but they can be used to interpret observations of the nearly constant $Q$ in a frequency range of interest.

\section{Viscoacoustic wave equations}
The viscoacoustic wave equation for a general dissipative model can be formulated from the constitutive relations, the relationship between stress and pressure, the relationship between the cubical dilatation and the particle displacement (or the strain), and the equation of motion. The viscoacoustic wave equation for a general dissipative model is expressed as:
\begin{equation} \label{eq:VAWEgen}
\frac{\partial^2 P}{\partial t^2} = \phi(t) \odot (\nabla^2 P) 
+ S,
\end{equation}
where $\phi(t)=\psi(t)/\rho$ denotes the density-normalized relaxation function, $P$ denotes the pressure, and $\rho$ and $S$ denote density and source, respectively. $\nabla^2 = \partial^2/\partial x^2 + \partial^2/\partial y^2 + \partial^2/\partial z^2$ denotes the Laplacian operator, where $x$, $y$ and $z$ denote the Cartesian coordinates. 

The viscoacoustic wave equation \ref{eq:VAWEgen} is essentially an integral-differential equation. For the first- and second-order nearly constant $Q$ models, we may transform it to differential form, which may be solved efficiently by multiple time-domain methods as mentioned above.

From equation \ref{eq:W_Wr}, we rewrite the term $W(\omega)-W(\omega_{0})$ as:
\begin{equation} \label{eq:WWr}
W(\omega) - W_{R}(\omega_{0}) = g - h(\omega)  ,
\end{equation}
with
\begin{align}
\label{eq:g}
& g = \sum_{l=1}^{L} 
\frac{\frac{\tau_{\epsilon l}}{\tau_{\sigma l}}-1}
{1+\omega_{0}^2\tau_{\sigma l}^2} , \\
\label{eq:h}
& h(\omega) = \sum_{l=1}^{L}
\frac{\frac{\tau_{\epsilon l}}{\tau_{\sigma l}}-1}
{1-i\omega\tau_{\sigma l}} .
\end{align}

We substitute the moduli \ref{eq:M1st} and \ref{eq:M2nd} for the first- and second-order nearly constant $Q$ models into the frequency-domain viscoacoustic wave equation, and further substitute equation \ref{eq:WWr} with equations \ref{eq:g} and \ref{eq:h}. We next adopt the first of the frequency-domain methods in \cite{hao.greenhalgh:2019} to derive the viscoacoustic wave equations in differential form. The derivation is given in detail in Appendix D. In fact, these wave equations may also be obtained by the time-domain methods in \cite{hao.greenhalgh:2019}. The viscoacoustic wave equations are summarized below.
  
For the first-order nearly constant $Q$ model, the viscoacoustic wave equations are given by:
\begin{equation} \label{eq:VWE1}
\begin{split}
& \frac{\partial^2 P}{\partial t^2} = v_{U}^2 \nabla^2 P
- v_{H}^2 \sum_{l=1}^{L} r_{l} 
+ S , \\
& \frac{\partial r_{l}}{\partial t} = s_{l} \nabla^2 P
- \frac{1}{\tau_{\sigma l}} r_{l} ,
\end{split}
\end{equation}
with
\begin{align}
& v_{U}^2 = v_{0}^2 \left(
1 + \frac{g}{Q_{0}} \right),              \\
& v_{H}^2 = \frac{v_{0}^2}{Q_{0}},       \\
\label{eq:sl}
& s_{l} = \frac{1}{\tau_{\sigma l}} 
\left(\frac{\tau_{\epsilon l}}{\tau_{\sigma l}} - 1 \right) , 
\end{align}
where $v_{0} = \sqrt{M_{0}/\rho}$ denotes the reference velocity corresponding to $Q_{0}=\infty$ (no attenuation). Quantity $v_{U}$ denotes the unrelaxed velocity corresponding to $\omega=\infty$.  Quantity $v_{H}$ denotes the velocity corresponding to the coefficient in front of $h(\omega)$ in equation \ref{eq:M1st_hser} of Appendix D. Quantity $S$ denotes the source term. Quantity $g$ is given in equation \ref{eq:g}. 

For the second-order nearly constant $Q$ model, the viscoacoustic wave equations are written as:
\begin{equation} \label{eq:VWE2}
\begin{split}
& \frac{\partial^2 P}{\partial t^2} = \tilde{v}_{U}^2 \nabla^2 P
- \tilde{v}_{H1}^2 \sum_{l=1}^{L} r_{l}^{(1)} 
+ \tilde{v}_{H2}^2 \sum_{l=1}^{L} r_{l}^{(2)} 
+ S , \\
& \frac{\partial r_{l}^{(1)}}{\partial t} = s_{l} \nabla^2 P
- \frac{1}{\tau_{\sigma l}} r_{l}^{(1)} , \\
& \frac{\partial r_{l}^{(2)}}{\partial t} = s_{l} \sum_{l=1}^{L} r_{l}^{(1)} 
- \frac{1}{\tau_{\sigma l}} r_{l}^{(2)} ,
\end{split}
\end{equation}
with
\begin{align}
& \tilde{v}_{U}^2 = v_{0}^2 \left(
1 + \frac{g}{Q_{0}} + \frac{g^2}{2Q_{0}^2} \right), \\
& \tilde{v}_{H1}^2 = \frac{v_{0}^2}{Q_{0}} \left(1 + \frac{g}{Q_{0}} \right), \\
& \tilde{v}_{H2}^2 = \frac{v_{0}^2}{2Q_{0}^2} ,
\end{align}
where quantity $\tilde{v}_{U}$ denotes the unrelaxed velocity for the second-order nearly constant $Q$ model, corresponding to $\omega=\infty$. Quantities $\tilde{v}_{H1}$ and $\tilde{v}_{H2}$ denote the velocities corresponding to the coefficients in front of $h(\omega)$ and $h^2(\omega)$ in equation \ref{eq:M2nd_hser}, respectively. Quantities $g$ and $s_{l}$ are given in equations \ref{eq:g} and \ref{eq:sl}, respectively. Ignoring the terms associated with $1/Q_{0}^2$, the viscoacoustic wave equations \ref{eq:VWE2} for the second-order nearly constant $Q$ model reduce to the viscoacoustic wave equations \ref{eq:VWE1} for the first-order nearly constant $Q$ model.

\section{Numerical examples of wave propagation}
In the first example, we analyze the dissipative waves generated by a point source in the Kjartansson model, the Kolsky model, the first- and second-order nearly constant $Q$ models and the generalized SLS model determined by the $\tau$ method \cite[]{blanch:1995,bohlen:2002}. The point-source solution of the acoustic wave equation can be found in \cite{aki.richards:1980} and \cite{pujol:2003}, and its frequency-domain version can be obtained by the Fourier transform (equation \ref{eq:Fourier}). According to the correspondence principle \cite[]{ben-menahem.singh:1981}, we may replace the real modulus in the frequency-domain point-source solution of the acoustic wave equation by the complex modulus, to obtain the solution of the viscoacoustic wave equation. As an alternative, we may directly simplify the point-source solution of the viscoacoustic anisotropic wave equation in \cite{hao.alkhalifah:2019} to the isotropic case. 

\begin{figure}[H]
\centering
\subfloat[Wavelet]{\includegraphics[width=4.5cm]{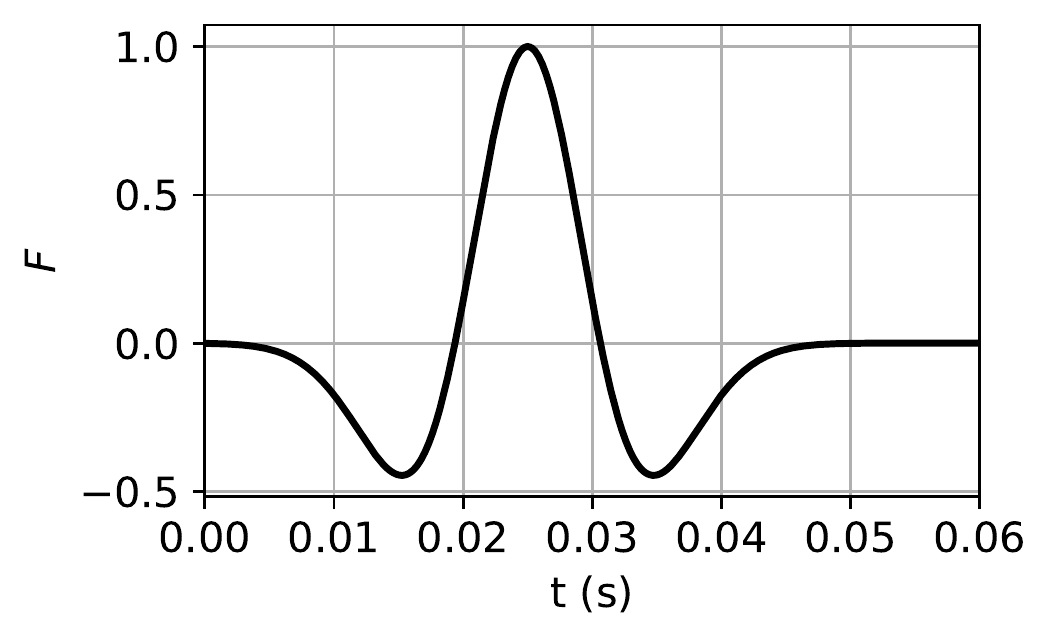}}
\qquad
\subfloat[Spectrum]{\includegraphics[width=4.5cm]{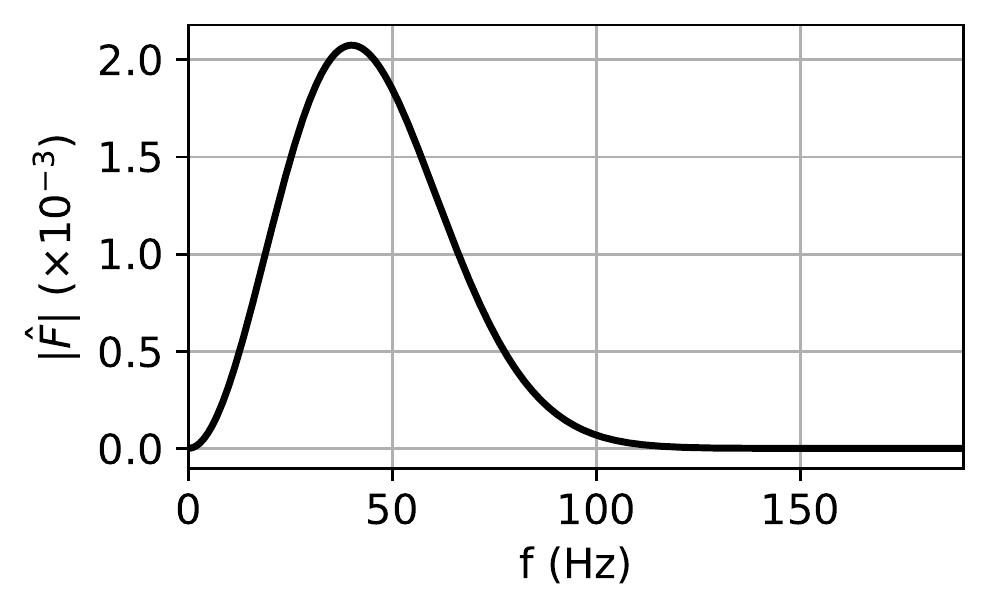}}
\caption{
A Ricker wavelet and its amplitude spectrum. The dominant frequency of the Ricker wavelet is 40~Hz. 
}
\label{fig:fig5}
\end{figure}

The time-domain viscoacoustic wave equation is given in equation \ref{eq:VAWEgen}. We denote the source term as $S = F\delta(\mathbf{x})$, where $F$ denotes the source wavelet in the time domain and its dimension is set as $10^9$~Pa. This is to make the plot ordinate amplitudes clearer and more reasonable by eliminating the effect of the distance dimension (km=$10^3$~m) and the velocity dimension (km/s=$10^3$~m/s) squared in the denominator term on the magnitude of the point-source solution. Quantity $\delta(.)$ denotes the Dirac delta function, and $\mathbf{x}$ denotes the Cartesian coordinate vector. As illustrated in Figure \ref{fig:fig5}, the source function is a Ricker wavelet with a unit peak amplitude and dominant frequency of 40~Hz. The amplitude spectrum of the source wavelet is distributed over frequencies much less than 200~Hz, which is the upper bound of the frequency range of interest for the relaxation time parameters in Table \ref{tab:tabl4}. Referring to equations \ref{eq:tau_scale}, we choose the scaling factor $\xi = 0.65$ to scale the parameters shown in Table \ref{tab:tabl4} valid for the frequency range $[1, 200]$~Hz to those valid for the frequency range $[0.65, 130]$~Hz. The amplitude spectrum of the source wavelet is completely concentrated inside the frequency range after scaling. In all these dissipative models, the reference frequency is set as $f_{0} = 40$~Hz, the corresponding reference angular frequency is known from $\omega_{0} = 2\pi f_{0}$, and the reference velocity is set as $v_{0} = 3$~km/s. To sufficiently analyze the effect of dissipation on wave propagation, we consider the following four attenuation cases: (1) weak attenuation ($Q_{0}=100$); (2) moderate attenuation ($Q_{0}=60$); (3) strong attenuation ($Q_{0}=30$); (4) extremely strong attenuation ($Q_{0}=5$). Here, $Q_{0}$ is the reference quality factor in a considered dissipative model. 

\begin{figure}[H]
\centering
\subfloat[$Q_{0} = 100$]
{\includegraphics[width=0.23\textwidth]{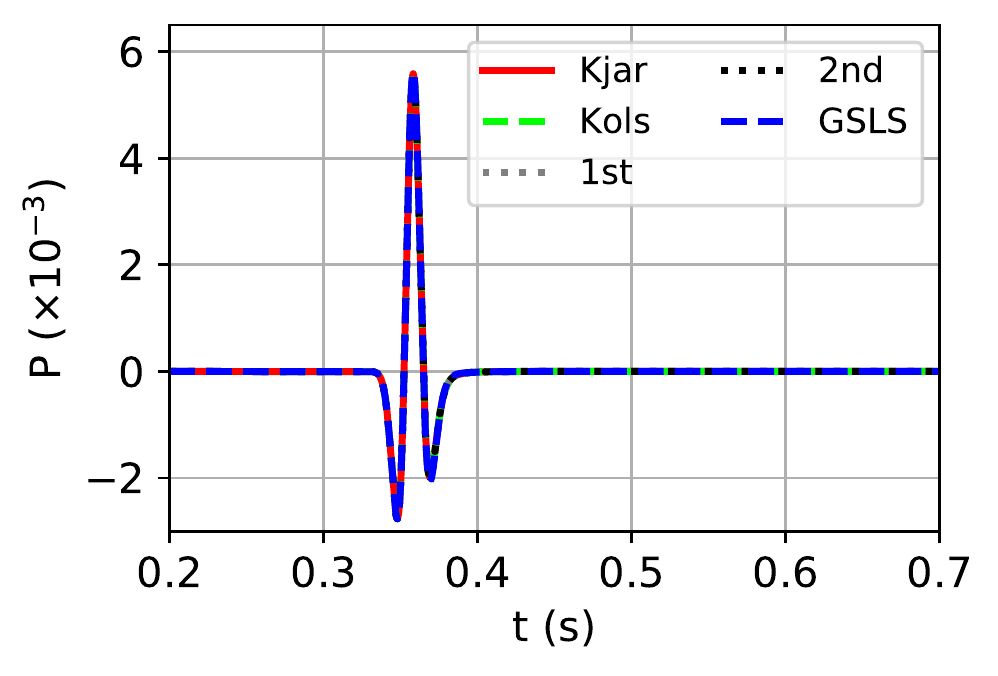}
\label{fig:waves_Qc100_fc40_r1}}
\subfloat[$Q_{0} = 60$]
{\includegraphics[width=0.23\textwidth]{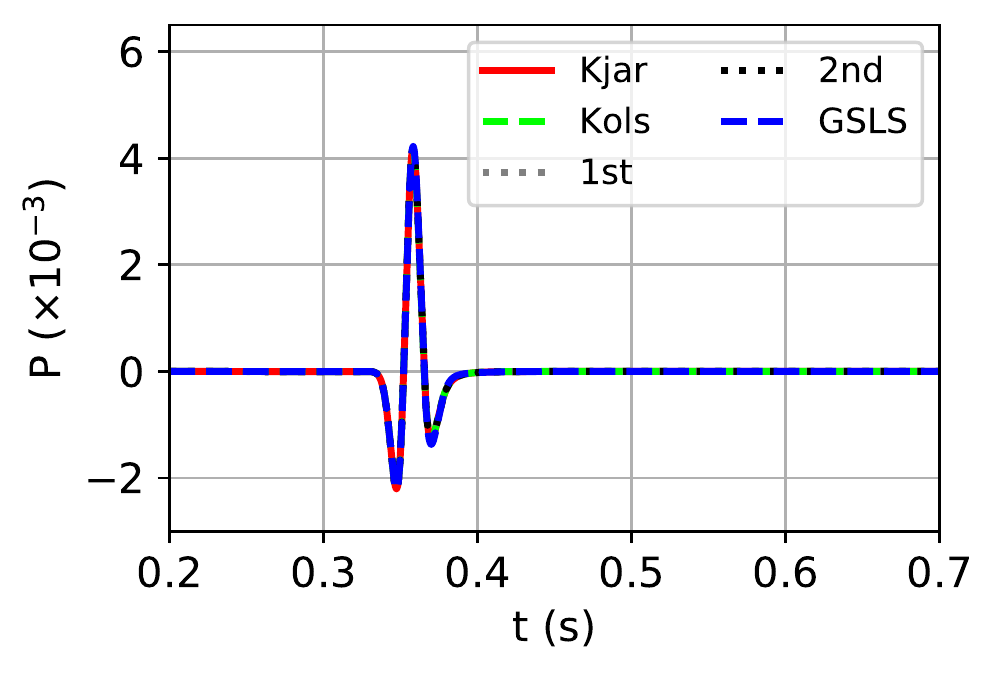}
\label{fig:waves_Qc60_fc40_r1}}
\subfloat[$Q_{0} = 30$]
{\includegraphics[width=0.23\textwidth]{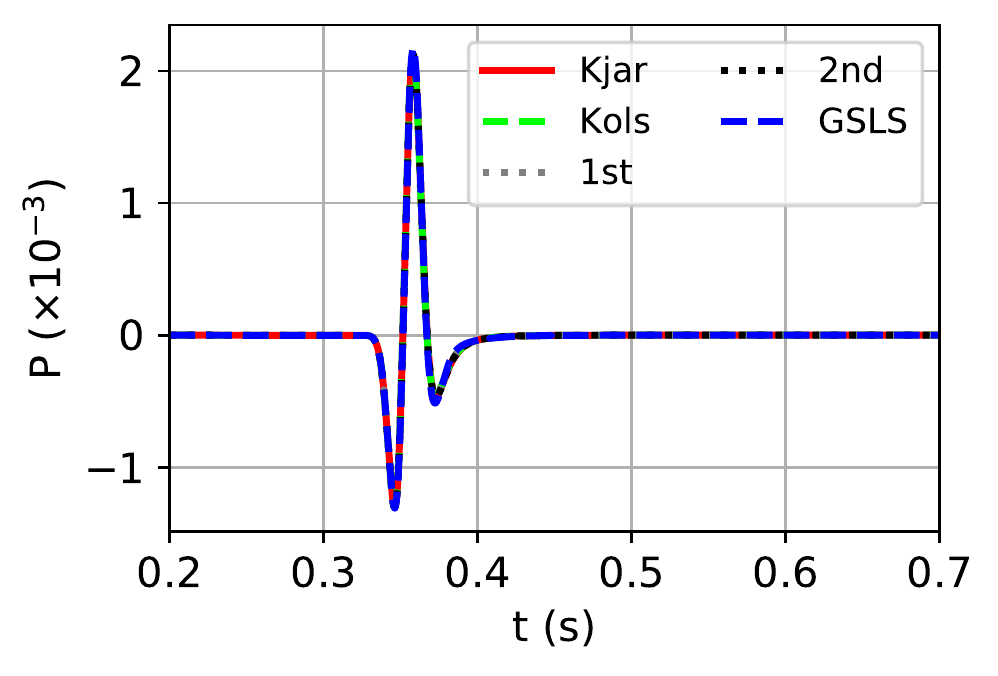}
\label{fig:waves_Qc30_fc40_r1}}
\subfloat[$Q_{0} = 5$]
{\includegraphics[width=0.23\textwidth]{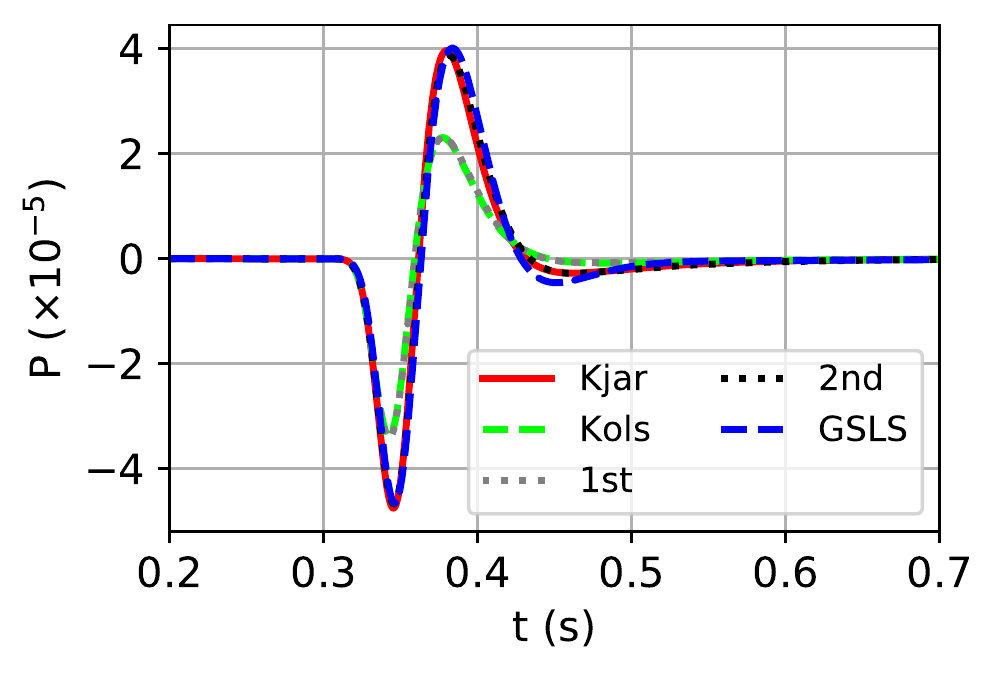}
\label{fig:waves_Qc5_fc40_r1}}
\caption{
Waveforms at a propagation distance $r=1$~km in (a) the weak attenuation case ($Q_{0} = 100$), (b) the moderate attenuation case ($Q_{0} = 60$), (c) the strong attenuation case ($Q_{0} = 30$) and (d) the extremely strong attenuation case ($Q_{0} = 5$). The legend abbreviations ``Kjar'', ``Kols'', ``1st'', ``2nd'' and ``GSLS'' denote the same as those in Figure \ref{fig:fig1}. 
}
\label{fig:fig6}
\end{figure}

\begin{figure}[H]
\centering
\subfloat[$Q_{0} = 100$]
{\includegraphics[width=0.23\textwidth]{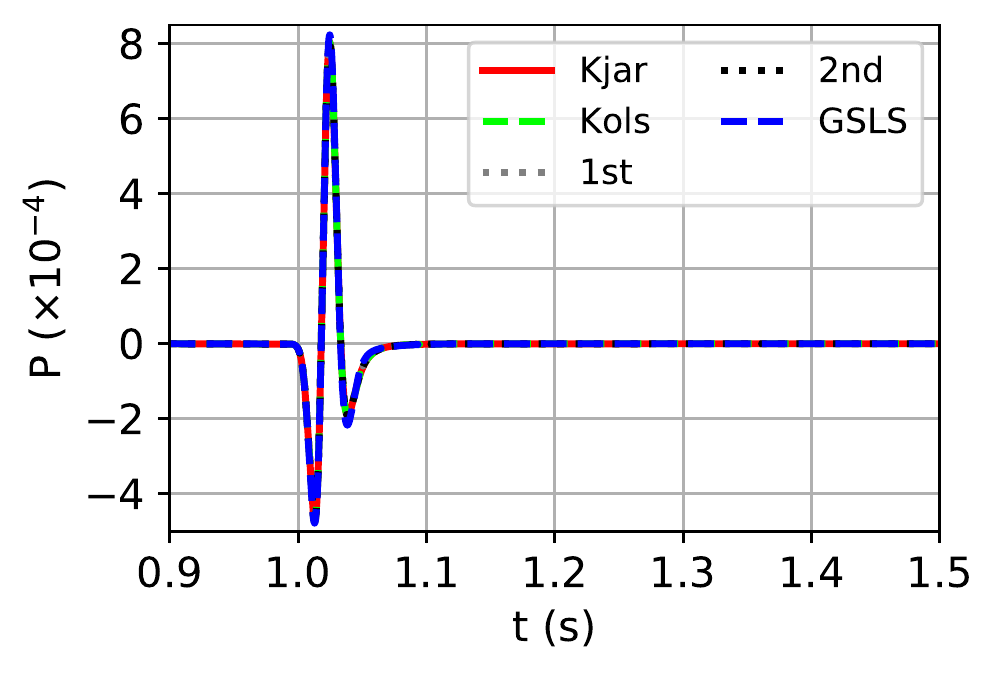}
\label{fig:waves_Qc100_fc40_r3}}
\subfloat[$Q_{0} = 60$]
{\includegraphics[width=0.23\textwidth]{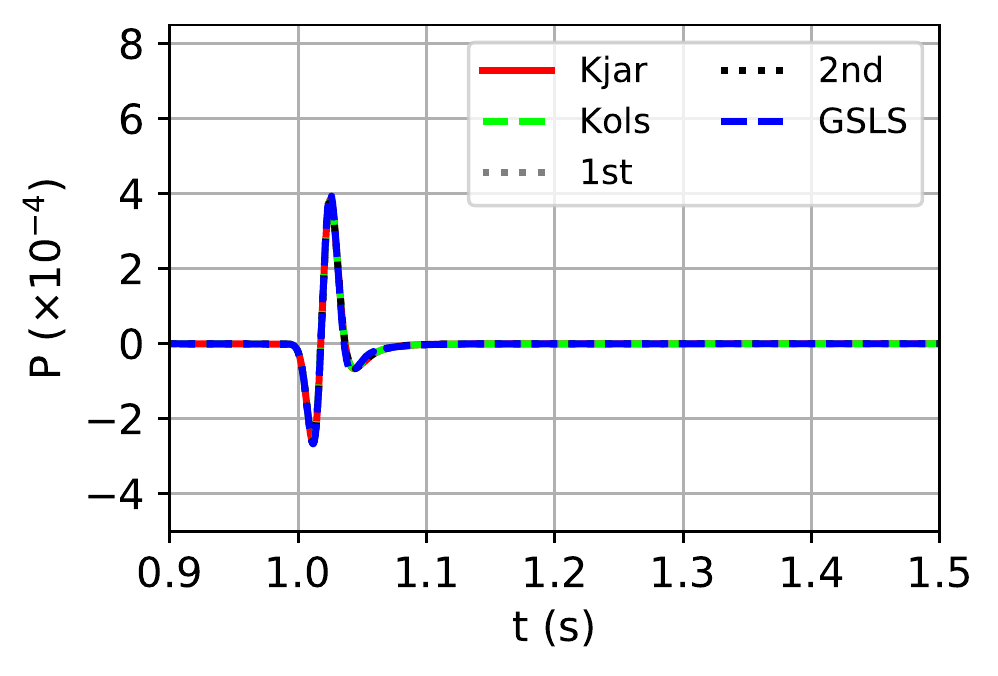}
\label{fig:waves_Qc60_fc40_r3}}
\subfloat[$Q_{0} = 30$]
{\includegraphics[width=0.23\textwidth]{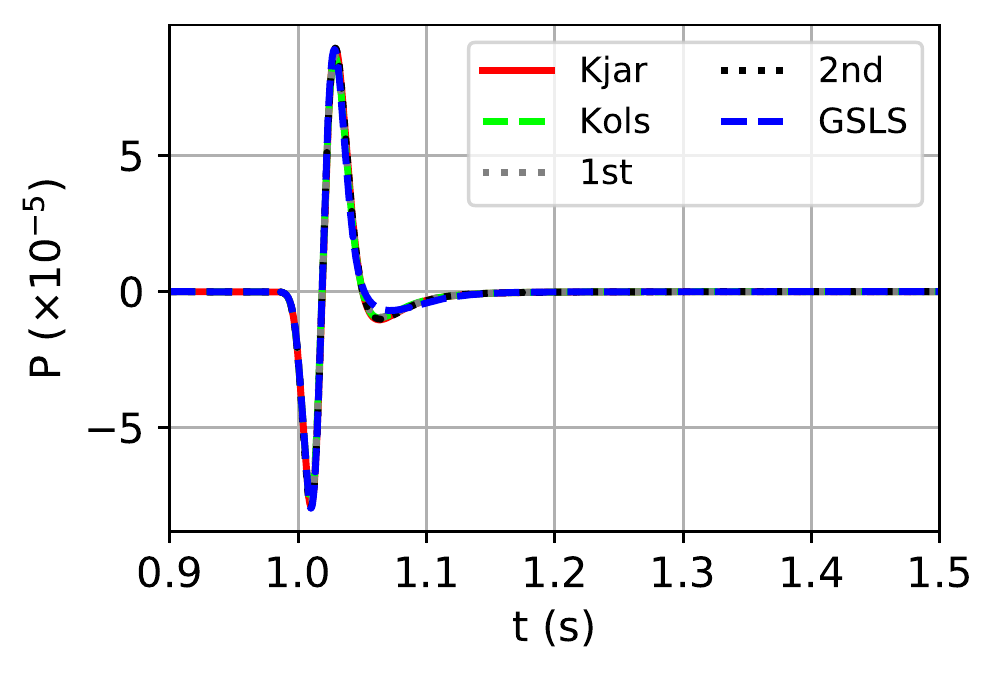}
\label{fig:waves_Qc30_fc40_r3}}
\subfloat[$Q_{0} = 5$]
{\includegraphics[width=0.23\textwidth]{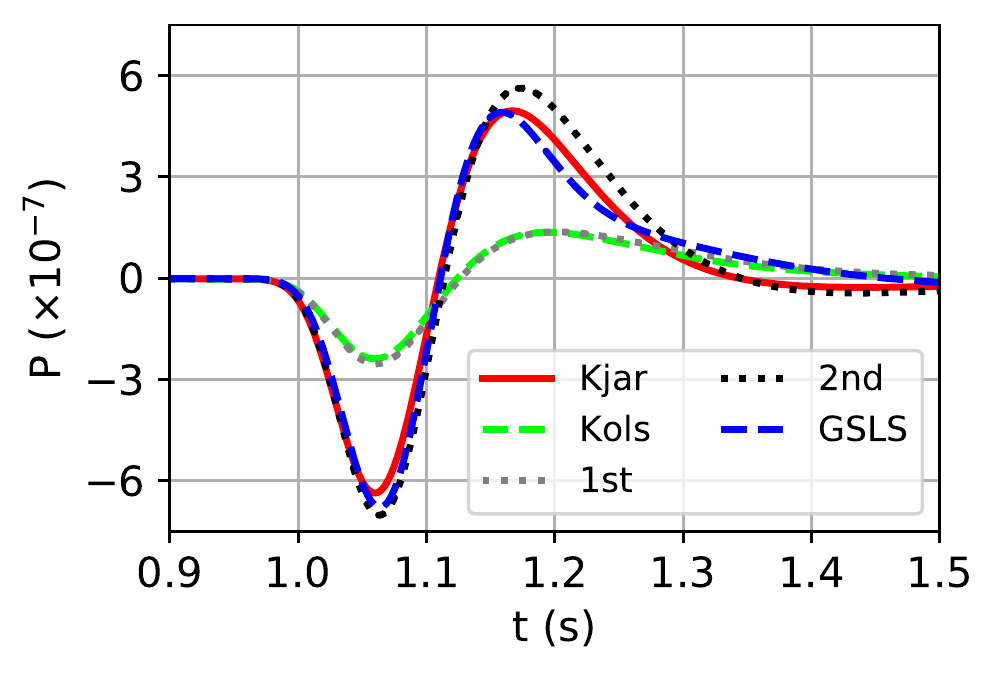}
\label{fig:waves_Qc5_fc40_r3}}
\caption{
Similar to Figure \ref{fig:fig6} but at a propagation distance $r=3$~km. 
}
\label{fig:fig7}
\end{figure}

\begin{figure}[H]
\centering
\subfloat[$Q_{0} = 100$]
{\includegraphics[width=0.23\textwidth]{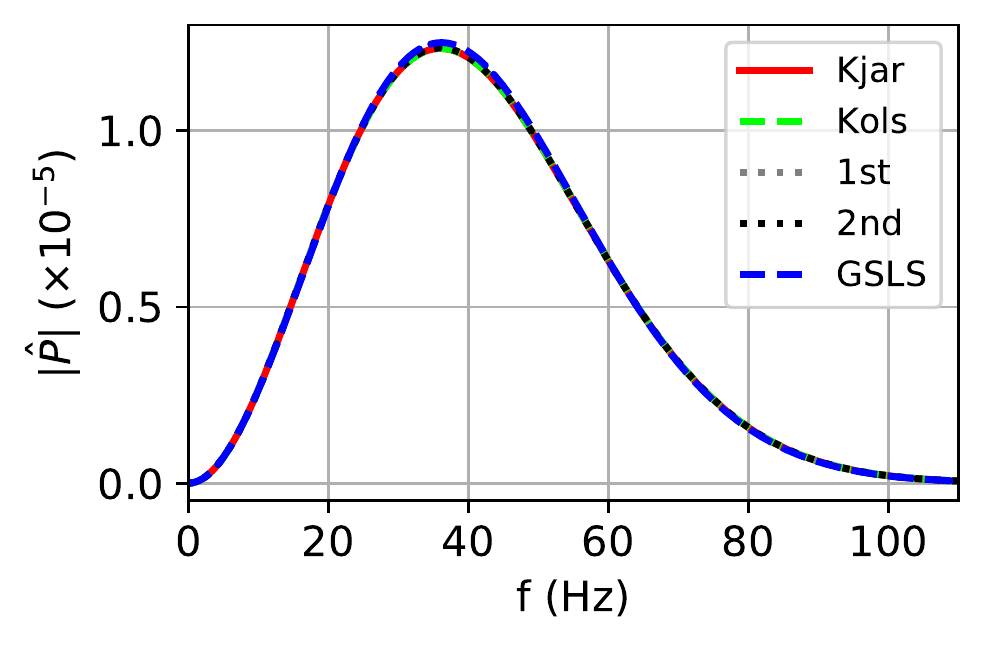}}
\subfloat[$Q_{0} = 60$]
{\includegraphics[width=0.23\textwidth]{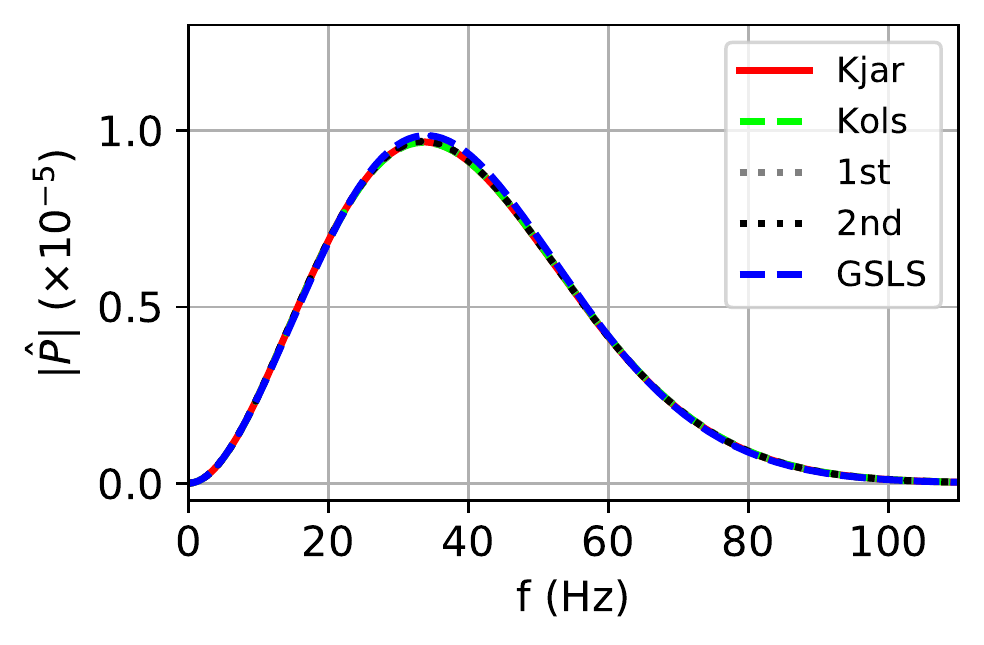}}
\subfloat[$Q_{0} = 30$]
{\includegraphics[width=0.23\textwidth]{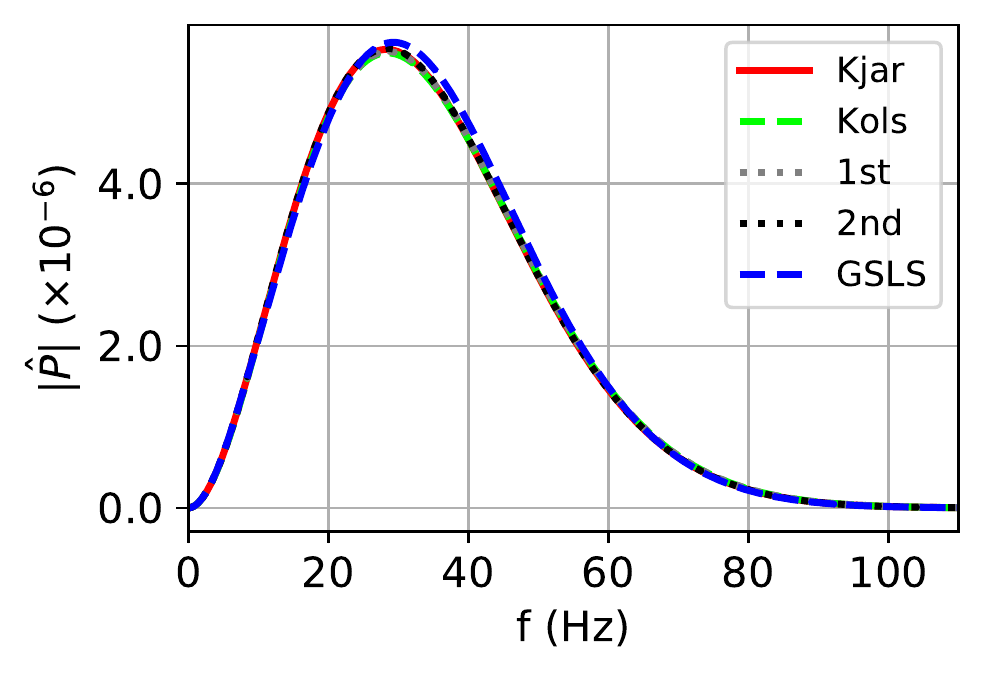}}
\subfloat[$Q_{0} = 5$]
{\includegraphics[width=0.23\textwidth]{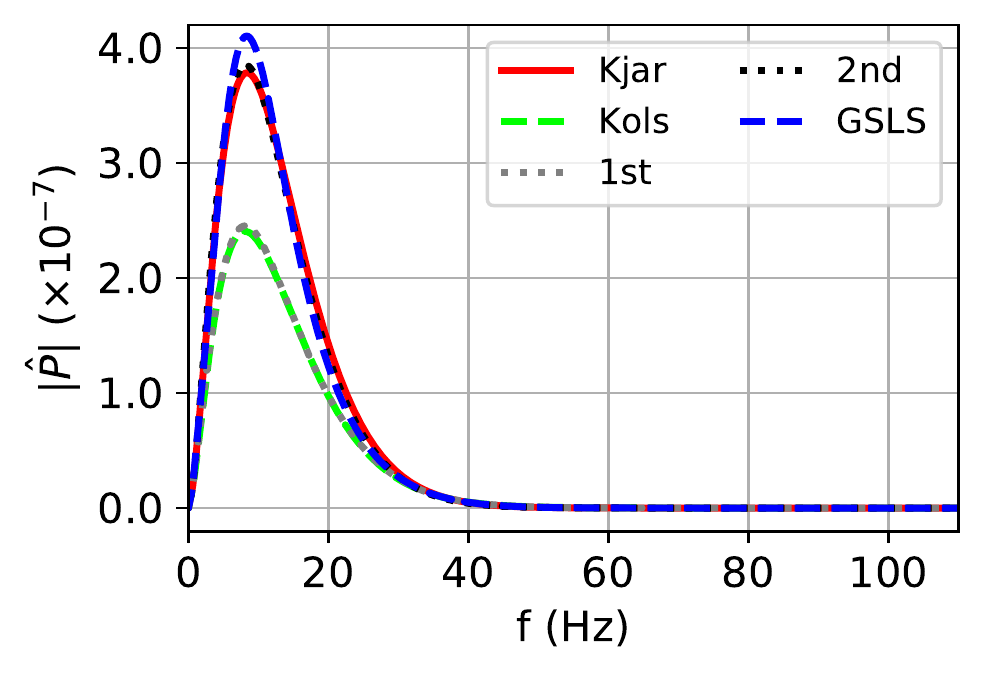}}
\caption{
The amplitude spectra of the waveforms at a propagation distance $r=1$~km in (a) the weak attenuation case ($Q_{0} = 100$), (b) the moderate attenuation case ($Q_{0} = 60$), (c) the strong attenuation case ($Q_{0} = 30$) and (d) the extremely strong attenuation case ($Q_{0} = 5$). The corresponding waveforms are shown in Figure \ref{fig:fig6}.  
}
\label{fig:fig8}
\end{figure}

\begin{figure}[H]
\centering
\subfloat[$Q_{0} = 100$]
{\includegraphics[width=0.23\textwidth]{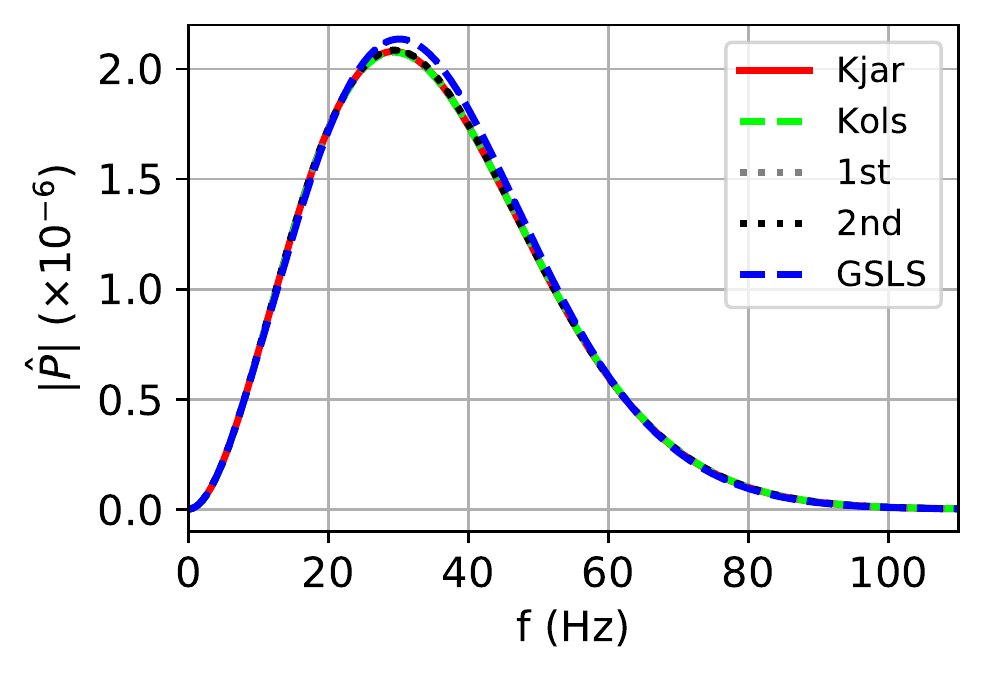}}
\subfloat[$Q_{0} = 60$]
{\includegraphics[width=0.23\textwidth]{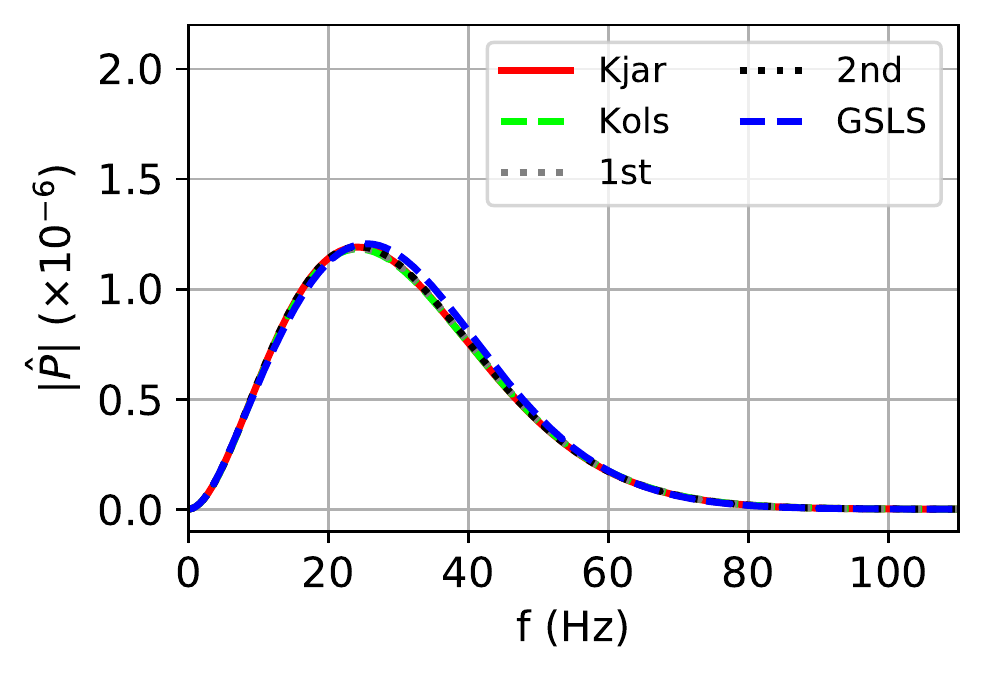}}
\subfloat[$Q_{0} = 30$]
{\includegraphics[width=0.23\textwidth]{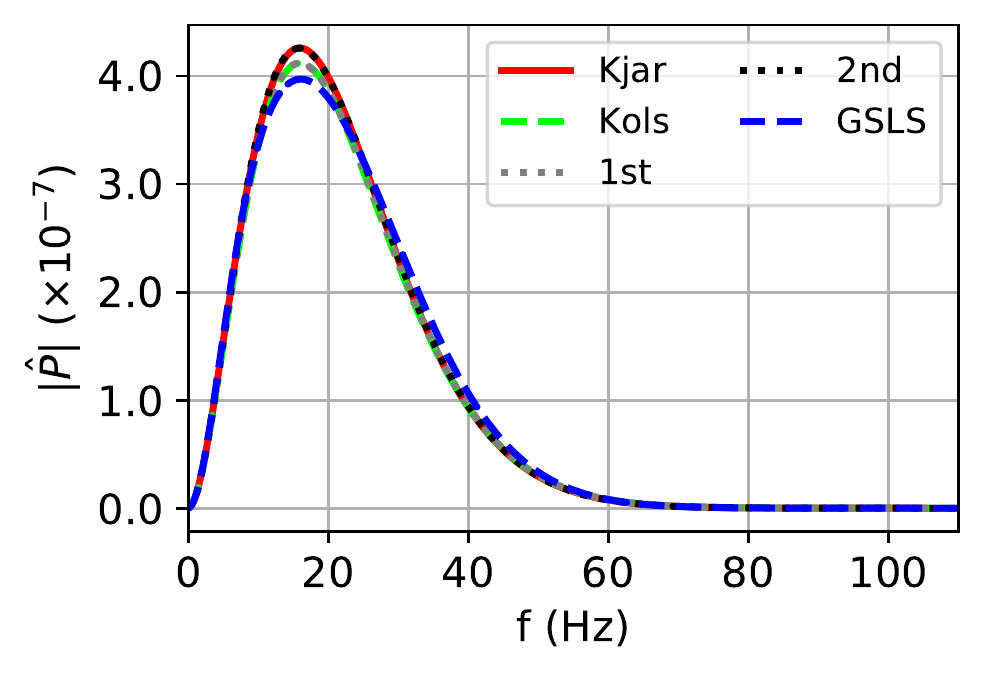}}
\subfloat[$Q_{0} = 5$]
{\includegraphics[width=0.23\textwidth]{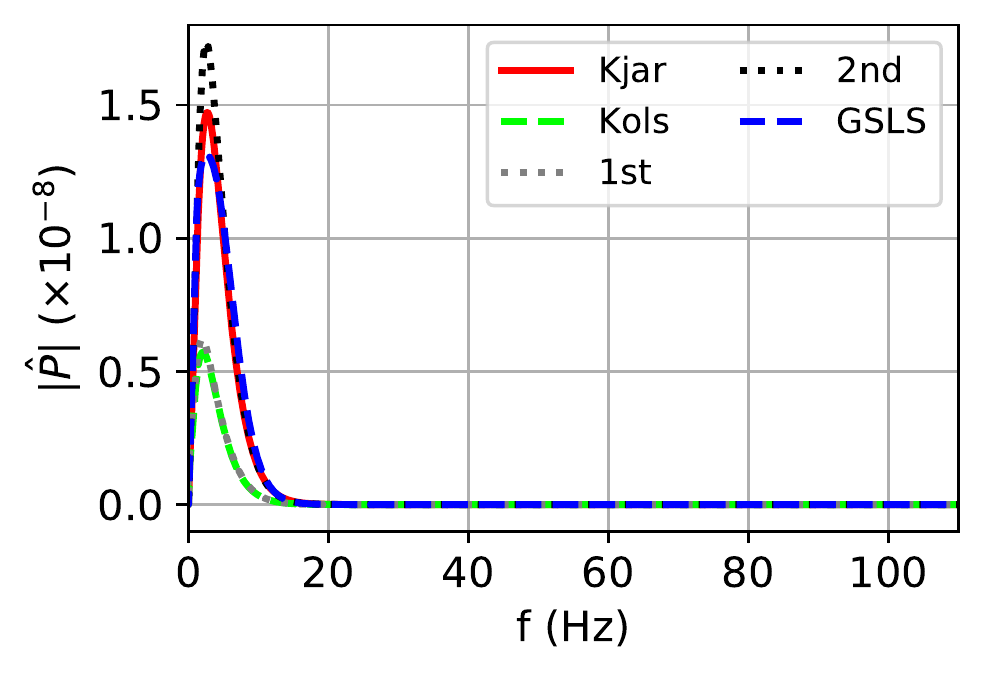}}
\caption{
Similar to Figure \ref{fig:fig8} but at a propagation distance $r=3$~km and corresponding to the waveforms in \ref{fig:fig7}. 
}
\label{fig:fig9}
\end{figure}

We next calculate the waveforms and their spectra in the dissipative models. 
Figures \ref{fig:fig6} and \ref{fig:fig7} show that the waveforms from the first- and second-order nearly constant $Q$ models and the generalized SLS model for nearly constant $Q$ fit well with those from the Kjartansson and Kolsky models, in the attenuation cases from weak to strong ($Q_{0}=100, 60$ and $30$), but not for the extremely strong attenuation case ($Q_{0} = 5$). 
Reasonable fits in the amplitude spectra of the waveforms are also observed, as illustrated in Figures \ref{fig:fig8} and \ref{fig:fig9}. This implies that even in a strongly dissipative medium ($Q_{0} = 30$) the first- and second-order nearly constant $Q$ models are good substitutes for the Kolsky and Kjartansson models and the first-order nearly constant $Q$ model is enough to satisfy the need of constant $Q$. The amplitude spectra of the waveforms from the generalized SLS model deviate slightly from those from the Kjartansson model. We now analyze the results in the extremely strong attenuation case ($Q_{0} = 5$). As illustrated in Figure \ref{fig:waves_Qc5_fc40_r1}, in the extremely strong attenuation case, the waveforms at $r=1$~km from the first- and second-order nearly constant $Q$ models fit with those from the Kolsky and Kjartansson models, respectively. Except for the late-arrival trough, the  waveform from the GSLS model typically fits that from the Kjartansson model. Comparing Figure \ref{fig:waves_Qc5_fc40_r1} with Figure \ref{fig:waves_Qc5_fc40_r3} shows that the waveform difference between the second-order nearly constant $Q$ model and the Kjartansson model varies substantially at distances of 1~km and 3~km in the extremely strong attenuation case ($Q_{0} = 5$). A similar phenomenon can be found in the waveform difference between the GSLS model and the Kjartansson model. The wave amplitude decay caused by energy absorption is proportional to the factor $\text{exp}[-\omega r/(2VQ)]$, where $V$ and $Q$ denote the phase velocity and the quality factor, respectively, at a specific frequency. This factor shows that the amplitude error caused by errors in the phase velocity and the quality factor will be amplified with propagation distance. However, Figures \ref{fig:waves_Qc5_fc40_r1} and \ref{fig:waves_Qc5_fc40_r3} show that the waveform from the first-order nearly constant $Q$ model still fits quite well with that from the Kolsky model, which means that the first-order nearly constant $Q$ model can completely replace the Kolsky model in nearly constant $Q$ dissipative wave propagation even in the extremely strong attenuation case ($Q=5$). 

An overall analysis on the dissipative waveforms (Figures \ref{fig:fig6} and \ref{fig:fig7}) shows that (1) the decay of dissipative waveforms increases with medium attenuation strength (characterized by $1/Q_{0}$) and propagation distance $r$; (2) the late-arrival trough of the dissipative waveforms is attenuated more significantly than the early-arrival trough of the dissipative waveforms. This behavior becomes more and more obvious with increase in $1/Q_{0}$ and $r$, and is distinct from the behavior of the non-dissipative waveforms; (3) the dissipative waveforms are extended in time with increasing $1/Q_{0}$ and $r$, whereas the nondissipative waveforms show no broadening with increasing distance. An overall comparison between the dissipative waveforms (Figures \ref{fig:fig8} and \ref{fig:fig9}) indicates that (1) the central frequency of the dissipative amplitude spectra, which corresponds to the peak of an amplitude spectrum, shifts towards lower frequency with increasing $1/Q_{0}$ and $r$. This phenomenon is quite obvious in the case of strong attenuation and large propagation distance. However, the central frequency of the non-dissipative amplitude spectra does not vary with $r$; (2) apart from the change in magnitude, the dissipative amplitude spectra shift to lower frequency with increasing $1/Q_{0}$ and $r$. In fact, the above phenomena associated with the dissipative waveforms and amplitude spectra result mainly from the velocity dispersion, because the quality factors for the dissipative models discussed here are either exactly independent of frequency (i.e., the Kjartansson model) or nearly independent of frequency (i.e., the Kolsky model and the first- and second-order nearly constant $Q$ models), as known already from Figure \ref{fig:fig1}.

\begin{figure}[H]
\centering
\label{fig:v_noPML}
\subfloat[Velocity]{\includegraphics[width=0.3\textwidth]{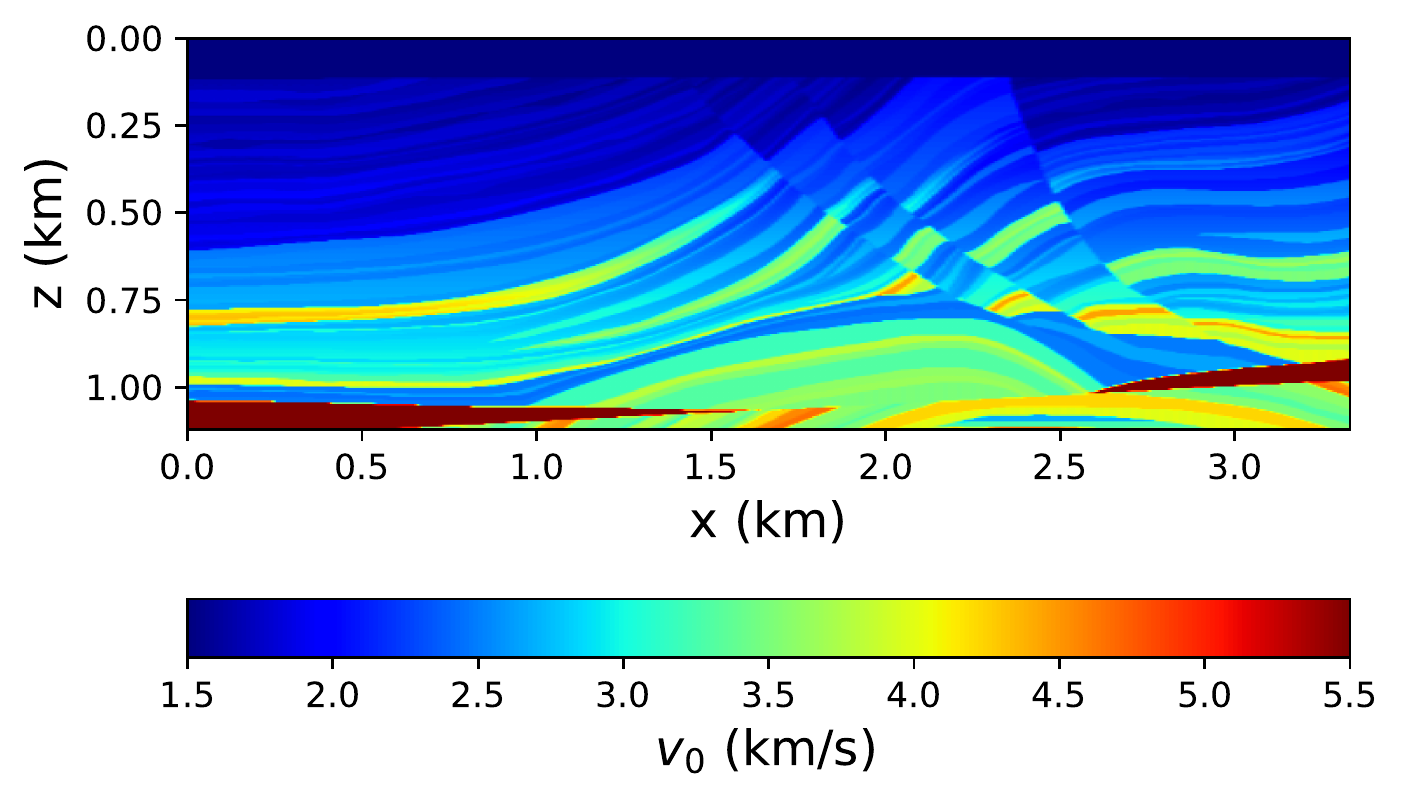}}
\qquad
\label{fig:iQ_noPML}
\subfloat[Quality factor]{\includegraphics[width=0.3\textwidth]{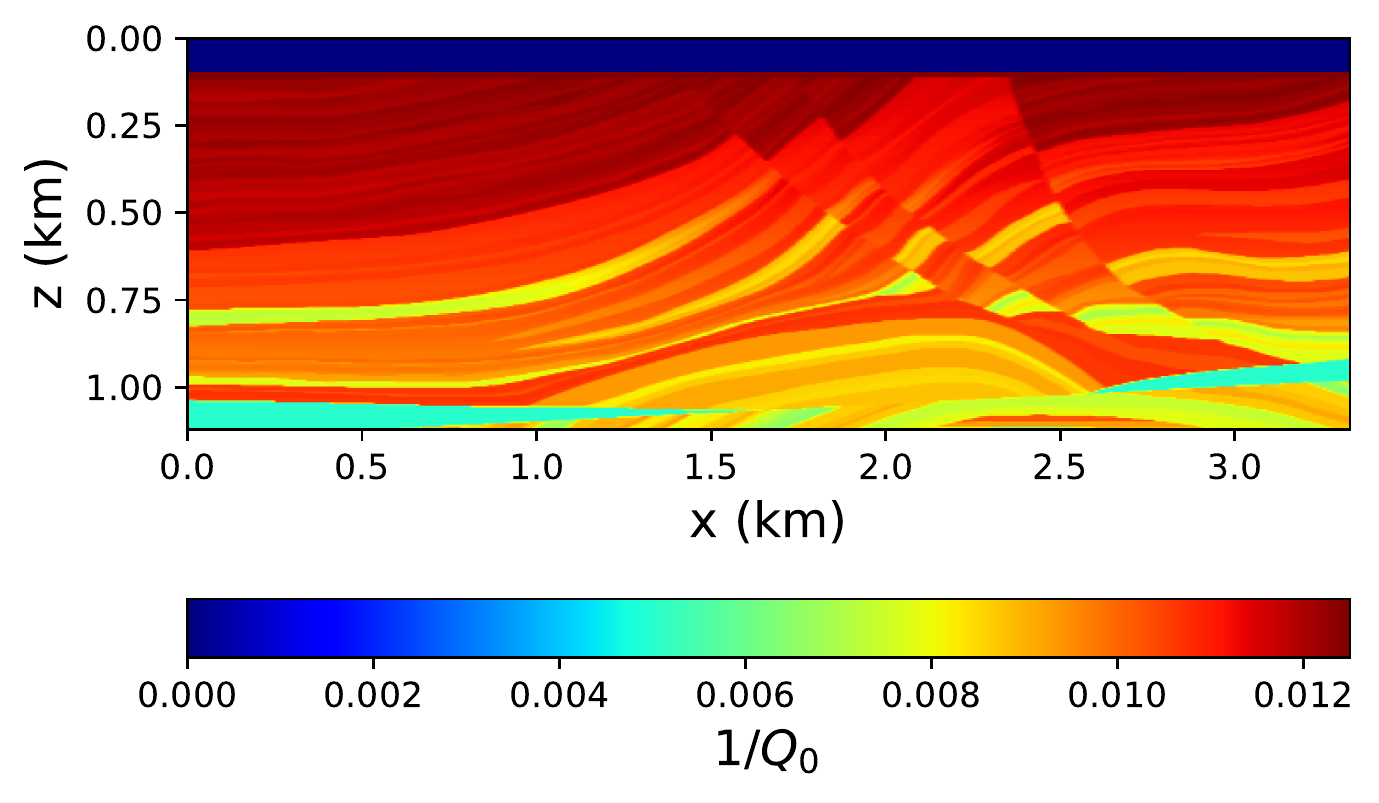}}
\caption{
The dissipative Marmousi model. The medium parameters $v_{0}=\sqrt{M_{0}/\rho}$ and $Q_{0}$ denote the reference velocity and quality factor in the Kjartansson model, where $M_{0}$ denotes the velocity corresponding to $Q_{0}=\infty$, and $\rho$ denotes density.  
}
\label{fig:fig10}
\end{figure}

\begin{figure}[H]
\centering
\subfloat[$Q_{0} = 100$]{\includegraphics[width=0.25\textwidth]{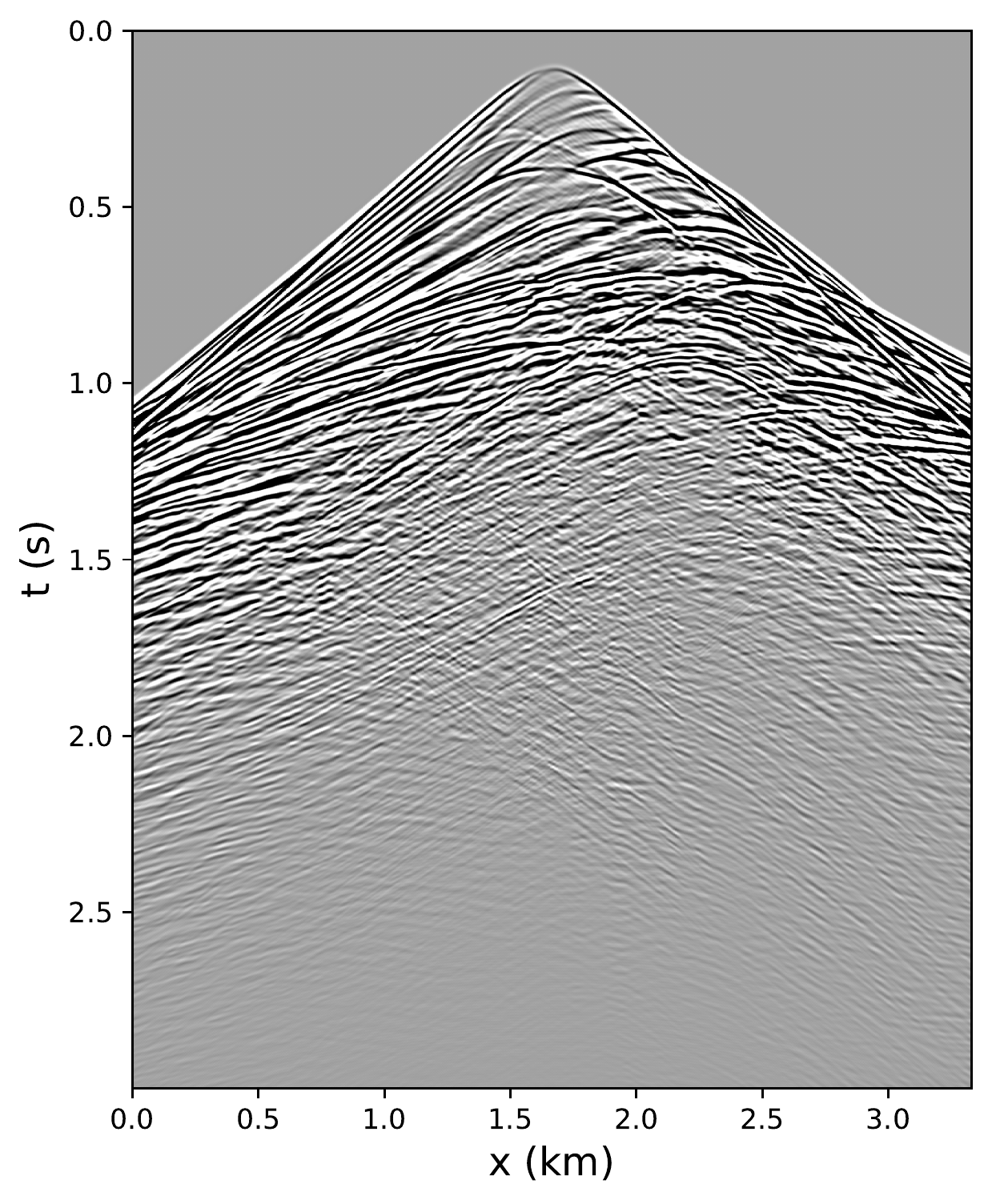}}
\quad
\subfloat[$Q_{0} = 60$]{\includegraphics[width=0.25\textwidth]{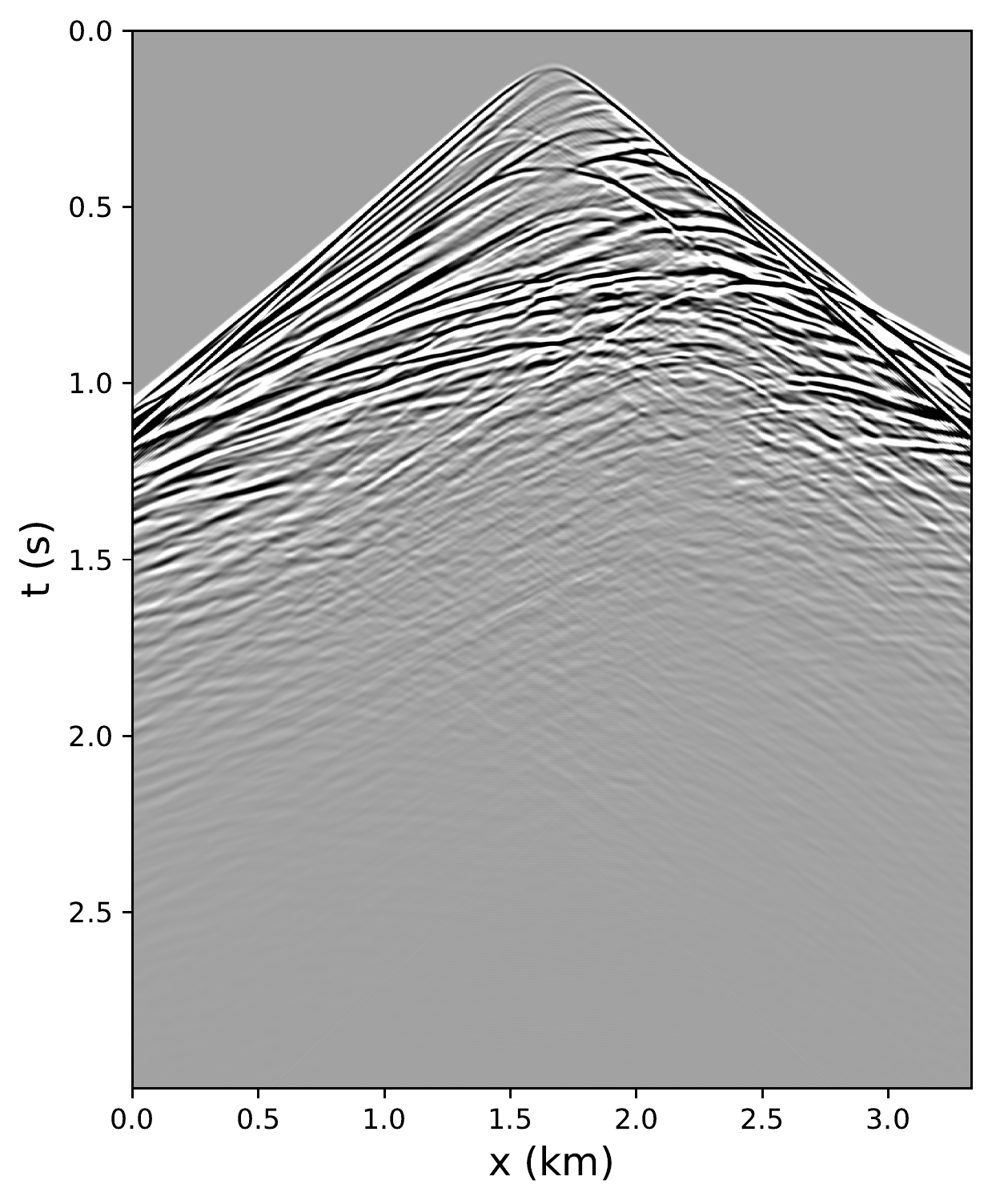}}
\quad
\subfloat[$Q_{0} = 30$]{\includegraphics[width=0.25\textwidth]{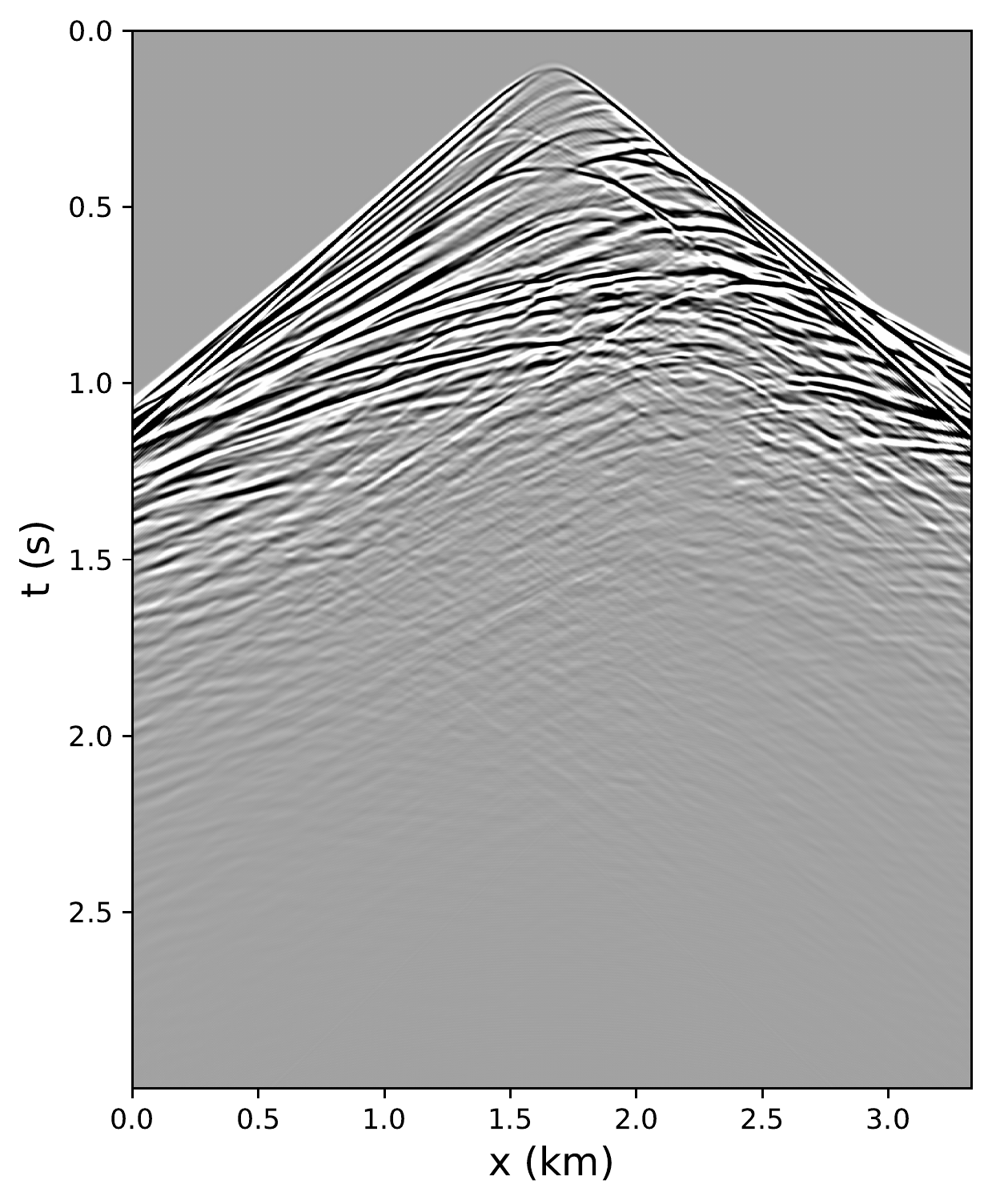}}
\caption{
A comparison between the seismograms from (a) the acoustic wave equation, (b) the viscoacoustic wave equations for the first-order nearly constant-$Q$ model, and the viscoacoustic wave equations for the second-order nearly constant $Q$ model. In these plots, the direct-arrivals are removed already. The gain function $t^{0.8}$ is applied to the seismic data, where $t$ denotes time.  
}
\label{fig:fig11}
\end{figure}

\begin{figure}[H]
\centering
\includegraphics[width=0.5\textwidth]{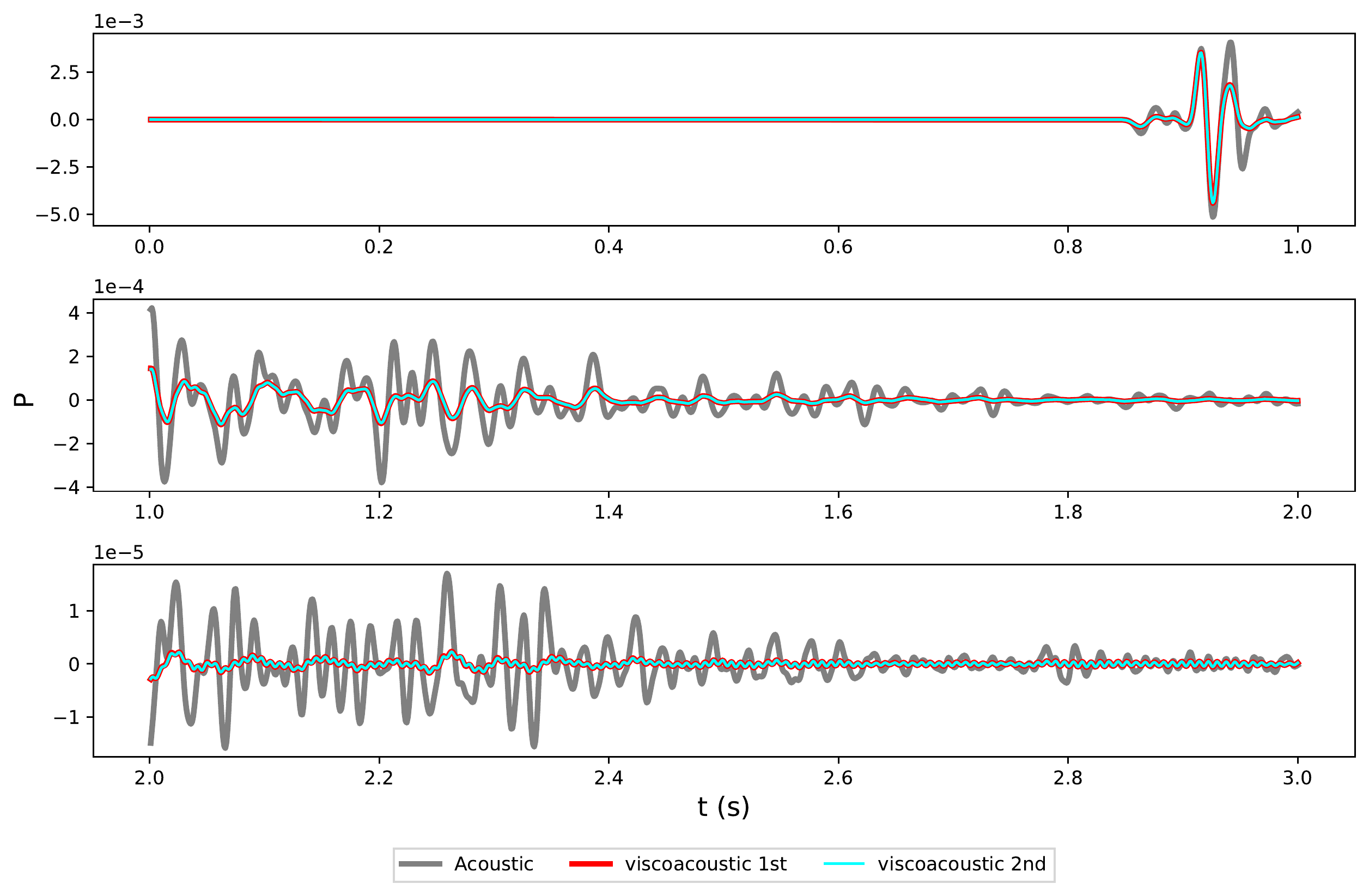}
\caption{
A comparison between the single-trace non-dissipative and dissipative seismograms recorded at $x=0.325$~km. These single-trace seismograms are extracted from the seismograms in Figure \ref{fig:fig7}. 
The black lines correspond to the acoustic waveforms. The red and cyan-dashed lines correspond to the dissipative waveforms from the viscoacoustic wave equations for the first- and second-order nearly constant-$Q$ models, respectively. The gain function $t^{0.8}$ is applied to the seismic data, where $t$ denotes time.
}
\label{fig:fig12}
\end{figure}

\begin{figure}[H]
\centering
\includegraphics[width=0.5\textwidth]{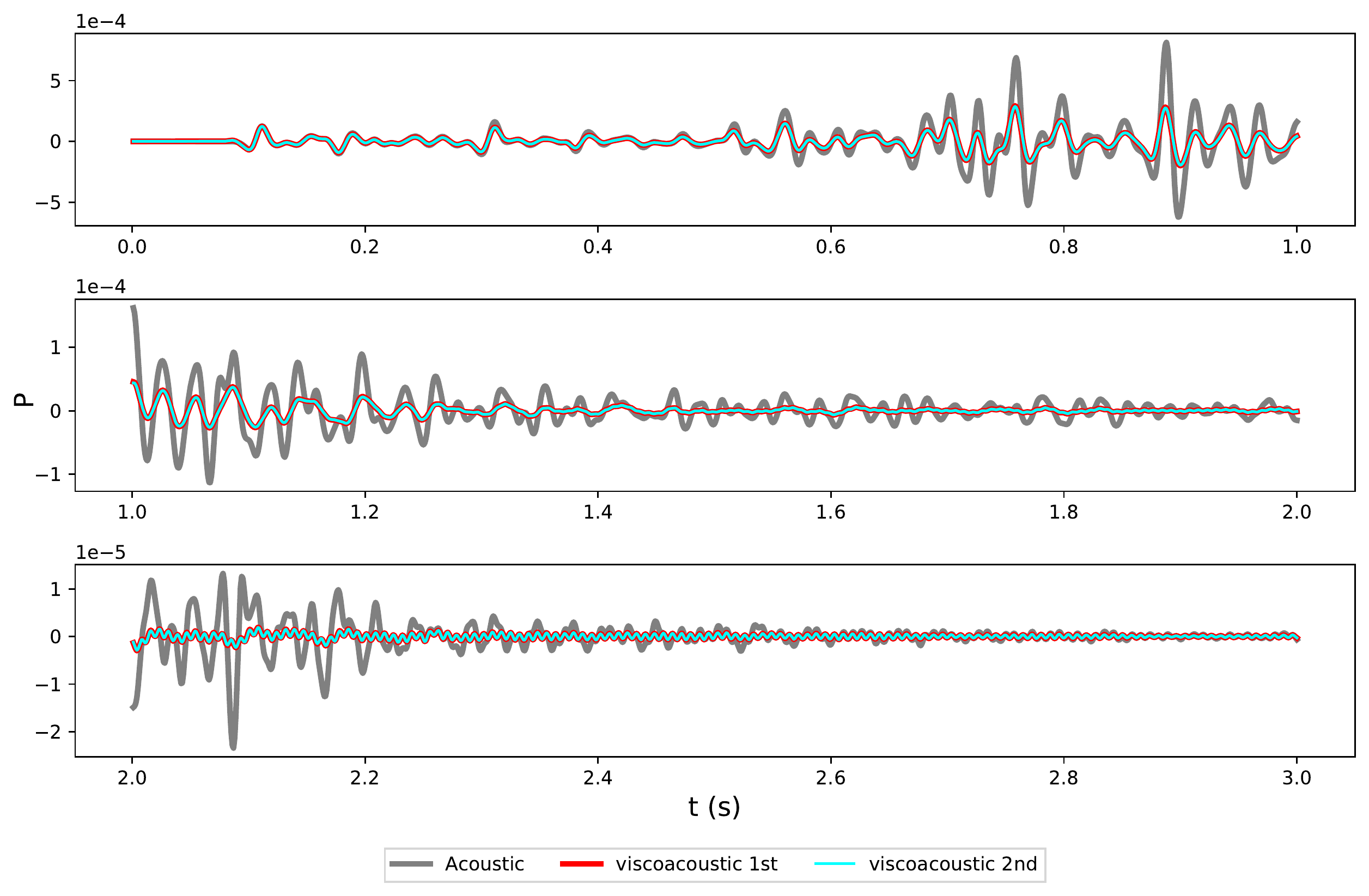}
\caption{
Similar to Figure \ref{fig:fig8}, but recorded at $x=1.665$~km (vertically above the source position). 
}
\label{fig:fig13}
\end{figure}

\begin{figure}[H]
\centering
\includegraphics[width=0.5\textwidth]{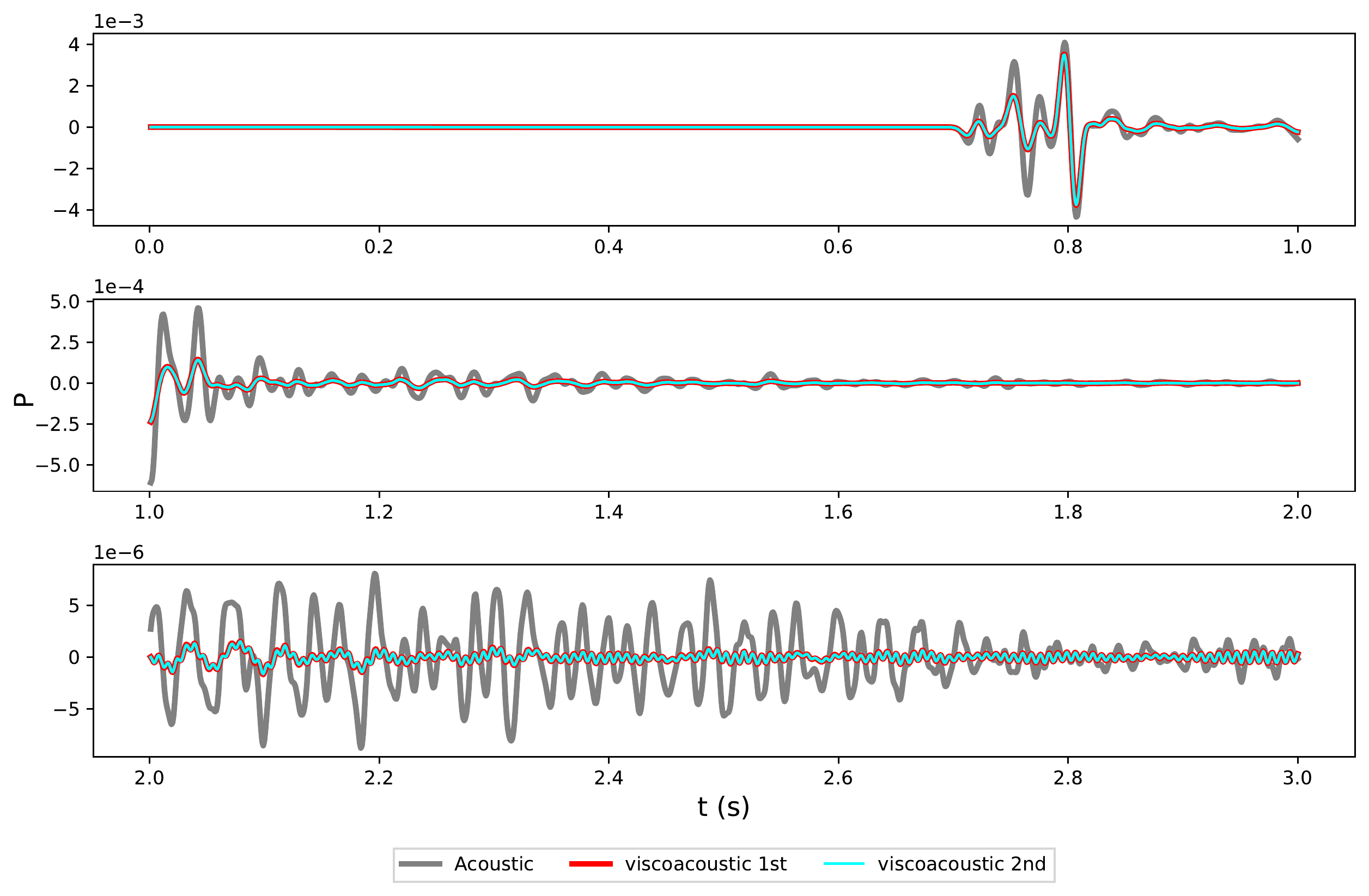}
\caption{
Similar to Figure \ref{fig:fig8}, but recorded at $x=2.825$~km. 
}
\label{fig:fig14}
\end{figure}

In the second example, we compare the reflection seismograms from the non-dissipative and dissipative Marmousi models. We implemented the finite-difference method \cite[]{carcione:2014} to solve the acoustic wave equation, and the viscoacoustic wave equations (equations  \ref{eq:VWE1} and \ref{eq:VWE2}). The second derivatives $ \partial^2/\partial x^2$ and $\partial^2/\partial z^2$ in the Laplacian operator are computed by applying the fourteenth-order staggered-grid finite-difference operator of first derivative twice. The finite-difference stencil weights can be found in Table 3 of \cite{chu.stoffa:2012}. Figure \ref{fig:fig10} shows the dissipative Marmousi model defined at the reference frequency $f_{0}=40$~Hz, where the velocity varies from $1.5$~km/s to $5.5$~km/s, and the quality factor varies from 80 to infinity. The non-dissipative Marmousi model shares the same velocity with the dissipative Marmousi model at the reference frequency. The top layer in the Marmousi model is a water layer of 100~m thickness. A point source with a 40~Hz Ricker wavelet (Figure \ref{fig:fig5}) is located in the center of the water layer, the $x$- and $z$-coordinates of which are 1.665~km and 0.05~km, respectively. The receivers are floating at the same depth as the source, and the receiver spacing is 0.05~km. As illustrated in Figure \ref{fig:fig11}, the dissipative seismograms include fewer high-frequency components than the non-dissipative seismograms. Figures \ref{fig:fig12}-\ref{fig:fig14} compare the seismograms from the acoustic wave equation, and the viscoacoustic wave equations for the first- and second-order nearly constant $Q$ models. Since this dissipative Marmousi model is only weakly lossy (the minimum quality factor is 80), the seismograms calculated by using the viscoacoustic wave equations for the first- and second-order constant $Q$ models are quite close to each other. This means that the first-order nearly constant $Q$ model and the corresponding wave equations are enough for modeling wave propagation in  weakly dissipative constant $Q$ media. Although the dissipative Marmousi model is not strongly dissipative, apart from the reflection from the water and solid interface at $z=100$~m, the reflection signals from the dissipative model are clearly weaker and flatter than the non-dissipative ones, because the velocity dispersion effect broadens the waveforms and the energy absorption effect decays the wave amplitudes in the dissipative model. 

\section{Discussion}
Regarding the novel weighting function method, a few relevant extensions and issues are discussed below.

\subsection{Extending the method to a class of dissipative models}
We used the $Q$-independent weighting function, which has a similar form as the complex modulus for the generalized SLS model, to build the first- and second-order nearly constant $Q$ models. In fact, this method can be extended to cater for a class of dissipative models, for which we want to obtain the wave equation in differential form. The complex modulus for this class of models is denoted by $M(\omega, Q_{0})$. Here, the quality factor parameter $Q_{0}$ controls the dissipation level of these models. 
The Maclaurin series expansion of the complex modulus is written as:
\begin{equation} \label{eq:M_class}
M(\omega,Q_{0})  = M_{0} \left[
1 +  \frac{a_{1}(\omega)}{Q_{0}} +  \frac{a_{2}(\omega)}{Q_{0}^2} + O\left(\frac{1}{Q_{0}^3}\right)
\right] ,
\end{equation}
where $M_{0}$ denotes the reference modulus in the nondissipative case ($Q_{0} = \infty$). Quantities $a_{1}$ and $a_{2}$ denote the first- and second-order coefficients normalized by the reference modulus, and hence they are dimensionless. We may deliberately choose the weighting functions to represent these coefficients, so that we derive the corresponding wave equations in differential form. 

Regarding the choice of the weighting function, it is well known that the Kelvin-Voigt model, the Maxwell model, the SLS model and its generalized version can yield the wave equations in differential form \cite[e.g.,][]{carcione:2014,hao.alkhalifah:2019,hao.greenhalgh:2019}. The same applies to a linear combination of these models. 

\subsection{Higher-order nearly constant \textit{Q} models}  
The proposed weighting function method was used to build the first- and second-order nearly constant $Q$ models and derive the corresponding viscoacoustic wave equations. In fact, higher-order nearly constant $Q$ models can be obtained in a similar way. We only need to retain more terms in the Maclaurin series expansion of the complex modulus for the Kjartansson model (equation \ref{eq:Mkjar_appr}) with respect to $1/Q_{0}$. The corresponding viscoacoustic wave equations can be obtained by referring to the derivation of the viscoacoustic wave equations for the second-order nearly constant $Q$ model (see Appendix D). Although higher-order models are closer to the Kjartansson model in the frequency range of interest, they lack practical value in seismology because the second-order model is sufficiently accurate in the case of quite strong attenuation. 

\subsection{Calibration of the model parameters}
We started with the Kjartansson model, which is characterized by the reference quality factor $Q_{0}$ and modulus $M_{0}$ (corresponding to $Q_{0}=\infty$), to obtain the Kolsky model and further proposed the first- and second-order nearly constant $Q$ models. However, for both nearly constant $Q$ models, the velocity and quality factor vary with frequency, although such variations with frequency are only mild in a weakly dissipative case. 
In practice, it is convenient to describe a dissipative medium by the parameters defined at the dominant frequency of a source wavelet. We define the medium parameters at the reference angular frequency $\omega_{c}=\omega_{0}$ as: $M_{c}$ (the real part of the complex  modulus) and $Q_{c}$ (the quality factor). 
For the first-order nearly constant $Q$ model, we may observe that $M_{0} = M_{c}$ and $Q_{0} = Q_{c}$ from equation \ref{eq:M1st} together with equation \ref{eq:WWRi}. For the second-order nearly constant $Q$ model, the reference modulus and quality factor can be expressed as:
\begin{align} 
\label{eq:M0}
& M_{0} = M_{c} \frac{2Q_{0}^2}{2Q_{0}^2 - 1} ,  \\
\label{eq:Q0}
& Q_{0} = \frac{1}{2}\left( Q_{c} + \sqrt{Q_{c}^2 + 2} \right) ,
\end{align}
where we have used the complex modulus $M_{c}(1 - i/Q_{c})$ to fit the modulus (equation \ref{eq:M2nd}) at the reference frequency $\omega_{0}$ and taken account of the approximation $W_{I}(\omega_{c}) \approx 1$ according to equation \ref{eq:WWRi}. 

\subsection{Extension of the proposed models to realistic (viscoelastic and/or anisotropic) media}
Although we only considered viscoacousticity in this paper, the first- and second-order nearly constant $Q$ models can be easily extended to viscoelasticity and anisotropy. The extension of the second-order nearly constant $Q$ model to its viscoelastic and anisotropic versions only requires changing the modulus and quality factor parameters to tensors. For example, the viscoelastic and anisotropic stiffness coefficients in the first-order nearly constant $Q$ model are written as  
\begin{equation} \label{eq:M2ndani}
M_{mnpq}(\omega) = M_{0,mnpq}\left \{
1 + \frac{1}{Q_{0,mnpq}} \left[W(\omega) - W_{R}(\omega_{0}) \right] 
\right \} ,
\end{equation} 
where $M_{mnpq}$ denotes the components of the complex stiffness coefficient tensor. $Q_{0,mnpq}$ denotes the components of the reference quality factor tensor, which are the ratios of the real parts of $M_{mnpq}$ to their imaginary parts. $M_{0,mnpq}$ denotes the components of the reference stiffness coefficient tensor corresponding to $Q_{0,mnpq}=\infty$. 

The viscoelastic and anisotropic versions of the second-order nearly constant $Q$ model can be obtained in a similar way. By taking account of the acoustic approximation \cite[]{hao.alkhalifah:2019}, we may obtain the viscoacoustic anisotropic (transversely isotropic and orthorhombic) versions of these two models. Furthermore, the corresponding viscoacoustic anisotropic wave equations in differential form can be obtained by referring to \cite{hao.greenhalgh:2019}.

\subsection{An alternative way of determining the weighting function}
As shown in equation \ref{eq:costF}, we build the cost function by taking account of the real and imaginary parts of the term in the square brackets in equation \ref{eq:Mkjar_appr}. Here, we provide an alternative way to determine the weighting function. It involves fitting the imaginary part of that term with the imaginary part of the weighing function. Hence, the cost function is written as: 
\begin{equation} \label{eq:costF2}
\displaystyle
\tilde{G} = \frac{1}{2(\omega_{U}-\omega_{L})} \int_{\omega_{L}}^{\omega_{U}} 
\left(
\sum_{l=1}^{L}
\frac{\omega \Delta \tau_{l}}{1+\omega^2 \tau_{\sigma l}^2} - 1
\right)^2 
d\omega .
\end{equation} 

\begin{figure}[H]
\centering
\subfloat[$Q_{0} = 100$]
{\includegraphics[width=0.23\textwidth]{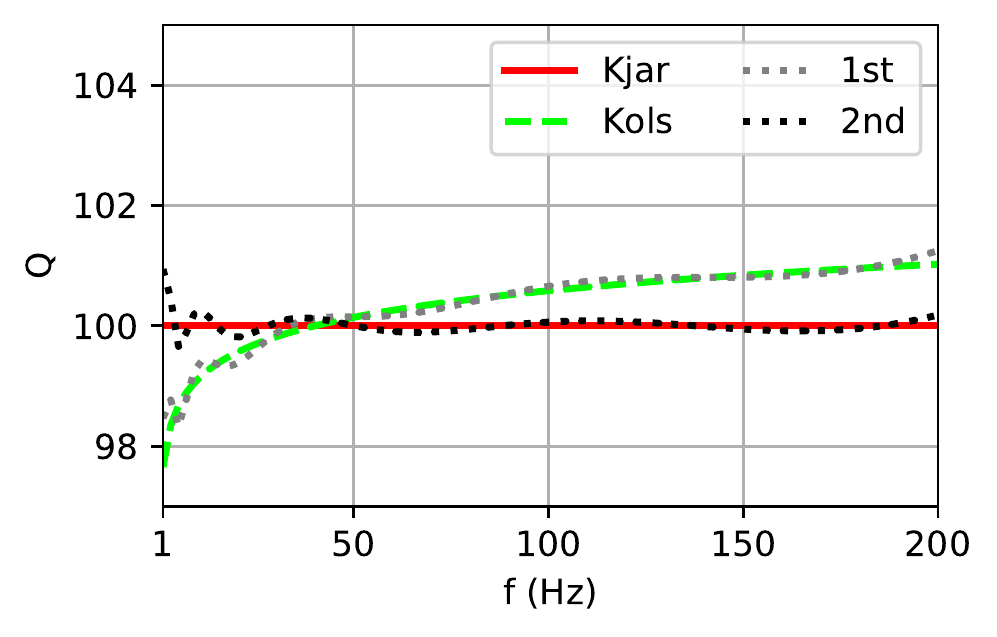}}
\subfloat[$Q_{0} = 60$]
{\includegraphics[width=0.23\textwidth]{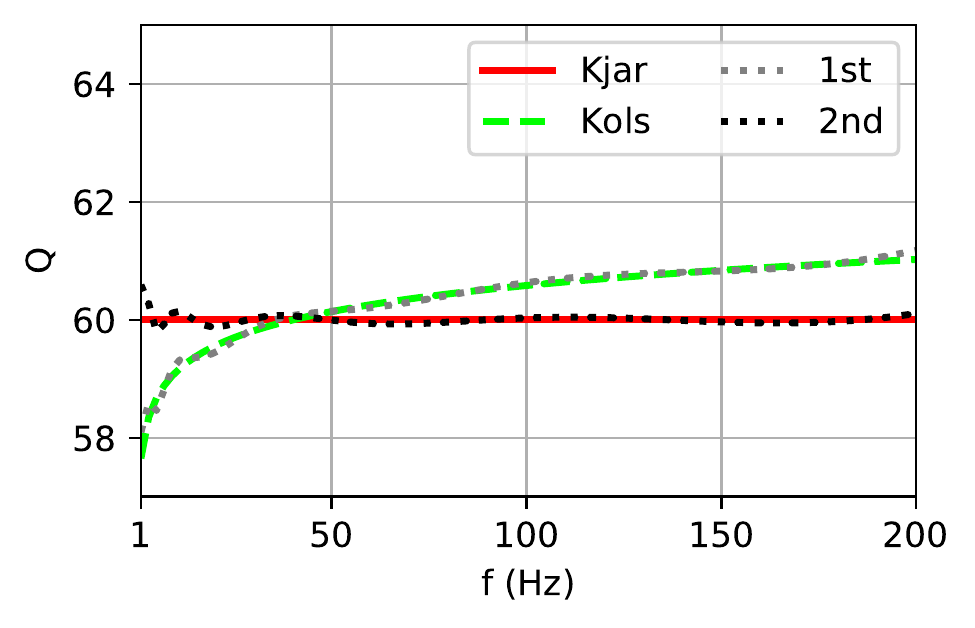}}
\subfloat[$Q_{0} = 30$]
{\includegraphics[width=0.23\textwidth]{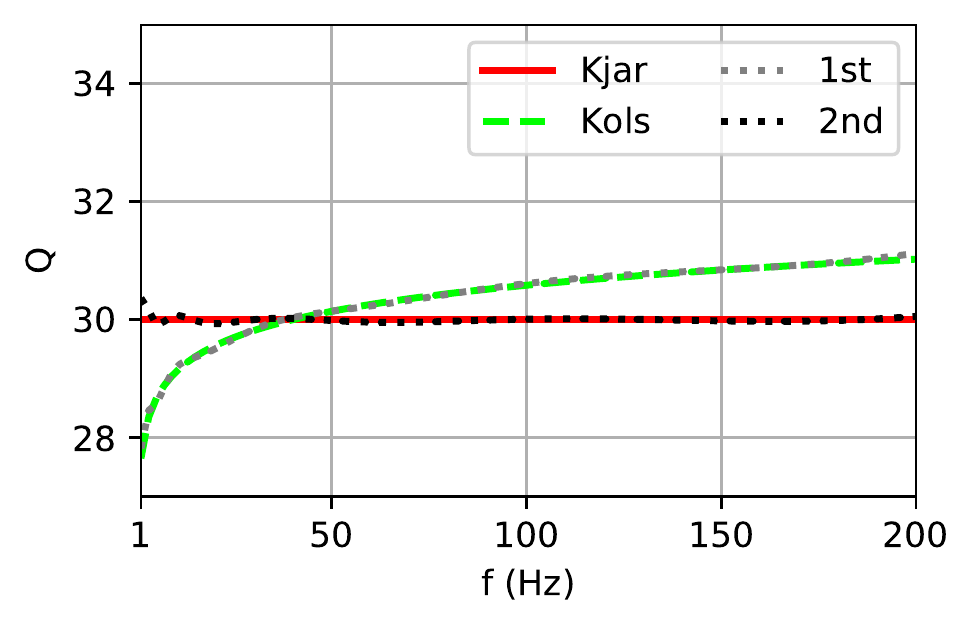}}
\subfloat[$Q_{0} = 5$]{\includegraphics[width=0.23\textwidth]{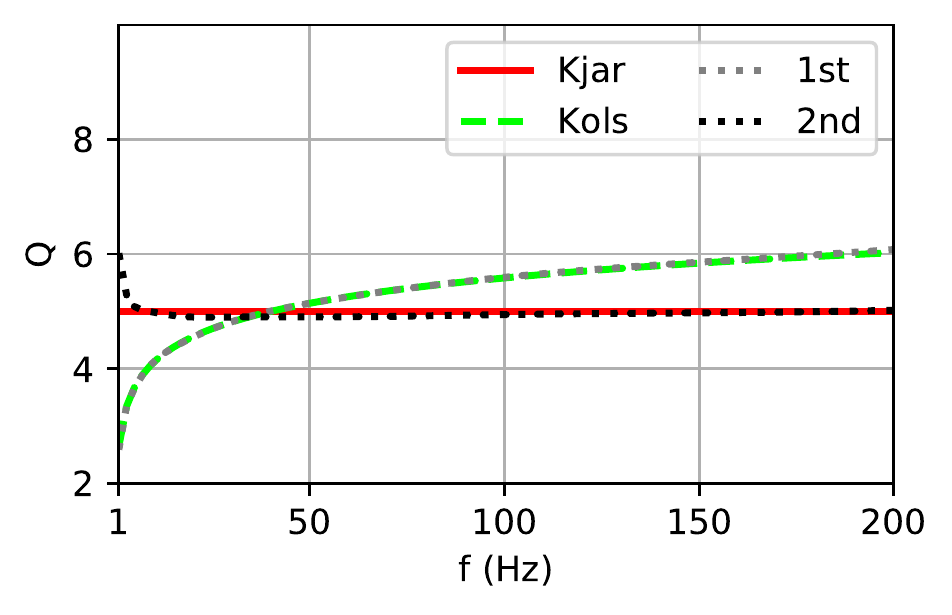}}
\caption{
The variation of quality factor with frequency. The plot order, the legend abbreviations and the model parameters are the same as those in Figure \ref{fig:fig1}.  
For the first- and second-order nearly constant $Q$ models, the weighting function here is determined by the parameters in Table \ref{tab:tabl11}, which is different from Figure \ref{fig:fig1}. 
}
\label{fig:fig15}
\end{figure}

\begin{figure}[H]
\centering
\subfloat[$Q_{0} = 100$]
{\includegraphics[width=0.23\textwidth]{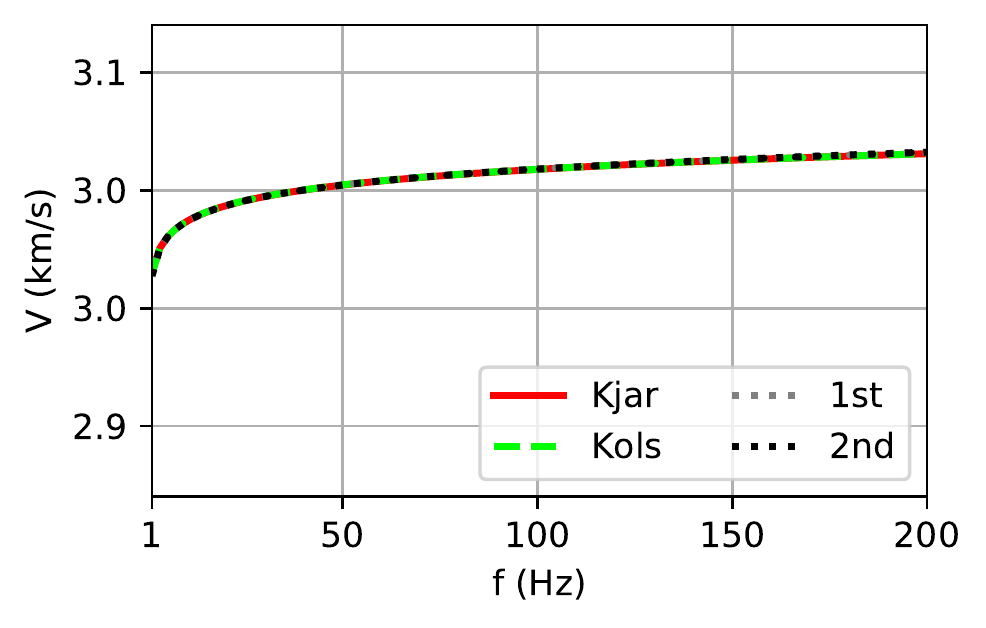}}
\subfloat[$Q_{0} = 60$]
{\includegraphics[width=0.23\textwidth]{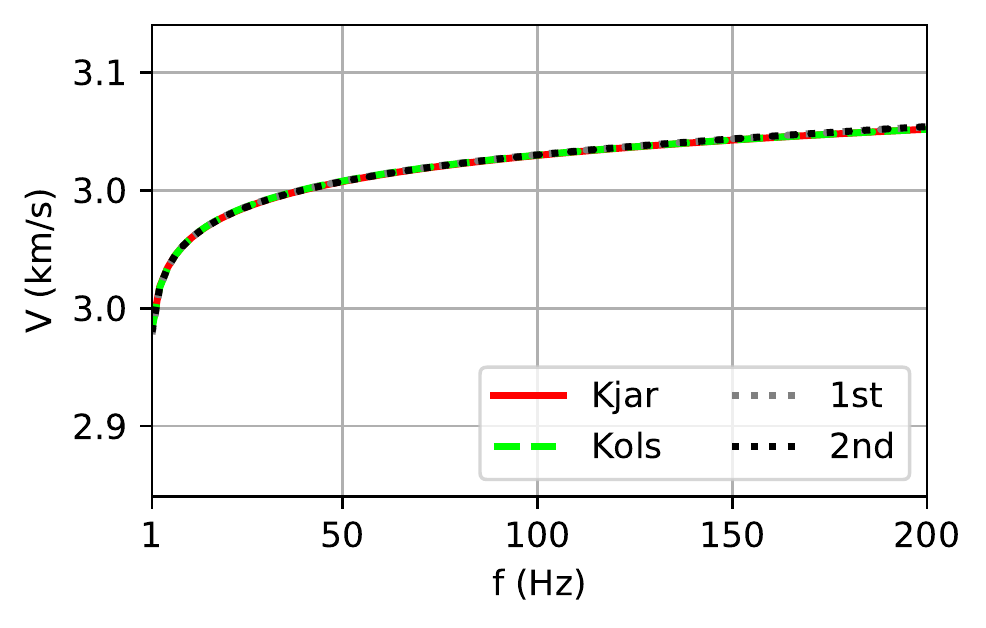}}
\subfloat[$Q_{0} = 30$]
{\includegraphics[width=0.23\textwidth]{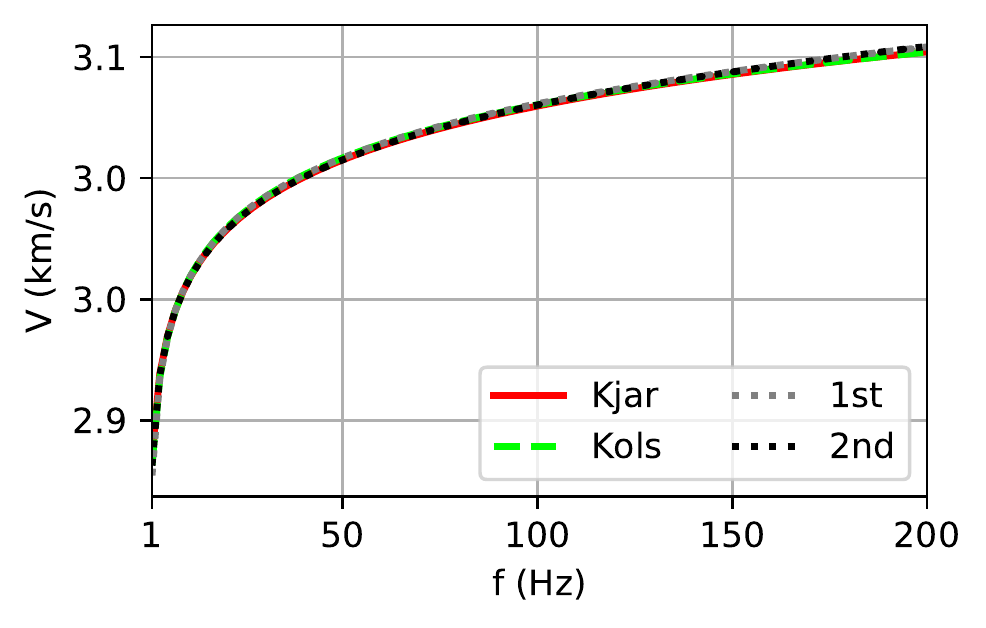}}
\subfloat[$Q_{0} = 5$]
{\includegraphics[width=0.23\textwidth]{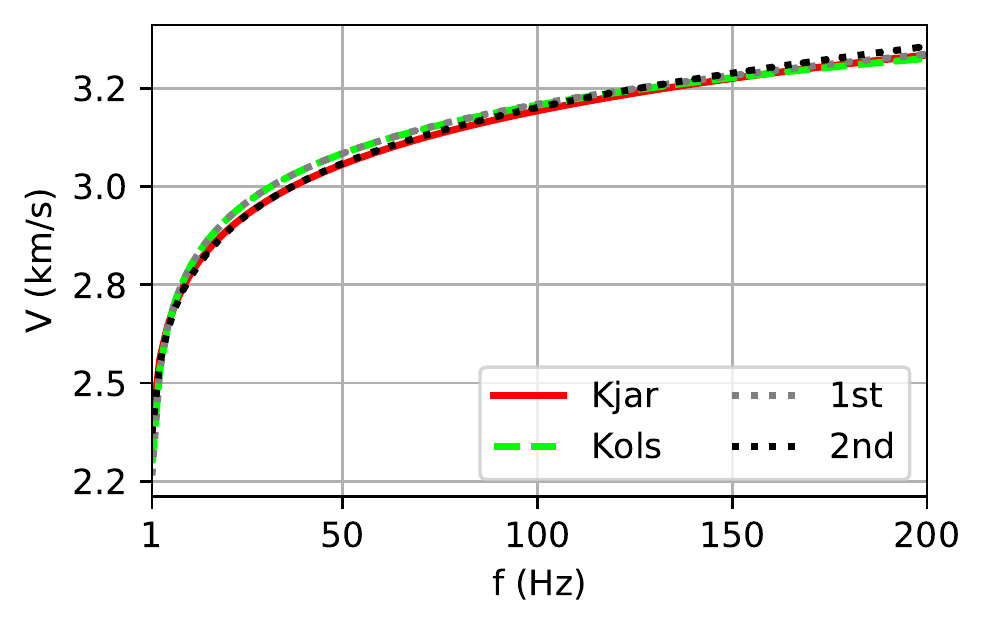}}
\caption{
The variation of phase velocity with frequency. The plot order, the legend abbreviations, the model parameters and the weighting function are the same as for Figure \ref{fig:fig15}.
}
\label{fig:fig16}
\end{figure}

We adopt the same optimization scheme as shown in the section ``$Q$-independent weighting function'' to minimize the cost function. 
Table \ref{tab:tabl11} lists the values of the optimal parameters $\tau_{\sigma l}$ and $\Delta \tau_{l} = \tau_{\epsilon l} - \tau_{\sigma l}$. We choose a reference velocity $v_{0}=3$~km/s and a density $\rho=10^3~\text{kg/m}^3$, which are the same as those used in Figures \ref{fig:fig1}-\ref{fig:fig4}. Substitution of these values into equations \ref{eq:M1st} and \ref{eq:M2nd} leads to the moduli expressions in the first- and second-order nearly constant $Q$ models. From the complex moduli we are able to compute the quality factor and the phase velocity. Figures \ref{fig:fig15} and \ref{fig:fig16} show that the quality factors and the phase velocities for the first- and second-order nearly constant $Q$ models fit well with those for the Kolsky and Kjartansson models, respectively, in quite strongly dissipative media. As mentioned already, we take account of only the imaginary part in the process of determining the weighting function, but surprisingly Figures \ref{fig:fig15} and \ref{fig:fig16} imply that the real part of the weighting function fits the real part of the term in the square brackets in equation \ref{eq:Mkjar_appr}. Comparing Figures \ref{fig:fig1} and \ref{fig:fig2} with Figures \ref{fig:fig15} and \ref{fig:fig16}, we observe that (1) only at frequencies (about $1-7$~Hz), which are quite close to the lower bound of the frequency range of interest, are the quality factors obtained using the first- and second-order nearly constant $Q$ models with the weighting function associated with the cost function \ref{eq:costF2} more accurate than those obtained from the cost function \ref{eq:costF}; (2) the velocities in the first- and second-order nearly constant $Q$ models obtained with the weighting function associated with the cost function \ref{eq:costF2} are as accurate as those from the cost function \ref{eq:costF}. In fact, our relevant numerical experience shows that it is not an accidental phenomenon that the weighting function determined by using only the imaginary part can always yield a comparable result with that determined by using both the real and imaginary parts. However, we still need to do more research to find the reason for this.  

\begin{table}[ht!]
\centering
\caption{The optimal parameters for the five-element weighting function, which are obtained by minimizing the new cost function (equation \ref{eq:costF2}), in the frequency range $[1,200]$~Hz.}
\label{tab:tabl11}
\begin{tabular}{c c c}
\toprule
$l$ & $\tau_{\sigma l}$ (s) & $\Delta \tau_{l} = \tau_{\epsilon l}-\tau_{\sigma l}$ (s) \\
\midrule
1 &	1.8230838 $\times 10^{-1}$ & 2.7518001 $\times 10^{-1}$ \\
2 &	3.2947348 $\times 10^{-2}$ & 3.0329269 $\times 10^{-2}$ \\
3 &	8.4325390 $\times 10^{-3}$ & 6.9820198 $\times 10^{-3}$ \\
4 &	2.3560480 $\times 10^{-3}$ & 1.9223614 $\times 10^{-3}$ \\
5 &	5.1033826 $\times 10^{-4}$ & 7.2390630 $\times 10^{-4}$ \\
\bottomrule   
\end{tabular}
\end{table}

\subsection{A failed nearly constant \textit{Q} model}
In this section we discuss a failed nearly constant $Q$ model which does not obey causality, which the readers should find both puzzling and instructive. The complex modulus for this model is defined as:
\begin{equation} \label{eq:Mfail}
M(\omega) = M_{0} \left[
1 - i \frac{W_{I}(\omega)}{Q_{0}}  
\right],
\end{equation} 
where $W_{I}(\omega)$ is the negative of the imaginary part of the weighting function defined in equation \ref{eq:W}, and it is required to approximate $\text{sgn}(\omega)$ in a frequency range of interest. The previous subsection showed that we can obtain the optimal parameters $\tau_{\sigma l}$ and $\tau_{\epsilon l}$ (see Table \ref{tab:tabl11}) in this case.  

Taking account of the weighting function \ref{eq:W}, equation \ref{eq:Mfail} is rewritten as:
\begin{equation} \label{eq:Mfail2}
M(\omega) = M_{0} \left \{
1 + \frac{1}{2Q_{0}}\left[ W(\omega) - W^{*}(\omega) 
\right]
\right \} ,
\end{equation}
where the superscript $*$ denotes the complex conjugate. Referring to \cite{hao.greenhalgh:2019}, the complex modulus $M(\omega) = W(\omega)$ (equation \ref{eq:W}) corresponds to the relaxation function given by:
\begin{equation} \label{eq:psi1}
\tilde{\psi}^{(1)}(t) = \sum_{l=1}^{L} 
\left[
1 - \left(1 - 
\frac{\tau_{\epsilon_l}}{\tau_{\sigma_l}} 
\right)
e^{-\frac{t}{\tau_{\sigma_l}}}
\right] H(t) .
\end{equation}
The relaxation function corresponding to the complex modulus $M(\omega) = W^{*}(\omega)$ may be derived from equation \ref{eq:Momega_gen}. We use the relation between $W(\omega)$ and $W^{*}(\omega)$, and the relation between the complex modulus $M(\omega) = W(\omega)$ (equation \ref{eq:W}) and the corresponding relaxation function $\psi(t)$ (equation \ref{eq:psi1}). It follows that the relaxation function corresponding to the complex modulus $M(\omega) = W^{*}(\omega)$ is given by:
\begin{equation} \label{eq:psi2}
\tilde{\psi}^{(2)}(t) = -\sum_{l=1}^{L}  
\left[
1 - \left(1 - 
\frac{\tau_{\epsilon_l}}{\tau_{\sigma_l}} 
\right)
e^{\frac{t}{\tau_{\sigma_l}}}
\right] H(-t) .
\end{equation}

Taking into account the correspondence relation between equations \ref{eq:const_t1} and \ref{eq:const_f1}, the relaxation function corresponding to the complex modulus \ref{eq:Mfail2} is given by:
\begin{equation} \label{eq:psifail}
\psi(t) = M_{0} \left \{
H(t) + \frac{1}{2Q_{0}} \left[
\tilde{\psi}^{(1)}(t) - \tilde{\psi}^{(2)}(t)
\right]
\right \} .
\end{equation}

As mentioned in the section ``The time- and frequency-domain constitutive relations'', the relaxation function has the physical interpretation as the stress response corresponding to a unit step function in strain, starting at $t=0$, which implies that the relaxation function is necessarily causal. However, equation \ref{eq:psifail} indicates $\psi(t) \neq 0$ for $t < 0$. It shows that the model breaks the causality requirement, i.e., no effect before a cause. 
Referring to equations \ref{eq:const_t1} and \ref{eq:phitheta}, such a relaxation function implies that the stress at the current time depends on the future values of the strain. Hence, this model is non-physical, which is why we call it ``the failed nearly-constant $Q$ model''. The failure of the model can also be verified by numerical modeling of wave propagation. By analogy with the derivation of viscoacoustic wave equations \ref{eq:VWE1}, we may derive the viscoacoustic wave equations corresponding to this non-physical model. A simple finite-difference modeling scheme applied to this model shows that the amplitude of waves increases with time, which demonstrates that it is non-physical. 

Replacing the term $W_{I}(\omega)$ in equation \ref{eq:Mfail} by $\text{sgn}(\omega)$ leads to the complex modulus for the constant $Q$ model proposed by \cite{knopoff:1956}. His model is the limiting case of the failed model, equivalent to using the limit $W_{I}$ with an infinite number of elements in equation \ref{eq:Mfail} to fit $\text{sgn}(\omega)$ for all frequencies. Similar to the analysis in the previous paragraph, the Knopoff model is incompatible with the causality condition. The violation of causality can also be found by applying the Kramers-Kronig dispersion relations \cite[e.g.][]{kronig:1926,futterman:1962,carcione:2014}, as mentioned in \cite{knopoff:1964,knopoff:1965}.

\section{Conclusions}
The newly derived first- and second-order constant $Q$ models originate from the Kolsky and Kjartansson models. The complex moduli for the first- and second-order nearly constant $Q$ models are essentially approximations to the complex modulus for the Kolsky model and the second-order Maclaurin series expansion of the complex modulus for the Kjartansson model, respectively. The key step of building such approximations is realized by the weighting function method, which chooses a weighting function to fit the $Q$-independent common coefficients in the complex modulus for the Kolsky model and the Maclaurin series expansion of the complex modulus for the Kjartansson model. The weighting function chosen in this paper is similar in form to the complex modulus for the generalized SLS model. However, the weighting function is itself dimensionless and independent of the quality factor. Determination of the weighting function requires numerically solving a nonlinear optimization problem, which is only dependent on the frequency range of interest and does not involve any model parameters of the Kolsky and Kjartansson models. 

The first- and second-order nearly constant $Q$ models are closely linked through the weighting function to three classic dissipative models: the Kolsky model, the Kjartansson model, and the generalized SLS model. The first- and second-order nearly constant $Q$ models are essentially the generalized SLS model and the quasi generalized SLS model, respectively. The first- and second-order nearly constant $Q$ models are physically distinct from the Kolsky and Kjartansson models, although their complex moduli in the frequency range of interest are quite close to those for the Kolsky and Kjartansson models. The major difference is that the moduli for the first- and second-order nearly constant $Q$ models are bounded and physically plausible for all frequencies, whereas the Kolsky and Kjartansson models become implausible as the frequency approaches zero or infinity. The advantage of these two proposed models is that they can always give rise to the dissipative wave equations in differential form, whereas the Kolsky and Kjartansson model cannot achieve this. Theoretically, these dissipative wave equations in differential form can be solved effectively by all existing time-domain wavefield numerical modeling techniques. The second-order nearly constant $Q$ model is closer to constant $Q$ than the first-order one. In reality, however, this does not mean that the second-order nearly constant $Q$ model is more plausible than the first-order one because in the Introduction we mentioned several observations of the frequency dependence of $Q$ from real data.  

\section{Acknowledgements}
Q. Hao is funded by the CPG project SF19010 at KFUPM. We are grateful to Dr. Tong Bai for providing the data in Table \ref{tab:tabl10}.

\section{Data and Materials Availability}
Data and high-quality figures are available online at \url{https://github.com/xqihao/constQ}.

\appendix
\section{The complex modulus for the Kolsky model} 
In this appendix we derive the complex modulus for the Kolsky model. 

Referring to \cite{kolsky:1956}, the phase velocity and attenuation coefficient for this model are given by:
\begin{align} 
\label{eq:V}
& V = v_{0} \left( 1 + \frac{1}{\pi Q_{0}} 
\text{ln} \left| 
\frac{\omega}{\omega_{0}} 
\right| \right) , \\
\label{eq:alpha}
& \alpha = \frac{|\omega|}{2v_{0} Q_{0}} ,
\end{align}
where $V$ and $\alpha$ denote the phase velocity and the attenuation coefficient, respectively. Quantities $v_{0}$ and $Q_{0}$ denote the reference velocity and quality factor at the reference angular frequency $\omega_{0}$, respectively. It is noteworthy that \cite{kolsky:1956} used the loss tangent (also called loss factor) $\tan \delta$ instead of $1/Q_{0}$ in the phase velocity and attenuation coefficient formulas, where $\delta$ denotes the loss angle and it measures the phase lag between the stress and the strain for a dissipative medium under the action of a steady-state stress varying sinusoidally with time. The loss tangent is identical to the ratio between the imaginary part of the complex modulus and its real part \cite[]{lakes:2009}. It follows that the quality factor and the loss tangent satisfy the relation $\tan \delta = 1/Q_{0}$, referring to the quality factor expression in equation \ref{eq:Qdef}.   

The dispersion equation is given by:
\begin{equation} \label{eq:k}
k = \frac{\omega}{v} = \frac{\omega}{V} + i \text{sgn}(\omega) \alpha,
\end{equation}
where $k$ and $v$ denote the complex wavenumber and velocity, respectively. The real and imaginary parts of the complex wavenumber are odd and even functions of frequency, respectively. The plus sign ``+'' in front of the imaginary unit is due to the sign convention in the exponential term of the Fourier transform (equation \ref{eq:Fourier}).

Substitution of equations \ref{eq:V} and \ref{eq:alpha} into equation \ref{eq:k} gives rise to the complex velocity, namely
\begin{equation} \label{eq:v}
v \approx v_{0} \left \{
1 + \frac{1}{2Q_{0}} 
\left[
\frac{2}{\pi} \text{ln}\left| \frac{\omega}{\omega_{0}} \right|
- i \text{sgn}(\omega)
\right] 
\right \} ,
\end{equation}
where we have taken into account the Maclaurin series expansion of the complex velocity with respect to $1/Q_{0}$ up to the first order. This equation can also be found in \cite{aki.richards:1980}, who summarized Azimi et al.'s (\citeyear{azimi:1968}) research work on using the Hilbert transform to obtain a pair of phase velocity and attenuation coefficient for nearly constant $Q$.

From equation \ref{eq:v}, the complex modulus for the Kolsky model is written as:
\begin{equation}
M(\omega) = M_{0} \left \{
1 + \frac{1}{Q_{0}}  
\left[
\frac{2}{\pi} \text{ln} \left|\frac{\omega}{\omega_{0}}\right| - i \text{sgn}(\omega) \right]
\right \} ,
\end{equation}
where we ignore the second- and higher-order terms with respect to $1/Q_{0}$. Quantity $M_{0} = \rho v_{0}^2$ denotes the reference modulus, where $\rho$ denotes the density.

%\appendix
%%\begin{appendix}
\section{The derivative of the cost function} 
In this appendix we provide the first partial derivatives of the cost function \ref{eq:costF} with respect to the unknown parameters $\tau_{\sigma l}$ and $\Delta \tau_{l} = \tau_{\epsilon l} - \tau_{\sigma l}$. 

The cost function is rewritten as:
\begin{equation} 
G = \frac{1}{(\omega_{U}-\omega_{L})} \int_{\omega_{L}}^{\omega_{U}} 
\left[
G_{1}(\omega) + G_{2}(\omega)
\right]
d\omega ,
\end{equation}
where $G_{1}$ and $G_{2}$ are given by:
\begin{align}
& G_{1}(\omega) = \frac{1}{2}
\left[\pi \sum_{l=1}^{L} \frac{\omega^2 \tau_{\sigma l} \Delta \tau_{l}}
{(1+\omega^2 \tau_{\sigma l}^2)^2} - 1 \right]^2, \\
& G_{2}(\omega) = \frac{1}{2}
\left(
\sum_{l=1}^{L}
\frac{\omega \Delta \tau_{l}}{1+\omega^2 \tau_{\sigma l}^2} - 1
\right)^2 .
\end{align} 
The first partial derivative of the cost function with respect to $\tau_{\sigma l}$ is given by:
\begin{equation} 
\frac{\partial G}{\partial \tau_{\sigma l}} = \frac{1}{(\omega_{U}-\omega_{L})} \int_{\omega_{L}}^{\omega_{U}} 
\left[
\frac{\partial G_{1}(\omega)}{\partial \tau_{\sigma l}} +
\frac{\partial G_{2}(\omega)}{\partial \tau_{\sigma l}}
\right]
d\omega ,
\end{equation}
with
\begin{align}
& \frac{\partial G_{1}(\omega)}{\partial \tau_{\sigma l}} =
\pi \frac{\omega^2 \Delta \tau_{l}(1-3\omega^2 \tau_{\sigma l}^2) }
{(1+\omega^2\tau_{\sigma l}^2)^3}
\left[
\pi \sum_{l=1}^{L} 
\frac{\omega^2 \tau_{\sigma l} \Delta \tau_{l}}{(1+\omega^2\tau_{\sigma l}^2)^2}
-1
\right] , \\
& \frac{\partial G_{2}(\omega)}{\partial \tau_{\sigma l}} =
-\frac{2\omega^3 \tau_{\sigma l} \Delta \tau_{l}}
{(1+\omega^2\tau_{\sigma l}^2)^2}
\left(
\sum_{l=1}^{L} 
\frac{\omega \Delta \tau_{l}}{1+\omega^2\tau_{\sigma l}^2} -1
\right) .
\end{align} 
The first partial derivative of the cost function with respect to $\Delta \tau_{l}$ is given by:
\begin{equation} 
\frac{\partial G}{\partial \Delta \tau_{l}} = \frac{1}{(\omega_{U}-\omega_{L})} \int_{\omega_{L}}^{\omega_{U}} 
\left[
\frac{\partial G_{1}(\omega)}{\partial \Delta \tau_{l}} +
\frac{\partial G_{2}(\omega)}{\partial \Delta \tau_{l}}
\right]
d\omega ,
\end{equation}
with
\begin{align}
& \frac{\partial G_{1}(\omega)}{\partial \Delta \tau_{l}} =
\pi \frac{\omega^2 \tau_{\sigma l}}
{(1+\omega^2\tau_{\sigma l}^2)^2}
\left[
\pi \sum_{l=1}^{L} 
\frac{\omega^2 \tau_{\sigma l} \Delta \tau_{l}}{(1+\omega^2\tau_{\sigma l}^2)^2}
-1
\right] , \\
& \frac{\partial G_{2}(\omega)}{\partial \Delta \tau_{l}} = 
\frac{\omega}{1+\omega^2\tau_{\sigma l}^2}
\left(
\sum_{l=1}^{L} 
\frac{\omega \Delta \tau_{l}}{1+\omega^2\tau_{\sigma l}^2} -1
\right) .
\end{align}

\section{The relaxation and creep functions for the first- and second-order nearly-constant \textit{Q} models } 
In this appendix we derive the relaxation and creep functions for the first- and second-order nearly-constant $Q$ models. 

\subsection{Two special cases}
As a preliminary, we analyze two special cases, where we ignore the dimensions of the complex modulus, the relaxation function and the creep function. In the first case, we analyze the relation between the complex modulus and relaxation function. Equations \ref{eq:Momega} suggests that a linear combination of two complex moduli leads to the same combination of the corresponding relaxation functions. Complex modulus $M(\omega) = W(\omega)$ can be viewed as a sum of the complex moduli for $L$ SLS elements with a relaxed modulus of unity, where $W(\omega)$ is the weighting function given in equation \ref{eq:W}. The corresponding relaxation functions can be found in \cite{carcione:2014}. In addition, equation \ref{eq:Momega} can verify that a constant (frequency-independent) modulus corresponds to the relaxation function equal to the product of this constant and the Heaviside function. 
Let $\zeta(t)$ as an intermediate variable denote the relaxation function corresponding to the complex modulus $M(\omega) = W(\omega) - W_{R}(\omega_{0})$. It follows that $\zeta(t)$ is given by
\begin{equation} \label{eq:zeta}
\zeta(t) = - W_{R}(\omega_{0}) H(t) 
+ 
\sum_{l=1}^{L} 
\left[
1 - \left(1 - 
\frac{\tau_{\epsilon_l}}{\tau_{\sigma_l}} 
\right)
e^{-\frac{t}{\tau_{\sigma_l}}}
\right] H(t). 
\end{equation}
In the second case, we analyze the relation between the complex compliance and the creep function. Equation \ref{eq:Comega} is mathematically similar to equation \ref{eq:Momega}. Hence, the above analysis also applies to the relaxation between the complex compliance and the creep function. If the complex compliance is taken as $J(\omega) = W(\omega) - W_{R}(\omega_{0})$, we conclude that the creep function is identical to $\zeta(t)$ in equation \ref{eq:zeta}.

\subsection{The first- and second-order nearly constant-$Q$ models}
We next analyze the first-order nearly constant-$Q$ model. As shown in equation \ref{eq:M1st}, the complex modulus for the first nearly constant-$Q$ model is given by:
\begin{equation} \label{eq:M1st_v2}
M(\omega) = M_{0} \left \{
1 + \frac{1}{Q_{0}} \left[W(\omega) - W_{R}(\omega_{0}) \right]
\right \} .
\end{equation}
We use the correspondence relation between the time- and frequency-domain constitutive equations \ref{eq:const_t1} and \ref{eq:const_f1}, and the result in the previous subsection ``Two special cases''. Finally, the relaxation function for the first order nearly constant-$Q$ model is given by:
\begin{equation} \label{eq:psi1_ele}
\psi(t) = M_{0} \left[
H(t) + \frac{1}{Q_{0}} \zeta(t)
\right] .
\end{equation}
Taking the inverse of the complex modulus \ref{eq:MC}, the complex compliance for the first-order nearly constant-$Q$ model is written as:
\begin{equation} 
J(\omega) = J_{0} 
\left \{
1 + \frac{1}{Q_{0}} \left[W(\omega) - W_{R}(\omega_{0}) \right]
\right \}^{-1} ,
\end{equation}
where $J_{0}=1/M_{0}$ denotes the reference compliance.

The Maclaurin series expansion of the complex compliance with respect to $1/Q_{0}$ is written as:
\begin{equation}
\frac{J(\omega)}{J_{0}} = 1 + 
\sum_{n=1}^{\infty} \frac{(-1)^{n}}{Q_{0}^{n}}
\left[W(\omega) - W_{R}(\omega_{0}) \right]^{n} .
\end{equation}
Here, we already account for the condition $|W(\omega) - W_{R}(\omega_{0})| < Q_{0}$. In fact, this inequality is valid even for the extremely strong attenuation case (e.g. $Q_{0} = 5$), from the definition of $W(\omega)$ (equation \ref{eq:W}) with the relaxation times shown in Tables \ref{tab:tabl1}-\ref{tab:tabl8}. 

We take into account the correspondence relation between the time-domain constitutive equation \ref{eq:const_t2} and the frequency-domain constitutive equation \ref{eq:const_f2}. We also use the result in subsection ``Two special cases''. The creep function for the first-order model is given by:
\begin{equation} \label{eq:chi1}
\chi(t) = J_{0} \left[
H(t) + 
\sum_{n=1}^{\infty} \frac{(-1)^{n}}{Q_{0}^n} \zeta^{\langle n \rangle}(t)  
\right] ,
\end{equation}
where $\zeta^{\langle n \rangle}$ is defined as:
\begin{equation}
\zeta^{\langle n \rangle}(t) =
\begin{cases}
\underbrace{\zeta(t) \odot \zeta(t) \cdots \odot \zeta(t)}_{n}, & \text{if } n > 1, \\
\zeta(t), & \text{if } n = 1 .
\end{cases}
\end{equation}

We finally analyze the second order nearly constant-$Q$ model. As shown in equation \ref{eq:M2nd}, its complex modulus is given by:
\begin{equation} \label{eq:M2nd_ele}
M(\omega) = M_{0} \left \{
1 + \frac{1}{Q_{0}} \left[W(\omega) - W_{R}(\omega_{0}) \right] 
+ \frac{1}{2Q_{0}^2} 
\left[W(\omega) - W_{R}(\omega_{0}) \right]^2
\right \}.
\end{equation}
Following the same method used for the first-order nearly constant-$Q$ model, we derive the relaxation function for the second-order nearly constant-$Q$ model:
\begin{equation} \label{eq:psi2_tmp}
\psi(t) = M_{0} \left \{
H(t) + \frac{1}{Q_{0}} \zeta(t) 
+ \frac{1}{2Q_{0}^2} \zeta^{\langle 2 \rangle}(t) 
\right \}.
\end{equation}
From equation \ref{eq:M2nd_ele}, the complex compliance for the second-order nearly constant-$Q$ model is written as:
\begin{equation} \label{eq:C2} 
J(\omega) = J_{0} \left \{
1 + \frac{1}{Q_{0}} \left[W(\omega) - W_{R}(\omega_{0}) \right] 
+ \frac{1}{2Q_{0}^2} 
\left[W(\omega) - W_{R}(\omega_{0}) \right]^2
\right \}^{-1} .
\end{equation}
We expand the complex compliance \ref{eq:C2} into a Maclaurin series with respect to $1/Q_{0}$. We imitate the derivation of the creep function \ref{eq:chi1}. Finally, the creep function for the second-order nearly constant-$Q$ model is given by:
\begin{equation}
\begin{aligned}
\chi(t) & = J_{0} H(t) + 
J_{0} \sum_{n=1}^{\infty} 
\frac{(-1)^n}{2^{2n} Q_{0}^{4n}} \zeta^{\langle 4n \rangle}(t) 
+ J_{0} \sum_{n=0}^{\infty} 
\frac{(-1)^{n+1}}{2^{2n} Q_{0}^{4n+1}} \zeta^{\langle 4n+1 \rangle}(t)  \\
& + J_{0} \sum_{n=0}^{\infty} 
\frac{(-1)^n}{2^{2n+1} Q_{0}^{4n+2}} \zeta^{\langle 4n+2 \rangle}(t) .
\end{aligned}
\end{equation}

\section{Derivation of the viscoacoustic wave equations for the nearly constant-$Q$ model}
In this appendix, we adopt the first of the frequency-domain methods in \cite{hao.greenhalgh:2019} to derive the viscoacoustic wave equations for the first- and second-order nearly constant-$Q$ models. 

Taking into account the correspondence relation between equations \ref{eq:const_t1} and \ref{eq:const_f1}, the Fourier transform of the viscoacoustic wave equation for a general dissipative medium is written as:
\begin{equation} \label{eq:VAWEomega}
- \omega^2 \hat{P} = \frac{M}{\rho} \nabla^2 \hat{P} 
+ \hat{S},
\end{equation}
where $\rho$ and $M$ denotes the density and the complex modulus, respectively. $\hat{P}$ denotes pressure field in the frequency domain. $\hat{S}$ denotes the source term in the frequency domain.

As illustrated in equations \ref{eq:WWr}-\ref{eq:h}, the term $W(\omega)-W(\omega_{0})$ is written as:
\begin{equation} \label{WWR_tmp}
W(\omega) - W_{R}(\omega_{0}) = g - h(\omega)  ,
\end{equation}
with
\begin{align}
& g = \sum_{l=1}^{L} 
\frac{\frac{\tau_{\epsilon l}}{\tau_{\sigma l}}-1}
{1+\omega_{0}^2\tau_{\sigma l}^2} , \\
& h(\omega) = \sum_{l=1}^{L}
\frac{\frac{\tau_{\epsilon l}}{\tau_{\sigma l}}-1}
{1-i\omega\tau_{\sigma l}} .
\end{align}

Hence, the complex modulus (equation \ref{eq:M1st}) for the first-order nearly constant-$Q$ model is rewritten as:
\begin{equation} \label{eq:M1st_hser}
M(\omega) = M_{0} \left(1 + \frac{g}{Q_{0}} \right) -
M_{0} \frac{h(\omega)}{Q_{0}} .
\end{equation}

Substituting the complex modulus into the wave equation \ref{eq:VAWEomega} and then introducing the auxiliary variables yields  
\begin{equation} \label{eq:VWE1stomega}
\begin{split}
- & \omega^2 \hat{P} = v_{U}^2 \nabla^2 \hat{P} - v_{H}^2 \sum_{l=1}^{L} \hat{r}_{l} 
+ \hat{S}, \\
& \hat{r}_{l} =
\frac{\frac{\tau_{\epsilon l}}{\tau_{\sigma l}}-1}
{1-i\omega\tau_{\sigma l}} \nabla^2 \hat{P} ,
\end{split}
\end{equation}
with 
\begin{align}
& v_{U}^2 = v_{0}^2 \left(
1 + \frac{g}{Q_{0}} \right), \\
& v_{H}^2 = \frac{v_{0}^2}{Q_{0}}, 
\end{align}
where quantity $v_{0} = \sqrt{M_{0}/\rho}$ denotes the reference velocity for the Kjartansson model. Quantity $v_{U}$ denotes the unrelaxed velocity for the first-order nearly constant $Q$ model, corresponding to $\omega=\infty$.  Quantity $v_{H}$ denotes the velocity corresponding to the coefficient in front of $h(\omega)$ in equation \ref{eq:M1st_hser}. Quantity $\hat{r}_{l}$ denotes the auxiliary variable in the frequency domain.

The inverse Fourier transform of equations \ref{eq:VWE1stomega} gives rise to the viscoacoustic wave equation for the first order nearly constant-$Q$ model, namely
\begin{equation}
\begin{split}
& \frac{\partial^2 P}{\partial t^2} = v_{U}^2 \nabla^2 P
- v_{H}^2 \sum_{l=1}^{L} r_{l} 
+ S , \\
& \frac{\partial r_{l}}{\partial t} = s_{l} \nabla^2 P
- \frac{1}{\tau_{\sigma l}} r_{l} ,
\end{split}
\end{equation}
where $s_{l}$ is given by:
\begin{equation}
s_{l} = \frac{1}{\tau_{\sigma l}} 
\left(\frac{\tau_{\epsilon l}}{\tau_{\sigma l}} - 1 \right) . 
\end{equation}

Using equation \ref{WWR_tmp}, the complex modulus \ref{eq:M2nd} for the second-order nearly constant-$Q$ model is rewritten as:
\begin{equation} \label{eq:M2nd_hser}
M(\omega) = M_{0} \left(1 + \frac{g}{Q_{0}} + \frac{g^2}{2Q_{0}^2} \right) 
- \frac{M_{0}}{Q_{0}} \left(1 + \frac{g}{Q_{0}} \right) h(\omega) 
+ \frac{M_{0}}{2Q_{0}^2} h^2 (\omega).
\end{equation}
Substituting it into the wave equation \ref{eq:VAWEomega} and then introducing the auxiliary variables leads to the following equations:
\begin{equation} \label{eq:VWE2ndomega}
\begin{split}
- &\omega^2 \hat{P} = \tilde{v}_{U}^2 \nabla^2 \hat{P} 
- \tilde{v}_{H1}^2 \sum_{l=1}^{L} \hat{r}_{l}^{(1)} 
+ \tilde{v}_{H2}^2 \sum_{l=1}^{L} \hat{r}_{l}^{(2)} 
+ \hat{S}, \\
& \hat{r}_{l}^{(1)} = 
\frac{\frac{\tau_{\epsilon l}}{\tau_{\sigma l}}-1}
{1-i\omega\tau_{\sigma l}} \nabla^2 \hat{P} , \\
& \hat{r}_{l}^{(2)} =  
\frac{\frac{\tau_{\epsilon l}}{\tau_{\sigma l}}-1}
{1-i\omega\tau_{\sigma l}} \sum_{l=1}^{L} \hat{r}_{l}^{(1)} ,
\end{split}
\end{equation}
with
\begin{align}
& \tilde{v}_{U}^2 = v_{0}^2 \left(
1 + \frac{g}{Q_{0}} + \frac{g^2}{2Q_{0}^2} \right), \\
& \tilde{v}_{H1}^2 = \frac{v_{0}^2}{Q_{0}} \left(1 + \frac{g}{Q_{0}} \right), \\
& \tilde{v}_{H2}^2 = \frac{v_{0}^2}{2Q_{0}^2} ,
\end{align}
where quantity $\tilde{v}_{U}$ denotes the unrelaxed velocity for the second-order nearly constant $Q$ model, corresponding to $\omega=\infty$. Quantities $\tilde{v}_{H1}$ and $\tilde{v}_{H2}$ denote the velocities corresponding to the coefficients in front of $h(\omega)$ and $h^2(\omega)$ in equation \ref{eq:M2nd_hser}, respectively. Quantities $\hat{r}_{l}^{(1)}$ and $\hat{r}_{l}^{(2)}$ are the frequency-domain auxiliary variables.

The inverse Fourier transform of equations \ref{eq:VWE2ndomega} yields the viscoacoustic wave equation for the second-order nearly constant-$Q$ model, namely
\begin{equation}
\begin{split}
& \frac{\partial^2 P}{\partial t^2} = \tilde{v}_{U}^2 \nabla^2 P
- \tilde{v}_{H1}^2 \sum_{l=1}^{L} r_{l}^{(1)} 
+ \tilde{v}_{H2}^2 \sum_{l=1}^{L} r_{l}^{(2)} 
+ S , \\
& \frac{\partial r_{l}^{(1)}}{\partial t} = s_{l} \nabla^2 P
- \frac{1}{\tau_{\sigma l}} r_{l}^{(1)} , \\
& \frac{\partial r_{l}^{(2)}}{\partial t} = s_{l} \sum_{l=1}^{L} r_{l}^{(1)} 
- \frac{1}{\tau_{\sigma l}} r_{l}^{(2)} .
\end{split}
\end{equation}

\bibliographystyle{./macros/elsarticle-num}
\bibliography{./refs/refs20210121,./refs/qi_refs20191026}

\end{document}